\documentstyle[12pt,twoside,epsfig]{article}

\def\be{\begin{equation}}
\def\ee{\end{equation}}
\def\bea{\begin{eqnarray}}
\def\ena{\end{eqnarray}}  \def\eea{\end{eqnarray}}
\def\ba{\begin{array}}
\def\ea{\end{array}}
\def\unit{1 \hskip-.3em \raise2pt\hbox{$ \scriptstyle |$ } }

\def\a{\alpha}
\def\b{\beta}
\def\c{\chi}
\def\d{\delta}
\def\e{\epsilon}           % Also, \varepsilon
\def\g{\gamma}
\def\h{\eta}

\def\j{\psi}
\def\l{\lambda}
\def\m{\mu}
\def\n{\nu}
\def\o{\omega}  \def\w{\omega}
\def\p{\pi}                % Also, \varpi
\def\q{\theta}                    %     \vartheta
\def\r{\rho}                                     %     \varrho
\def\s{\sigma}                                   %     \varsigma
\def\t{\tau}

\def\x{\xi}
\def\z{\zeta}
\def\D{\Delta}

\def\G{\Gamma}

\def\del{\partial}

\def\ch{{\cal H}}

\def\cl{{\cal L}}

\catcode`@=11

\def\un#1{\relax\ifmmode\@@underline#1\else
$\@@underline{\hbox{#1}}$\relax\fi}

{\catcode`\'=\active \gdef'{{}^\bgroup\prim@s}}

\def\magstep#1{\ifcase#1 \@m\or 1200\or 1440\or 1728\or 2074\or 2488\or
       2986\fi\relax}   % to include magstep6

\catcode`@=12

\def\bop#1{\setbox0=\hbox{$#1M$}\mkern1.5mu
	\vbox{\hrule height0pt depth.04\ht0
	\hbox{\vrule width.04\ht0 height.9\ht0 \kern.9\ht0
	\vrule width.04\ht0}\hrule height.04\ht0}\mkern1.5mu}
\def\pa{\partial}                              % curly d
\def\>{\rangle} %right angle
\def\<{\langle} %left angle
\def\Dsl{D \hskip-.6em \raise1pt\hbox{$ / $ } }

\def\leftrightarrowfill{$\mathsurround=0pt \mathord\leftarrow \mkern-6mu
       \cleaders\hbox{$\mkern-2mu \mathord- \mkern-2mu$}\hfill
       \mkern-6mu \mathord\rightarrow$}
\def\dvec#1{\vbox{\ialign{##\crcr
       \leftrightarrowfill\crcr\noalign{\kern-1pt\nointerlineskip}
       $\hfil\displaystyle{#1}\hfil$\crcr}}}          % <--> accent
\def\hook#1{{\vrule height#1pt width0.4pt depth0pt}}
\def\leftrighthookfill#1{$\mathsurround=0pt \mathord\hook#1
       \hrulefill\mathord\hook#1$}
\def\underhook#1{\vtop{\ialign{##\crcr                 % |_| under
       $\hfil\displaystyle{#1}\hfil$\crcr
       \noalign{\kern-1pt\nointerlineskip\vskip2pt}
       \leftrighthookfill5\crcr}}}
\def\smallunderhook#1{\vtop{\ialign{##\crcr      % " for su'scripts
       $\hfil\scriptstyle{#1}\hfil$\crcr
       \noalign{\kern-1pt\nointerlineskip\vskip2pt}
       \leftrighthookfill3\crcr}}}
\def\fder#1{{\d \over \d {#1}}} % functional derivative
\def\sfrac#1#2{{\vphantom1\smash{\lower.5ex\hbox{\small$#1$}}\over
       \vphantom1\smash{\raise.4ex\hbox{\small$#2$}}}} % alt. fraction
\def\bfrac#1#2{{\vphantom1\smash{\lower.5ex\hbox{$#1$}}\over
       \vphantom1\smash{\raise.3ex\hbox{$#2$}}}}      % "
\def\afrac#1#2{{\vphantom1\smash{\lower.5ex\hbox{$#1$}}\over#2}}  %"
\def\on#1#2{{\buildrel{\mkern2.5mu#1\mkern-2.5mu}\over{#2}}}%acc.over
\def\ddt#1{\on{\hbox{\LARGE .\kern-2pt.}}#1}             % double dot
\def\tdt#1{\on{\hbox{\LARGE .\kern-2pt.\kern-2pt.}}#1}   % triple dot

\def\boxes#1{
       \newcount\num
       \num=1
       \newdimen\downsy
       \downsy=-1.5ex
       \mskip-2.8mu
       \bo
       \loop
       \ifnum\num<#1
       \llap{\raise\num\downsy\hbox{$\bo$}}
       \advance\num by1
       \repeat}
\def\boxup#1#2{\newcount\numup
       \numup=#1
       \advance\numup by-1
       \newdimen\upsy
       \upsy=.75ex
       \mskip2.8mu
       \raise\numup\upsy\hbox{$#2$}}

\newskip\humongous \humongous=0pt plus 1000pt minus 1000pt

\newif\ifdtup

\def\NPB#1#2#3{{\it Nucl. Phys.} {\bf B#1} (#2) #3}

\def\PLB#1#2#3{{\it Phys. Lett.} {\bf #1B} (#2) #3}

\def\RMP#1#2#3{{\it Rev. Mod. Phys.} {\bf #1} (#2) #3}

\def\PAMS#1#2#3{{\it Proc. Am. Math. Soc.} {\bf #1} (#2) #3}

\def\Phy#1#2#3{{\it Physica} {\bf #1} (#2) #3}

\parskip=\medskipamount  % space between paragraphs (TeX)
\def\baselinestretch{1.2}% magnification for line spacing (LaTeX)
\thicklines              % thick straight lines for pictures (LaTeX)
\def\border{
 }
\def\headpic{
 }% SB paper head
\def\title#1#2#3#4{\begin{document}
       \border
       \headpic
       {\hbox to\hsize{#4 \hfill ITP-SB-#3}}\par
       \begin{center}\vskip.8in minus.1in
       {\Large\bf #1}\\[.5in minus.2in]{#2}
       \vskip1.4in minus1.2in {\bf ABSTRACT}\\[.1in]\end{center}
       \begin{quotation}\par}
\def\author#1#2{#1\\[.1in]{\it #2}\\[.1in]}
\def\ITP{\footnote{Work supported by National Science Foundation
  grant PHY 89-08495.}\\[.1in] {\it Institute for Theoretical Physics\\
  State University of New York, Stony Brook, NY 11794-3840}\\[.1in]}
\def\endtitle{\par\end{quotation}\vskip3.5in minus2.3in\newpage}

\def\camera#1#2{
       \topmargin=.46in
       \textheight=22cm
       \textwidth=15cm
       \hsize=15cm
       \oddsidemargin=.28in
       \evensidemargin=.28in
       \marginparsep=0in
       \parindent=1.15cm
       \pagestyle{empty}
       \def\rm{\sf}
       \begin{document}
       \begin{center}{\Large\bf #1}\\[.5in minus.2in]{\bf #2}
       \vskip1in minus.8in {ABSTRACT}\\[.1in]\end{center}
       \renewcommand{\baselinestretch}{1}\small\normalsize
       \begin{quotation}\par}
\def\endabstract{\par\end{quotation}
       \renewcommand{\baselinestretch}{1.2}\small\normalsize}

\def\xpar{\par}                                       % \par in loops
\def\header{
 }
\def\letterhead{
 \header
 \font\sflarge=helvetica at 14pt   % if you have helvetica
 \leftskip=2.8in\noindent\phantom m\\[-.54in]
       {\large\sflarge STATE UNIVERSITY OF NEW YORK}
       {\scriptsize\sf INSTITUTE FOR THEORETICAL PHYSICS\\[-.07in]
       STONY BROOK, NY 11794-3840\\[-.07in]
       Tel: (516) 632-7979}
 \vskip.3in\leftskip=0in}
\def\letterneck#1#2{\par{\hbox to\hsize{\hfil {#1}\hskip 30pt}}\par
       \begin{flushleft}{#2}\end{flushleft}}
\def\letterhat{\parskip=\bigskipamount \def\baselinestretch{1}}% textile
\def\head#1#2{\letterhat\begin{document}\letterhead\letterneck{#1}{#2}}
\def\multihead#1#2{\thispagestyle{empty}\setcounter{page}{1}
       \letterhead\letterneck{#1}{#2}}        % multiple letters
\def\shead#1#2{\letterhat\begin{document}\sletterhead
       \letterneck{#1}{#2}}
\def\multishead#1#2{\thispagestyle{empty}\setcounter{page}{1}
       \sletterhead\letterneck{#1}{#2}}
\def\multisig#1{\goodbreak\bigskip{\hbox to\hsize{\hfil Kind regards,
       \hskip 30pt}}\nobreak\vskip .5in\begin{quote}\raggedleft{#1}
       \end{quote}}
\def\sig#1{\multisig{#1}\end{document}}

\def\watch{
 \newcount\hrs
 \newcount\mins
 \newcount\merid
 \newcount\hrmins
 \newcount\hrmerid
 \hrs=\time
 \mins=\time
 \divide\hrs by 60
 \merid=\hrs
 \hrmins=\hrs
 \divide\merid by 12
 \hrmerid=\merid
 \multiply\hrmerid by 12
 \advance\hrs by -\hrmerid
 \ifnum\hrs=0\hrs=12\fi
 \multiply\hrmins by 60
 \advance\mins by -\hrmins
 \number\hrs:\ifnum\mins<10 {0}\fi\number\mins\space\ifnum\merid=0
 AM\else PM\fi}
\def\half{\frac{1}{2}}

\def\eq{\begin{equation}}
\def\eqe{\end{equation}}
\def\eqa{\begin{eqnarray}}
\def\eqae{\end{eqnarray}}

\def\be{\begin{equation}}
\def\ee{\end{equation}}
\def\bea{\begin{eqnarray}}
\def\ena{\end{eqnarray}}

\def\to{\rightarrow}

\def\tj{\tilde{\j}}
\def\td{\tilde{D}}
\def\tv{\tilde{\varphi}}
\def\rld{\rlap{\,/}D}
\def\dv{\dot{\varphi}}
\def\dj{\dot{\j}}
\def\bv{\bar{\varphi}}
\def\bj{\bar{\j}}
\def\rld{\rlap{\,/}D}
\def\rla{\rlap{\,A}\bigcirc}
\def\1ov4{{1\over 4}}
\def\bc{\bar{\chi}}
\def\dox{\dot{x}}
\def\dc{\dot{\chi}}
\def\trld{\tilde{\rlap{\,/}D}}

\def\vecnab{\vec{\nabla}}
\def\vx{\vec{x}}
\def\vy{\vec{y}}
\def\arrowk{\stackrel{\rightarrow}{k}}
\def\kbar{k\!\!\!^{-}}
\def\karrow{k\!\!\!{\rightarrow}}
\def\arrowl{\stackrel{\rightarrow}{\ell}}

\def\SAmpl{ \langle z,\bar\eta | \exp \left( - { \b \over
\hbar } \hat H \right) | y,\chi \rangle }
\def\SAmplb{ \langle z | \exp \left( - { \b \over
\hbar } \hat H \right) | y \rangle }
\def\tr{{\rm tr}}
\def\Tr{{\rm Tr}}

\begin{document}

\thispagestyle{empty}
\begin{flushright}
{\sc ITP-SB}-95-32\\ {\sc QMW-PH}-95-34
\end{flushright}
\vspace{1mm}
\setcounter{footnote}{0}
\begin{center}
{\LARGE\sc{Loop calculations in quantum mechanical non-linear
sigma models with fermions and applications
to anomalies\footnote{This research was supported in part by NSF grant no
PHY9309888.}
    }}\\[8mm]

\sc{Jan de Boer\footnote{e-mail: deboer@insti.physics.sunysb.edu},
    Bas Peeters\footnote{e-mail: peeters@insti.physics.sunysb.edu,
  address after oct. 1st:
  Department of Physics, Queen Mary and
  Westfield College, University of
  London, Mile End Road, London, U.K.},
    Kostas Skenderis\footnote{e-mail: kostas@insti.physics.sunysb.edu} and
    Peter van Nieuwenhuizen\footnote{e-mail:
                          vannieu@insti.physics.sunysb.edu},}\\[5mm]
{\it Institute for Theoretical Physics\\
State University of New York at Stony Brook\\
Stony Brook, NY 11794-3840, USA}\\[11mm]

{\sc Abstract}\\[2mm]
\end{center}

\noindent

{\small
We construct the path integral for one-dimensional non-linear sigma models,
starting from a given Hamiltonian operator and states in a Hilbert space.
By explicit evaluation of the discretized propagators and vertices we find
the correct Feynman rules which differ from those often assumed. These
rules, which we previously derived in bosonic systems \cite{paper1},
are now extended to
fermionic systems. We then generalize the work of Alvarez-Gaum\'e and
Witten \cite{alwi} by developing a framework to compute anomalies of
an $n$-dimensional quantum field theory by evaluating perturbatively
a corresponding quantum mechanical path integral.
Finally, we apply this formalism to various chiral and trace anomalies,
and solve a series of technical problems:
$(i)$ the correct treatment of Majorana fermions in path
integrals with coherent states (the methods of fermion doubling and fermion
halving yield equivalent results when used in applications to anomalies),
$(ii)$ a complete
path integral treatment of the ghost sector of chiral Yang-Mills anomalies,
$(iii)$ a
complete path integral treatment of trace anomalies,
$(iv)$ the supersymmetric extension of the Van Vleck determinant,
and
$(v)$ a derivation of the spin-$3\over 2$ Jacobian
of Alvarez-Gaum\'{e} and Witten for Lorentz anomalies.}

\vfill

\newpage

\section{Introduction}

%Anomalies of $n$-dimensional quantum field theories  can be computed
%by evaluating quantum mechanical path integrals for corresponding
%nonlinear sigma models. This was first shown for
%chiral anomalies by Alvarez-Gaum\'e and Witten \cite{alwi}, and later
%extended to trace anomalies by Bastianelli and
%van Nieuwenhuizen \cite{bapvn}.
%Here we present the general case. In addition we solve several problems
%which were left unanswered in ref. \cite{alwi,bapvn}.

In this article we discuss Euclidean path integrals for one-dimensional
systems with a Hamiltonian which is more general than
$H(\hat{p},\hat{x})=T(\hat{p})+V(\hat{x})$\footnote{Hats denote
operators, but where no confusion arises we will omit them.},
and their applications to anomalies.
Namely we shall
consider non-linear sigma models whose classical action is of the
form\footnote{Curved indices in space-time will be denoted by
$\m,\n,\ldots$. In the non-linear sigma model the corresponding
indices will be denoted by $i,j,\ldots$}
$\int_{-\b}^0 \frac{1}{2} g_{ij}
\frac{dx^i}{dt}
\frac{dx^j}{dt} dt$, plus fermionic extensions, which may be,
but need not be, supersymmetric. This is the area of quantum
mechanical path integrals in curved space, a difficult and
controversial subject \cite{witt,sch}.

Quantum mechanical path integrals are of importance as toy
models for path integrals for field theories. Because the former
are finite, no renormalization is necessary and subtle issues can better
be studied. In addition, quantum mechanical path integrals are
useful because some quantities of $n$-dimensional quantum field theories
can be calculated in a much simpler way by using the corresponding
one-dimensional path integrals. The prime example are anomalies,
which we shall discuss in detail in the second part of this article.
In the first part we give a careful derivation of quantum mechanical
path integrals and the Feynman rules to which they give rise.

This article is an extension of a previous article \cite{paper1}
on bosonic
quantum mechanical path integrals to the fermionic case and to
applications to anomalies. In section 2.1 we briefly
review the bosonic path integrals, but most emphasis in section 2
and beyond is on the fermionic case. At the end of this introduction
we shall state which results are new, but we shall start with a
general introduction in which we discuss bosonic and fermionic
systems on equal footing.

The basic problem we solve is the following. Given a Hamiltonian
$\hat{H}(\hat{p},\hat{x},\hat{\j}^{\dagger},\hat{\j})$ with arbitrary
but a priori fixed ordering of the bosonic operators $\hat{p}_i$,
$\hat{x}^i$ $(i=1\ldots n)$ and fermionic operators\footnote{
For Majorana fermions, we shall formulate one approach where
$a=1,\dots n$ and another where $a=1,\ldots,n/2$. We shall
find it convenient to use fermionic operators satisfying
anti-commutation relations without $\hbar$, $\{\j^a,\j_b^{\dagger}\}=
\d^a_b$, while $[\hat{x}^i,\hat{p}_j]=i\hbar\d^i_j$ as usual.}
$\hat{\j}^a,\hat{\j}^{\dagger}_a$ $(a=1\ldots n)$, find a
path integral representation for the transition element
$\SAmpl$ where $|y,\c \>$ is an eigenket of $\hat{x}^i$
and $\hat{\j}^a$ with eigenvalues $y^i$ and $\c^a$, while
$\<z,\bar{\h}|$ is an eigenbra of $\hat{x}^i$ and
$\hat{\j}^{\dagger}_a$ with eigenvalues $z^i$ and $\bar{\h}_a$.
Following Dirac \cite{Dirac} and Feynman \cite{Feyn} we shall insert $N-1$
complete sets of $x$-eigenfunctions, $N$ complete sets of
$p$ eigenstates and $N$ complete sets of coherent states
\be \label{i1}
\int |x\> \sqrt{g(x)} \< x| d^n x = \int |p\>\<p| d^n p =
\int |\h\> e^{-\bar\h_a \h^a} \<\bar\h|
\prod_{a=1}^n (d\bar\h_a d\h^a)={\bf 1}
\ee
to obtain a phase space path integral in the limit $N\rightarrow\infty$
of the form
\be \SAmpl \sim \int dx^i dp_i d\bar\h_a d\h^a
e^{-\frac{1}{\hbar} \int_{-\b}^0 L dt }
\ee
where $L=-ip_j\dot{x}^j + \hbar \bar\h_a \dot{\h}^a+
H(p,x,\bar\h,\h)$. However several questions arise:
\begin{itemize}
\item[(i)] which is the relation between the operators
$\hat{H}(\hat{p},\hat{x},\hat{\j}^{\dagger},\hat\j)$ and the functions
$H(p,x,\bar\h,\h)$? Different operator orderings of $\hat{H}$ must
lead to different functions $H$. Are there particular orderings of
$\hat{H}$ for which $H$ is equal to the naive classical Hamiltonian?
After integrating out the momenta, is the action in
the path integral invariant under the usual Einstein (general
co-ordinate) transformations or supersymmetry transformations if
the corresponding Hamiltonian $\hat{H}$ commutes with the
generators of these symmetries? (The answer is no).
\item[(ii)] What are the Feynman rules needed to evaluate the
path integral perturbatively?
\item[(iii)] What is the precise meaning of the measure
$dx^idp_id\bar{\eta}_a d \eta^a$?
Is there a normalization constant in front of the path integral,
or even factors $g(z)^{\a}g(y)^{\b}$ where $g=\det(g_{ij})$?
(The answer is yes)
\item[(iv)] By adding couplings to external sources, one
obtains propagators. These propagators are not the usual
translationally invariant propagators because they must
satisfy boundary conditions at $t=-\b$ and $t=0$. What
boundary conditions for $p_i(t)$ and the fermions must be imposed?
\item[(v)] When one is dealing with Majorana fermions, how should
one define coherent states, and what is the Hilbert space in which
$\hat{H}$ is supposed to act. Should one impose boundary conditions
both at $t=-\b$ and $t=0$ for Majorana fermions,
and if so, how is this compatible with the fact that
the equations of motion are linear in time derivatives?
\end{itemize}

These are some preliminary questions. Not all of them are new, but
we shall automatically get answers by following our derivation of
path integrals. These answers will be summarized in the conclusions.
The most important question we solve has to do with
Feynman rules. The propagators $\<x^i(\s) x^j(\t)\>$
and $\<\j^a(\s)\j^{\dagger}_b(\t)\>$ for configuration space path
integrals (and in addition $\< x^i(\s) p_j(\t) \>$ and $\< p_i(\s)
p_j(\t) \>$ for phase space path integrals)
with $-1\leq \s,\t \leq 0$ contain factors
$\theta(\s-\t)$, where $\s$ and $\t$ are the time
divided by $\b$. In configuration
space, there are contractions of $\dot{x}^i(\s)$ with
$\dot{x}^j(\t)$, which contain a factor $\d(\s-\t)$. When one
computes Feynman graphs, one has to evaluate integrals over
products of these distributions, for example
\be I=\int_{-1}^0 \int_{-1}^0 \d(\s-\t) \theta(\s-\t)
\theta(\t-\s) d\s d\t
\label{i3}
\ee
In addition, there are in general equal-time contractions with factors
$\d(0)$ and $\theta(0)$.
How should one evaluate such integrals? The function $\theta(\s-\t)
\theta(\t-\s)$ is nonvanishing only at $\s=\t$, a set of zero measure,
so any smoothing of $\d(\s-\t)$ while keeping the
product of the two $\theta$'s intact leads directly to zero.
A more natural way would seem to be to expand $\d(\s-\t)$ and
$\theta(\s-\t)$ into an infinite series, and to cut off these series at
some large $N$ (`mode regularization'). (The series expansion
of $\theta(\s-\t)$ could for example be defined
so that $\partial_{\s} \theta(\s-\t)=\d (\s-\t)$).
Performing the whole
calculation for finite $N$, one would expect to  obtain
the correct answer by taking the limit $N\rightarrow\infty$
at the end of the calculation. This is incorrect as we shall
see but first we must answer a fundamental question: what
does
the expression `the correct answer' mean?

Several authors have tried to give meaning to continuum path integrals
in curved space, in particular configuration space integrals of the
form $\int [dx^i] \exp( -\frac{1}{\hbar} S)$, by freely inventing
definitions which maintain Einstein invariance at intermediate steps.
The transition element $\SAmpl$ being in principle known from either
heat kernel methods \cite{witt2,DeWia}
or direct operator approaches\footnote{Writing
$\SAmplb$ as $\int\< z|\exp(-\frac{\b}{\hbar} \hat{H} |p\> \<p|y\>
d^n p$, it is clear that by expanding the exponent and moving all
$\hat{p}_i$ to the right and all $\hat{x}^i$ to the left, keeping
track of commutators,
coefficients of a given power of $\b$  are
finite and unambiguous.
Similarly when fermions are present.}, the loop
calculation based on these path integrals should in the end
reproduce the results for the transition element. Here problems
arise: no prescription for $L$ is known which keeps covariance
at all stages and which gives the correct result at two-
or higher loops.

Our point of view is the following: we define the continuum limit
path integral as the limit of the discretized path integral obtained
from (\ref{i1}). Hence, for us `the correct answer' means: the
answer which reproduces the results of the Hamiltonian approach.
Integrals over products of discretized propagators, vertices and
equal-time contractions
are well defined and finite and by {\it taking the continuum
limit, the correct Feynman rules automatically emerge}.
The results are that $\d(\s-\t)$ should still be considered as a
Kronecker delta function, even in the continuum case. So, for
example, the correct value of $I$ in (\ref{i3}) is $1/4$. To
bring out the surprising consequences of these new Feynman rules,
consider another integral
\be I=\int_{-1}^0 \int_{-1}^0 \d(\s-\t) \theta(\s-\t)
\theta(\s-\t) d\s d\t
\label{i4}
\ee
If one were to require that $\d(\s-\t)=\del_{\s} \theta(\s-\t)$ even
at the regularized level, one would find $1/3$ for this integral, whereas
the correct answer (and the answer obtained from treating $\d(\s-\t)$
as a Kronecker delta) is $1/4$. This demonstrates that mode regularization
is incorrect.

No expressions with higher powers of $\delta(\s-\t)$ arise
if one introduces in configuration space path
integrals the extra ghosts of \cite{bapvn}. These ghosts are necessary to
exponentiate the factors $g^{1/2}$ which arise if one integrates out
the momenta. These factors $g^{1/2}$ were first found by Lee and
Yang in a study of non-linear deformations of the harmonic oscillator
\cite{LeeYang} and were written by them as extra terms
in the action of the form $g^{1/2}
\d(0)$. We find it much more convenient for higher
loop calculations to replace them by local terms with ghosts,
similarly to the familiar Faddeev-Popov ghosts in gauge theories.
The phase space path integrals are always finite since the
propagators do not have any $\d(\s-\t)$ singularities, and with
these ghosts also the configuration space path integrals are
finite\footnote{Power counting would seem to indicate that there
are linear divergences due to the double derivative interactions.
This would seem to contradict the theorem that quantum mechanics
is finite. The ghosts save the theorem.}.

In the first part of this article we shall give a careful derivation
of the new Feynman rules. We begin by {\it rewriting} the Hamiltonian
$\hat{H}$ in Weyl-ordered form $(H)_W$, and then use that
matrix elements of $\exp (-\frac{\e}{\hbar} (H)_W)$
with $\e=\b/N$ can be immediately evaluated to order $\e$ by
using the `midpoint rule'. (Weyl ordering is discussed in section 2.3.)
Namely one may replace
$\exp (-\frac{\e}{\hbar} \hat{H})$ by
$\exp (-\frac{\e}{\hbar} (H)_W)$ in the kernels of the path integral
\be
\int \< x_k, \bar\h_k |
\exp (-\frac{\e}{\hbar} \hat{H}) | p_k, \xi_{k-1} \>
e^{-\bar\xi_{k-1} \xi_{k-1}} \< \bar\xi_{k-1},p_k | x_{k-1}, \h_{k-1} \>
dp_{k,i} d\bar\xi_{k-1,a} d\xi_{k-1}^a
\label{i5}
\ee
and then one may replace in $(H)_W$ the operator $\hat{p}_i$
by $(p_k)_i$, $\hat{x}^i$ by $\half(x^i_k+x^i_{k-1})$,
$\hat{\j}^{\dagger}_a$ by $\bar\xi_{k-1,a}$, and
$\hat\j^a$ by $\half(\xi^a_{k-1}+\h^a_{k-1})$. (Or
$\hat\j^{\dagger}_a$ by $\half(\bar\h_{k,a}+\bar\xi_{k-1,a})$ and
$\hat\j^a$ by $\xi^a_{k-1}$.)  The net effect is that one can extract
the function $H(p_k,\half(x_k+x_{k-1}),\bar{\xi}_{k-1},\half(\xi_{k-1}+
 \h_{k-1}))$.
For linear sigma models (with $H=T+V$)  rigorous proofs
based on Banach spaces and the `Trotter formula'
\cite{Trot} exist \cite{sch}.
These do not apply to non-linear sigma models,
but we have found a simple derivation of (\ref{i5}) which is precise
enough for our taste. Note that no matter whether $\hat{H}$ is
gauge invariant or not, this midpoint rule holds; it is a
purely algebraic result.

By using the background field formulation and coupling quantum
deviations to external sources and decomposing the action $S$ in
a suitable free part $S^{(0)}$ and an interaction part, we find
discretized propagators and vertices in closed form. The bosonic
discretized propagators were already found in
\cite{paper1},
while fermionic
propagators for Dirac fermions were already found in
\cite{Gava}.

We shall first consider Dirac fermions, but then
an even number of
Majorana (real)
fermions $\psi^a(t)$ and operators $\hat\j^a$ satisfying the
Dirac brackets $\{\hat\j^a, \hat\j^b \}=\d^{ab}$. To define
coherent states we need creation and annihilation operators,
and these we shall construct in two different ways: by
doubling the number of fermions by adding a second set $\psi_2^a$
of free fermions, or by `halving the number of fermions' and
constructing $\j^A$ and $\bar\j_A$ from pairs $\j^{2a-1}$ and
$\j^{2a}$ as $(\j^{2a-1} \pm i\j^{2a} ) /\sqrt{2}$. In either case
we Weyl order, use the fermionic midpoint rule, and find
propagators. The propagators $\< \j^a(\s) \j^b(\t) \>$ are
different depending on whether one doubles or halves the
Majorana fermions, and also the action in the path integral
and the transition elements differ. In applications to anomalies,
however, these differences disappear. In the conclusions we explain
the reason for these results.

In all cases (bosonic systems, Dirac fermions, Majorana fermions either
with doubling or halving), the propagators can be factored as follows
\be \SAmpl = e^{-\frac{1}{\hbar} S_{\rm class}} \tilde{D}_S^{1/2}
e^{{\cal A}_n} \label{i6}
\ee
where $S_{\rm class}$ is the classical action, $\tilde{D}_S^{1/2}$
contains the Van Vleck determinant and takes care of the
one-loop
contributions, while to order $\b$ ${\cal A}_n$
contains the trace anomaly for $n=2$ dimensions (it is
proportional to the scalar curvature $R$).
This structure of the propagator
was proven for general bosonic systems in \cite{witt2}.
For fermionic systems
$\tilde{D}_S$ is actually the superdeterminant of
$-\frac{\del}{\del \Phi^I} S_{\rm class}
\frac{\overleftarrow{{\scriptstyle\del }}}{\del \Phi_J}$
where $\Phi^I=\{z^i,\bar\h_a\}$ and $\Phi_J = \{y^i,\c_a\}$.
The fact that the one-loop contributions should be equal to
the Van Vleck superdeterminant is a check on the
correctness of our Feynman rules.

In general, the action in the path integral contains extra terms of order
$\hbar$ and $\hbar^2$. These extra terms are due to rewriting the
Hamiltonian in Weyl ordered form.
In our case, we shall only encounter $\hbar^2$ terms.
Hints that such terms might be necessary in
the path integral were first found by
DeWitt \cite{witt2}.
Schwinger, who
studied the Poincar\'e operator algebra for Yang-Mills
theory in the Coulomb gauge (which is a non-linear sigma model in
4 dimensions), found that one had to add extra non-naive
terms of higher order in $\hbar$ to the Hamiltonian in
order that the algebra closes \cite{schw}.
Subsequently
many others have studied these extra terms
\cite{geji,chlee}. To
check our new Feynman rules and also check that no
further modifications at order $\hbar^3$ are present,
we perform a 3-loop calculation in appendix A.1.

Our methods also apply to systems with more than two momenta
(higher derivative theories). We consider in appendix A.2 such
a system, and check that the results of the phase space approach
agree with those of the configuration space approach. This is
Matthews' theorem \cite{Matt}. It provides another test on the
correctness of our results.

The second part of this article contains applications of these
quantum mechanical path integrals to anomalies. Anomalies of an
$n$-dimensional field theory can according to Fujikawa be written
as \cite{Fuji}
\be \label{eqqj1}
 {\cal A}_n = {\rm Tr} (\hat{J} e^{-\b \hat{R}} ) \ee
where $\hat{J}$ is the Jacobian for a symmetry transformation of
the path integral, and $\hat{R}$ is a regulator. A general method
to construct consistent regulators $\hat{R}$ which maintain
a given set of symmetries is given in \cite{diaz}. It was first proposed
by Alvarez-Gaum\'e and Witten \cite{alwi} to consider a corresponding
linear or non-linear sigma model in one dimension, for
which $\hat{R}$ becomes the Hamiltonian $\hat{H}$.
The basic idea is that the Fujikawa trace can be viewed as the trace over
the Hilbert space of a quantum mechanical system with a finite number
of operators $x^{\m},\partial/\partial x^{\m}$, Dirac matrices $\g^a$
and, if present, internal symmetry generators $T_a$. After representing
the Dirac matrices and internal symmetry generators
(by means of Majorana fermions and an auxiliary ghost system)
as operators in the same Hilbert space, the Fujikawa trace can be
rewritten in terms of a suitable quantum mechanical path integral.
The path integral can be used to compute matrix elements, and in
terms of those the anomaly becomes
\be
{\cal A}_n = \int dx^i \sqrt{g(x)} d\h^a d\bar\h_a  e^{\bar\h_a
 \h^a} \<x,\bar\h | J e^{-\frac{\b}{\hbar} H} | x, \h\>
\label{i10}
\ee
and by inserting a complete set
of states between $J$ and $\exp(-\b H/\hbar)$
one obtains a product of the transition
element and the matrix element of $J$. The anomaly is the $\b$-independent
term. Depending on the anomaly, i.e., depending on the matrix element of
$J$, there are different factors of $\b$ in front of this expression, and
since $\b$ counts the number of loops,
different anomalies require a different number of world-line loops to be
evaluated.

The simplest anomalies are the chiral anomalies. For these (\ref{i10})
is actually $\b$ independent, due to the topological nature of chiral
anomalies, and, as we shall show, as a result one only needs to evaluate
one-loop or tree graphs. In fact, Alvarez-Gaum\'e and Witten wrote the whole
expression in (\ref{i10}) again as a path integral of the same kind
as we consider for the transition element, but now with periodic
boundary conditions both for the bosons and for the fermions. The
one-loop contribution for such path integrals can then easily be
written as the determinant of the
kinetic operator of
deviations
about classical solutions. We shall obtain, of course, the same
results but will start from the transition element we obtained previously,
and then simply do the rest of the integrals in (\ref{i10}).
In our approach the ghosts for internal symmetries are part of the
complete path integral, and are not treated by operator methods and by
projecting on one-particle states as in \cite{alwi}.
The results
of Alvarez-Gaum\'e and Witten were extended to trace anomalies in
\cite{bapvn}.
In this case the product of the Jacobian and the transition element
could again be written as a path integral with now anti-periodic
boundary conditions for the fermions, but since the background
fermions are constant if they satisfy the equations of motion, the
authors of \cite{bapvn} had problems in finding the correct boundary
conditions at $t=-\b$ and $t=0$ for Majorana fermions. Rather, they
used an operator formalism for the fermions, and operator-valued actions.
We shall present a complete path integral formulation. Since the
trace anomaly receives contributions from higher loop graphs, the details
of the path integral do matter very much. For example, forgetting the extra
$\hbar^2$ terms in the action or the measure factor, one obtains incorrect
results. We shall also give a derivation of trace anomalies.
Although we will not do so here, one can use the same framework
to derive the gravitational anomalies due to spin-$1/2$ and spin-$3/2$
fields. The only subtlety is the question precisely which Jacobian one should
use in (\ref{eqqj1}) for spin-$3/2$,
and this is discussed in detail in appendix A.3.

We conclude this introduction by stating which of our results are new. The
Feynman rules for products of distributions\footnote{Mathematically, it
is possible to multiply objects called generalized functions that contain
the set of distributions. However, to determine which generalized function
corresponds to a given distribution still requires an `underlying physical
principle' (see e.g. \cite{colo} and references therein), and our rules can be
seen as an example of such a principle.}
are new; for the
bosonic case they were obtained in \cite{paper1}
and for the fermionic case here.
Our treatment of coherent states for fermions (first considered in
\cite{klau})
follows \cite{textb}
and is somewhat simpler (we believe)
than those treatments in the literature which use four kinds of coherent
states (namely bras and kets which are eigenstates of $\hat\j$ or
$\hat\j^{\dagger}$). For a good discussion of the latter see
\cite{soho}.
The careful treatment of Majorana fermions (doubling
and halving), in particular the fact that the transition elements are
different, is new, as is the proof that in all cases the one-loop
contributions sum up to a superdeterminant. New in section three
is the complete path integral treatment of chiral and trace
anomalies, with no need to introduce matrix-valued Hamiltonians or
to perform certain projections on the ghost states
by hand. Also new is the complete
diagrammatic evaluation of these anomalies in section 3 and appendix
A.4. The final new result is
the correct incorporation of `Lee-Yang' ghosts for higher
derivative theories in appendix A.2.

Besides all these new results, we have
spent a great deal of time to make the whole subject of quantum
mechanical path integrals for non-linear sigma models accessible to a
large audience, and we hope that this article will
also be a useful review of known results.

\section{Path integrals for finite time}

In this section we establish the framework we need in order to
perform calculations using one-dimensional path integrals for finite
time. Given a quantum mechanical system defined by a Hamiltonian with
a certain ordering prescription, we derive the corresponding path
integral formulation. This includes both the action to be used in the
path integral, as well as the Feynman rules, the latter being not only
a set of expressions for the propagators and vertices, but also the
correct prescription how to evaluate integrals over products of these,
as they occur in actual loop calculations.
We will first discuss the bosonic non-linear sigma
model, and derive the correct rules for configuration space path
integrals.
For phase space path integrals for bosonic systems, see \cite{paper1}.
Then we will derive similar results for the extension to
complex (Dirac) fermions, for which we will need to introduce coherent states.
Finally we shall discuss the modifications which one must make for
Majorana fermions.  Along the way, we will present a variety of checks that
our Feynman rules are the correct ones, by evaluating various transition
elements and comparing these with the results obtained from operator
methods. Further convincing evidence is provided in appendix A.1 and A.2.

\subsection{Bosonic non-linear sigma model}

We will start by computing the transition element for the following
quantum mechanical Hamiltonian
\be \label{eqq9}
\hat H = {1\over 2} g^{-1/4} p_i
\sqrt{g} g^{ij} p_j g^{-1/4}
\ee
This operator in Einstein invariant if $p_i$ is hermitian. (The $p_i$
transform under Einstein (=general co-ordinate) transformations as
$p'_i = {1\over 2}
\left\{ {\partial x^j \over \partial x'^i} , p_j \right\}$,
if the inner product is defined by (\ref{i1}),
from which the Einstein invariance of $\hat{H}$ follows \cite{witt}).
The more general case, where
also a scalar and vector potential are present, can easily be found at
each stage in the computation by covariantizing the expressions, and
including the scalar potential in the interactions. The cases of
other Hamiltonians, for example Hamiltonians whose operator ordering
is different from (\ref{eqq9}), will also be clear.

One can evaluate the corresponding transition element,
\be
T(z,y;\b)  \equiv \<z| \exp \left( - \frac{\b}{\hbar} \hat{H} \right) |y\>,
\ee
for finite (Euclidean) time $\b$ through a direct, but
rather tedious, computation in the operator formalism
\cite{gra,paper1,bas}.
Namely, by writing $T$ as $\int d^n p
\<z| \exp ( - \frac{\b}{\hbar} \hat{H} )|p\> \<p |y\>$ and
expanding the exponent, moving all $\hat{p}_i$ to the right and all
$\hat{x}^i$ to the left, keeping track of all terms with
up to two commutators,
we find the following
result (correct
through order $\b$, counting $z-y$ as being of order $\b^{1/2}$)
\be
T=(2\pi\hbar\b)^{-n/2} \exp \left( - {1\over\hbar}
S_{cl}^B[z,y;\b] \right) \tilde{D}^{1/2}
 \,\exp \left( -{\b\hbar\over 12}R \right)
\label{trans}
\ee
where $(2\pi\hbar\b)^{-n/2}$ is the usual Feynman factor,
$S_{cl}^B[z,y;\b]$ is the classical action for a geodesic with
$x(-\b)=y$, $x(0)=z$, $-{1\over 12}R$ yields the trace anomaly in
$n=2$ dimensions\footnote{
Our convention for the curvatures are
$R(\Gamma)_{\rho\sigma\mu}{}^{\nu}=
\partial_{\r} \Gamma_{\s\m}{}^{\n}+ \G_{\r\t}{}^{\n} \G_{\s\m}{}^{\t}
 - (\r \leftrightarrow \s)
=
R(\omega)_{\rho\sigma a}{}^{b} e^b_{\m} e^{\n}_a$
with $R_{\r\s ab}=\partial_{\r} \omega_{\s ab}
+\omega_{\r a }{}^c \omega_{\s cb} - (\r \leftrightarrow \s)$
and $R_{\m\n}=
R(\Gamma)_{\m\s\n}{}^{\s}$, so $R_{\m b}=R(\omega)_{\m\r a b } e^{a\r}$,
while
$R=g^{\m\n}R_{\m\n}=(g^{\m\n}g^{\r\s} - g^{\m\r} g^{\n\s}) \partial_{\m}
\partial_{\n} g_{\r\s}+\ldots$.},
and $\tilde{D}^{1/2}$ is proportional to the
square root of the Van Vleck determinant \cite{Vlec}
which gives the one-loop corrections to the transition element
\bea
\tilde{D}^{1/2} &=& \b^{n/2} g^{-1/4}(z) \det \left( - {\partial\over\partial
z^i} {\partial\over\partial y^j} S_{cl}[z,y;\b] \right)^{1/2}
g^{-1/4}(y) \nonumber \\ &=& 1 - {1\over 12} R_{ij}(z) (z-y)^i (z-y)^j
+ {\cal O}(\b^{3/2})
\label{defD}
\eea
The results in (\ref{trans}) and (\ref{defD}) agree with DeWitt's classic
paper \cite{witt2} in which he uses heat kernel methods.
Note that if one views $T$ for $\b\rightarrow 0$ as the kernel of a
continuum path integral action, then $\tilde{D}^{1/2}$
corresponds to a non-local
term. Each of the factors in (\ref{trans}) is a general
co-ordinate bi-scalar (a
scalar both in $y$ and in $z$).
We stress that the
answer for $T$ in (\ref{trans}) is finite and unambiguous.

We will now construct the path integral whose loop expansion
reproduces (\ref{trans}). First we rewrite the Hamiltonian in
Weyl-ordered form \cite{wey1}, which for any monomial
in $p$ and $x$ is
defined by $(n+m)!(p^nx^m)_W = \partial_a^n \partial_b^m
(a\hat{p}+b\hat{x})^{m+n}$ \cite{miz,sato,wey2}.
For the Hamiltonian in (\ref{eqq9}) we find the well known result
\cite{miz,omo2}
\be
\hat H = \left( {1\over2} g^{ij} p_i p_j \right)_W + {\hbar^2\over 8}
\left( g^{ij} \G^k_{il} \G^l_{jk} + R \right)
\label{HWeyl}
\ee
Then we use the correspondence between Weyl-ordering and the midpoint
rule \cite{ber,miz}
\be
\int\! d^n p \,\langle x_{k+1} | G_W |p\rangle \langle p | x_k\rangle
= \int\! d^n p \, G(p,x_{k+1/2}) \langle x_{k+1}|p\rangle \langle p |
x_k\rangle
\label{midp}
\ee
where we defined $x_{k+1/2} = (x_k+x_{k+1})/2$, and $G$ is any
function of $p,x$. Clearly, $\exp \left( - {\e\over\hbar} H_W
\right) = (\exp\left( - {\e\over\hbar} H \right))_W + {\cal O}(\e^2)$,
so using (\ref{midp}) we find
\bea
\int\! d^n p \,\langle x_{k+1} | \exp\left( - {\e\over\hbar} H_W
\right) |p\rangle \langle p | x_k\rangle
\qquad\qquad\qquad\qquad\qquad\qquad \nonumber \\
\qquad =\int\! d^n p \,\exp\left(- {\e\over\hbar} H(p,x_{k+1/2}) \right)
\langle x_{k+1}|p\rangle \langle p | x_k\rangle + {\cal O}(\e^2)
\label{Hmidp}
\eea
Note that $H(p,x_{k+1/2})$ in (\ref{Hmidp}) contains the terms in
(\ref{HWeyl}) of order $\hbar^2$.
We argued in  \cite{paper1} that the terms denoted by ${\cal O}(\e^2)$ will not
contribute to the path integral, and can therefore be neglected.
This is the only point in our derivation of the path integral that
is not mathematically completely rigorous.
What is subtle is the meaning of ${\cal O}(\e^2)$. For example, one can
view $p$ either as being of order 1 or of order $\e^{-1/2}$
 (since there are Gaussian integrals with
 $\exp(-p^2/2\e)$). In the latter case the terms denoted by
${\cal O}(\epsilon^2)$ are actually of order $\epsilon$.
Each case leads
to a different kernel. We argued in \cite{paper1}, using the
effective potential trick of \cite{geji} and \cite{lasch},
that both kernels are
equivalent under the path integral. However, this is not a rigorous
argument. Ideally one should keep all terms on the right hand side of
(\ref{Hmidp}) that might contribute in the limit $N \rightarrow \infty$,
evaluate the path integral at the discretized level and then prove that in the
limit $N \rightarrow \infty$ all extra terms do indeed drop out.

We now insert
$N-1$ sets of $x$-eigenstates and $N$ sets of
$p$-eigenstates into $\langle z  | \exp - \frac{\b}{\hbar} \hat{H} | y
\rangle$, and
we arrive at the discretized phase space path integral using
$\int\! d^n x \, \sqrt{g(x)} | x\rangle \langle x| = \unit =
\int\! d^n p | p \rangle \langle p|$ and
$\langle z | p \rangle = (2 \p \hbar  )^{-n/2} (\exp {i\over \hbar} p \cdot
z) g^{-1/4} (z)$.
Integrating out the $N$ momenta we find the discretized configuration
space path
integral, with $N$ factors $g^{1/2}(x_{k+1/2})$ in the measure from the
$p$ integrals, $N$ products $g^{-1/4}(x_{k+1})g^{-1/4}(x_k)$ from the inner
products $\langle x|p\rangle$ and $N-1$ factors $g^{1/2}(x_k)$ from
the completeness
relation in $x$-space. The action is given by
\be
S=\sum_{k=0}^{N-1}\left[ \frac{1}{2\epsilon} g_{ij}(x_{k+1/2})
 (x_{k+1}-x_k)^i (x_{k+1}-x_k)^j  + \frac{\hbar^2\epsilon}{8} (
 \Gamma \Gamma + R)(x_{k+1/2}) \right]
\label{SBdisc}
\ee
where we define $x_N=z$ and $x_0=y$, and $\e=\b/N$.
We decompose $x_k^j$ into a
background and a quantum part, and $S$ into a free and interacting
part
\be
x_k^j = x_{{\rm bg},k}^j +q_k^j,
\hspace{8mm}  S=S^{(0)}+S^{({\rm int})};
\hspace{8mm} k=1,\ldots, N-1
\label{split}
\ee
where $S^{(0)}=\sum_{k=0}^{N-1} \frac{1}{2\epsilon}
g_{ij}(z) (q_{k+1}-q_k)^i (q_{k+1}-q_k)^j$.
We take the metric in $S^{(0)}$ at $z$ in order to facilitate comparison
with (\ref{trans}), although any other choice should give the same result.
(Of course, propagators and vertices will be different
if we make a different decomposition into a kinetic and interaction
part, and also the measure factor (see (\ref{path14})) will be different,
but the final results should not change). Since we take $x_{\rm bg}$
to be a solution of the classical equations of motion of $S^{(0)}$,
in general $S^{({\rm int})}$
contains in addition to the true interactions also
a pure background piece and
terms linear in $q^j_k$.
However, counting $z-y$ as being of order $\b^{1/2}$, we need only a
finite number of tree graphs and tadpoles at a given order in $\b$.
The $N$ factors $g^{1/2}(x_{k+1/2})$ are exponentiated following
\cite{bapvn} (for an alternative approach see \cite{Deck}) by
using anti-commuting ghosts $b$ and $c$ and a commuting ghost $a$
\bea
\sqrt{ \det g_{ij} (x_{k+1/2}) } &=& K
\int db_{k+1/2}^j dc_{k+1/2}^j da_{k+1/2}^j \nonumber \\ && \hspace{-3cm}
\exp\left( -\frac{\epsilon}{2\beta^2\hbar} g_{ij}(x_{k+1/2})
(b_{k+1/2}^i c_{k+1/2}^j + a_{k+1/2}^i a_{k+1/2}^j) \right)
\label{ghost}
\eea
Since the constant $K$ will cancel, we do not determine it, and
the reason for the particular normalization of the
ghost action will become clear later.
Introducing modes for the quantum
fluctuations $q$ by the orthonormal transformation
\be
q^j_k = \sum d^j_m \sqrt{\frac{2}{N}} \sin \left(\frac{km\pi}{N} \right) ;
\hspace{2cm} k,m=1,\ldots,N-1
\label{modes}
\ee
we may change $dx^j_k\rightarrow dq^j_k \rightarrow dd^j_m$. Obviously,
the Jacobian for this transformation is $1$.
The quantum part of the action $S^{(0)}$ becomes equal to
\be \label{aaa20}
-{1 \over \hbar} S^{(0)}(q) =
- \sum_{m=1}^{N-1} \frac{1}{\e \hbar} g_{ij}(z) d_m^i d_m^j
(1 - \cos \frac{m \p}{N}).
\ee
Next, we couple to external sources
\be
S^{({\rm source})}=-\e \sum_{k=0}^{N-1} \left(F_{k+1/2}
\frac{q_{k+1}-q_k}{\epsilon} +
 G_{k+1/2} q_{k+1/2} + \mbox{{\rm sources for ghosts}} \right)
\label{path13}
\ee
so that we can extract the exact discretized propagators in the usual way.
Completing squares and performing the final integration over
$d^j_m$, $b^j_{k+1/2}$, $c^j_{k+1/2}$, $a^j_{k+1/2}$ leads to $N-1$
factors $g^{-1/2}(z)$ and $N$ factors $g^{1/2}(z)$
as well as an overall factor $(2\p\hbar\b)^{-n/2}$.
(The factor $(2\p\hbar\e)^{-Nn/2}$ which comes from the $p$ integrations
and the normalization of the plane waves combines with the factor
$\prod_{m=1}^{N-1} (\p \e \hbar)^{n/2} (1 - \cos \frac{m \p}{N})^{-n/2}$
from the Gaussian integrations over $d_m^j$ to yield $(2\p\hbar\b)^{-n/2}$
since $\prod_{m=1}^{N-1}2(1 - \cos \frac{m \p}{N})=N$).
Hence
\be
T=\left(\frac{g(z)}{g(y)}\right)^{1/4} (2\p\hbar\b)^{-n/2}
\exp (-\frac{1}{\hbar} S^{({\rm int})} ) \exp ( -\frac{1}{\hbar}
 S^{({\rm source})}).
\label{path14}
\ee
The measure factor $(g(z)/g(y))^{1/4}$ is due to the split
$S=S^{(0)}+S^{({\rm int})}$ where in $S^{(0)}$ the metric
is taken at $z$, $g_{ij}(z)$ (cf. question (iii) in
the introduction). Different splits clearly lead
to different measures, but we continue with (\ref{path14}).
Such factors are often ignored but are crucial to obtain the correct
transition element.
Equation (\ref{path14}) is to be read as usual in path integral
formulations, namely $S^{{(\rm int})}$ contains only derivatives with
respect to the sources and background fields, while
 $S^{({\rm source})}$ is the function bilinear in
sources that appears after doing all the integrals,
and in the final result we are supposed to put all the sources
equal to zero. Thus,
$S^{({\rm int})}$ contains the discretized vertices while
$S^{({\rm source})}$
yields the discretized propagators. Defining $\dot{x}_{k+1/2}
=(x_{k+1}-x_k)/\epsilon$ and omitting superscripts and a factor
of $\hbar g^{ij}(z)$ for the time being, the propagators
 come out as follows
\bea
<q_{k+1/2} q_{l+1/2} > & = & -\frac{\epsilon}{4N} (2k+1)(2l+1)
+\frac{\epsilon}{2} (2\min(k,l)+
1-\frac{1}{2}\delta_{k,l}) \nonumber\\
<q_{k+1/2} \dot{q}_{l+1/2} > & = &
-\frac{k+1/2}{N} + \theta_{k,l} \nonumber\\
<\dot{q}_{k+1/2} \dot{q}_{l+1/2} > & = &
-\frac{1}{N\epsilon} + \frac{1}{\epsilon} \delta_{k,l} \nonumber \\
< b_{k+1/2} c_{l+1/2} > & = & -\frac{2}{\epsilon} \delta_{k,l}
\nonumber \\
< a_{k+1/2} a_{l+1/2} > & = & \frac{1}{\epsilon} \delta_{k,l},
\label{path15}
\eea
where $\theta_{k,l}$ is a discretization of the $\theta$ function:
$\theta_{k,l}=0$ if $k<l$, $\theta_{k,l}=1/2$ if $k=l$ and
$\theta_{k,l}=1$ if $k>l$.

As an example, let us give a more detailed derivation of the
$<q_{k+1/2} q_{l+1/2}>$ propagator. The partition function (we set the
sources $F$ equal to zero, and suppress internal indices) reads
\be
Z[G] = \int\! \prod\limits_{i=1}^{N-1}dq_i \, \exp  -\frac{\e}{\hbar} \left[
\sum\limits_{k=0}^{N-1} {1\over 2} \left( {q_{k+1} - q_k \over \e}
\right)^2 + G_{k+1/2} \left( {q_{k+1} + q_k \over 2} \right) \right]
\label{ZG}
\ee
Now make a change of variables as defined in (\ref{modes})
and (\ref{aaa20}), complete
the squares and do the Gaussian integrals over the modes $d_m$. Up to
an overall numerical factor which was already taken care
of in (\ref{path14}),
the result equals
\bea
Z[G] &\sim&  \exp \Biggl[ {\e^3 \over 2N\hbar} \sum\limits_{j=1}^{N-1}
{1\over 1-\cos \bigl(j\pi/N\bigr)} \nonumber \\ && \qquad \left(
\sum\limits_{k=0}^{N-1} G_{k+1/2}
\sin \Bigl((k+1/2) j \pi/N\Bigr) \, \cos \Bigl(j\pi/(2N)\Bigr)
\right)^2 \, \Biggr]
\label{ZG2}
\eea
Differentiating with respect to the sources $G$
 gives rise to the following expression for the propagator
\bea
<q_{k+1/2} q_{l+1/2} > &=& {\e\hbar\over 2N} \sum\limits_{j=1}^{N-1} \Biggl[
\cos^2\Bigl(j\pi/(2N)\Bigr) {\sin
\Bigl((k+1/2)j\pi/N \Bigr) \over \sin\Bigl( j\pi/(2N)\Bigr) }\nonumber \\
&& \qquad \qquad \qquad {\sin\Bigl( (l+1/2)j\pi/N \Bigr)
  \over \sin\Bigl( j\pi/(2N)\Bigr) } \Biggr]
\label{qqprop}
\eea
Each term can be written as a sum of powers of $e^{i\pi/(2N)}$, and
performing the sum over $j$ yields the propagator as given in
(\ref{path15}). The remaining discretized propagators can be found in
a similar fashion.
We require that $x_{{\rm bg},k}^{j}$ satisfies
the boundary conditions and the
equation of motion of $S^{(0)}$. In the continuum limit this
becomes $x^j_{{\rm bg}}(t)=z^j+(z-y)^j t/\b$, while $q^j(t)$
vanishes at the endpoints. In this limit the two-point functions
become (reinstating the superscripts, factors of $ g^{ij}(z)$, and
defining $t=\beta\tau$)
\bea
<q^i(\sigma) q^j(\tau) >  & = & -\beta \hbar g^{ij}(z)
\Delta(\sigma,\tau) \nonumber \\
<b^i(\sigma) c^j(\tau) > & = & -2 \beta \hbar g^{ij}(z)
 \partial_{\sigma}^2 \Delta(\sigma,\tau)
\nonumber \\
<a^i(\sigma) a^j(\tau) > & = &  \beta \hbar g^{ij}(z)
 \partial_{\sigma}^2 \Delta(\sigma,\tau)
\nonumber \\
\Delta(\sigma,\tau)& = & \sigma(\tau+1)\theta(\sigma-\tau) +
 \tau(\sigma+1) \theta(\tau-\sigma) .
\label{path16}
\eea
Note that $\Delta(\sigma,\tau)=\Delta_F(\sigma-\tau) + \sigma\tau
+\frac{1}{2}(\sigma+\tau)$, where $\Delta_F(\sigma-\tau)=
\frac{1}{2}(\sigma-\tau)\theta(\sigma-\tau)+
\frac{1}{2}(\tau-\sigma)\theta(\tau-\sigma)$
is the Feynman propagator,
and formally
$\partial_{\sigma}^2\Delta(\sigma,\tau)=\delta(\sigma-\tau)$ while
$\Delta(\sigma,\tau)=0$ at the boundaries. However, {\it the
$\delta(\sigma-\tau)$ is a Kronecker delta} and moreover
{\it the equal-time contractions\footnote{Equal-time contractions in quantum
field theory can in general only be fixed by imposing a symmetry
principle \cite{ref4}. In our case they are fixed by our requirement
that the path integral reproduces the Hamiltonian results.}
are unambiguously defined}. Kronecker
delta here means that $\int dx \delta(x) f(x)=f(0)$, even when $f$
contains a product of $\theta$ functions.
{}From (\ref{path15}) we further find in the continuum limit
\bea
<q^i(\sigma) \dot{q}^j(\tau)> & = &
-\beta \hbar g^{ij}(z) (\sigma+\theta(\tau-\sigma))
\nonumber \\
<\dot{q}^i(\sigma) \dot{q}^j(\tau) > & = &
-\beta \hbar g^{ij}(z) (1-\delta(\sigma-\tau) ).
\label{path17}
\eea
All propagators are now proportional to $\beta\hbar$ (this motivated
the normalization of the ghost action in (\ref{ghost})), and the
interactions are given by
\bea
\frac{1}{\hbar} S^{({\rm int})}  & =
& \frac{1}{\beta\hbar} \int_{-1}^0 \left[
\frac{1}{2} g_{ij} (x_{\rm bg}+q)
\left\{
( \dot{x}_{\rm bg} + \dot{q})^i
( \dot{x}_{\rm bg} + \dot{q})^j + b^ic^j + a^ia^j \right\} \right] d\tau
\nonumber \\
& & + \beta \hbar \int_{-1}^0 \frac{1}{8} (\Gamma \Gamma + R) d\tau
-\frac{1}{\hbar} S^{(0)}, \nonumber \\
\frac{1}{\hbar} S^{(0)}  & =
& \frac{1}{\beta\hbar} \int_{-1}^0 \left[
\frac{1}{2} g_{ij} (z)
\left\{ \dot{q}^i \dot{q}^j
 + b^ic^j + a^ia^j \right\} \right] d\tau
\label{path18}
\eea
Clearly, the interactions only depend on the combination $\b\hbar$.

To compute the configuration space path integral, we note that we must
expand the measure factor $g^{1/4}(z)/g^{1/4}(y)$ in (\ref{path14})
and evaluate all vacuum graphs with external $x_{\rm bg}$,
using the propagators in (\ref{path16}), (\ref{path17})
and the vertices in (\ref{path18}).
The $q$-independent
part of $S^{({\rm int})}$ does not yield the full $S^B_{\rm cl}$ of
(\ref{trans}) since $x_{\rm bg}$ is only a solution of the $S^{(0)}$ equation
of motion; rather, tree graphs with
two vertices from
$S^{({\rm int})}$ contribute to order $\b$
terms of the form $\frac{1}{\b} (\partial g)^2 (z-y)^4$, see (\ref{SBos}).
In the one-loop graphs with one vertex $S^{({\rm int})}$ one
finds equal-time contractions
proportional to $(z-y)^k \partial_k g_{ij}$ times
$<\dot{q}^i \dot{q}^j +
b^ic^j + a^ia^j>$ in which the $\d(0)$ cancel, yielding
$ \int_{-1}^0 \int_{-1}^0 \s ( \partial_\s \partial_\t \Delta
 + \partial_\s^2 \Delta) d\sigma d\tau=-\half$,
which cancels a similar contribution from the non-trivial measure
factor.
There are many other one-loop and two-loop graphs, and
the contribution of each corresponds to a particular term in (\ref{trans}).
In particular, the two-loop graph with one $\dot{q}\dot{q}$, one $\dot{q}q$ and
one $q\dot{q}$ propagator agrees with (\ref{trans})
only if one uses $\int^0_{-1} \int^0_{-1} \d (\s-\t) \theta (\s-\t)
\theta (\t-\s)  = {1\over 4}$,
in agreement with the discretized
expressions for the propagators in (\ref{path15}).
Adding all contributions we have found complete agreement. The non-covariant
vertices  $\frac{\b\hbar}{8}(\Gamma\Gamma + R)$ conspire with the non-covariant
vertices found by expanding $g_{ij}(x)$ and yield the Einstein invariant
expression (\ref{trans}). The Feynman rules one has to use in this
calculation follow from (\ref{path15}), and they amount to the following.
First, one writes down expressions for all Feynman diagrams using the
propagators given by (\ref{path16}) and (\ref{path17}).
Adding everything,
all divergences coming from products of delta functions will
cancel (the ghosts of \cite{bapvn} are crucial for this).
The resulting integrals should be worked out using the rules that
delta functions should really be seen as Kronecker deltas and
that $\theta(0)=1/2$. If there are explicit delta functions in the
integrals, one should be careful with partial integrations
and identities like $\int_a^b f'=f(b)-f(a)$, since these are not always
compatible with our Kronecker delta prescription \cite{paper1}. Luckily,
in practice we never need to partially integrate.

\subsection{The fermionic case}

We now repeat the analysis of the last section for the fermionic
case. We will work in a basis of coherent states, and the derivation
of the path integral is analogous to the one in phase space for the
case of the bosonic non-linear sigma model.
We consider operators $\hat\j^a$,$\hat\j^{\dagger}_{a}$, $a=1\ldots n$,
satisfying the
anticommutation relations $\{\hat\j^a,\hat\j^{\dagger}_{b} \}  =
\d^{a}_{b}$.
These operators $\hat\j$ are obtained from the canonical
variables by rescaling with a factor of $\hbar^{-1/2}$.
As a consequence, terms of the form $\hbar^2 R \psi^4$ or
$\hbar \omega \psi^2$ are terms of the classical action,
not higher-loop terms.
For cases such as the $N{=}1$ supersymmetric
non-linear sigma model, where only Majorana fermions are present, we
will need to replace these Majorana fermions by Dirac
fermions. This will be discussed later, here we will derive the general
expression for the path integral with Dirac fermions.

Coherent states are defined by
\be
|\h\rangle = e^{\hat\j^\dagger\h} |0\rangle \qquad ; \qquad \langle \bar\h
| = \langle 0| e^{\bar\h \hat\j}
\label{defcost}
\ee
satisfying $\hat\j|\h\rangle = \h|\h\rangle$ and
$\langle\bar\h|\hat\j^\dagger = \langle\bar\h| \bar\h$.
We could in addition also define coherent states build around a Dirac
vacuum (completely-filled Fermi sea)
$\langle \h |= \linebreak
(-1)^n \langle 0 |\hat{\j}^n \cdots \hat{\j}^1
\left( e^{\h\hat{\j}^\dagger} \right)$,
satisfying
$\langle \h | \hat\j^a = \langle \h | \h^a$,
and similarly
$| \bar\h \rangle = \linebreak
\left( e^{\hat\j^k \bar\h_k} \right) \hat\j_1^\dagger \ldots
\hat\j_n^\dagger | 0\rangle (-)^n$.
These states are often used in the literature for the
construction of fermionic path integrals in a way which closely mimicks the
$x\cdot p$ approach,
see e.g. \cite{soho}. In contrast, in our approach we only use the
coherent states
(\ref{defcost}); although both approaches are completely equivalent,
we believe ours is more economical.
The inner
product and decomposition of unity read formally the same as
for bosonic coherent states \cite{textb}
\be
\langle\bar\h | \xi \rangle = e^{\bar\h \xi} \qquad ; \qquad \unit =
\int\!d\bar\h d\xi\, | \xi\rangle e^{-\bar\h \xi} \langle \bar\h|
\ee
but note the ordering of the anticommuting variables. Our convention is
that $d\bar\h=d\bar\h_n \ldots d\bar\h_1$, while $d\xi=d\xi_1 \ldots d\xi_n$,
or equivalently, $d\bar\h d\xi = \prod_{k=1}^n (d\bar\h_k d\xi_k)$.
Hence $\int d\xi \xi_n \ldots \xi_1 = 1$, and $\int d\xi \prod_{k=1}^n
\xi_k= (-1)^{n/2}$ for even $n$.
With these conventions, the trace of an operator over the fermionic
Fock space is given by
\be \label{deftrace}
{\rm trace}(A)=\int d\xi d\bar\h e^{\bar\h \xi} \langle
\bar\h |A| \xi \rangle
\ee
Again, we define Weyl-ordering by $(n+m)!(\j^n \j^{\dagger m})_W
= \partial^n_{\bar\h} \partial^m_\h (\bar\h \hat\j + \h \hat\j^\dagger
)^{m+n} $, with the fermionic derivatives acting from the left.
For an arbitrary Weyl-ordered operator $\hat G$ we can now derive the
midpoint identity for coherent states
\bea
\langle \bar\h | \hat G |\h\rangle &=& \int\! d\bar\c d\c \,
G(\bar\c,{\c+\h \over 2}) e^{-\bar\c \c} \langle \bar\h | \c \rangle
\langle \bar\c | \h \rangle \nonumber \\
&=& \int\! d\bar\c d\c \,
G({\bar\c + \bar\h \over 2},\c) e^{-\bar\c \c} \langle \bar\h | \c \rangle
\langle \bar\c | \h \rangle \nonumber \\
&=& \int\! d\bar\c d\c \,
G(\l_1\bar\c + (1-\l_1)\bar\h ,\l_2\c+(1-\l_2)\h)
e^{-\bar\c \c} \langle \bar\h | \c \rangle
\langle \bar\c | \h \rangle,
\label{ferWO}
\eea
with $\l_1 \l_2=1/2$.
This formula can be proven in the following way (cf. \cite{Gava}). It
is easy to check
that it is valid for an operator $\hat G$ of the type
$(\hat\j^\dagger)^k$, which is automatically Weyl-ordered. Then one
uses that, for a Weyl-ordered operator $\hat A$, the operator $\hat B
= (\hat\j \hat A \pm \hat A \hat\j)/2$ (the sign depending on whether
$\hat A$ is
bosonic or fermionic) is also Weyl-ordered, and hence that any
Weyl-ordered operator can be obtained by repeatedly applying this
identity to an operator of the type $(\hat\j^\dagger)^k$.
We can now proceed by inserting unity in (\ref{ferWO}) for
 $\hat{G}=\hat B$,
using its validity for $\hat A$ as induction hypothesis.
We find
\bea
\langle\bar\h |\hat B | \h\rangle &=& \int\! d\bar\c d\c\,
e^{-\bar\c\c} {\c+\h\over 2}\langle \bar\h|\c\rangle \langle \bar\c| \hat A |
\h \rangle \nonumber \\
&=& \int\! d\bar\c d\c d\bar\xi d\xi \,
e^{-\bar\c\c} {\c+\h\over 2} \langle \bar\h|\c\rangle
A(\bar\xi,{\h+\xi\over 2}) e^{-\bar\xi\xi} \langle
\bar\c| \xi\rangle \langle \bar\xi |\h
\rangle \nonumber \\
&=& \int\! d\bar\xi d\xi\, {\xi + \h\over 2} A(\bar\xi, {\xi +\h \over
2}) e^{-\bar\xi \xi} \langle \bar\h | \xi \rangle \langle
\bar\xi | \h \rangle \nonumber \\
&=& \int\! d\bar\xi d\xi\, B(\bar\xi, {\xi +\h \over
2}) e^{-\bar\xi \xi} \langle \bar\h | \xi \rangle \langle
\bar\xi | \h \rangle
\label{cohmidp}
\eea
where we have used that $\int\!d\bar\c d\c\, e^{-\bar\c \c} f(\c) =
f(0)$. This completes the proof of the first two lines of
(\ref{ferWO}). The third line of (\ref{ferWO}) can be demonstrated in
a similar fashion.  We can now apply
this identity to $\hat G = \exp\left( - {\e\over\hbar}\hat H_W \right)$,
and use that $\hat G$ is Weyl-ordered to order ${\cal O}(\e^2)$. We
find, neglecting these higher order terms, after inserting unity $N-1$
times, with $\b=N\e$,
\bea
\langle \bar\h | \exp\left( - {\b\over\hbar} \hat H_W \right) | \c
\rangle &=& \nonumber \\ && \hskip-4cm \int \prod\limits_{k=1}^{N-1}
\left[  d\bar\xi_k d\xi_k \,
e^{-\bar\xi_k \xi_k} \langle \bar\xi_{k+1} | \exp\left( -
{\e\over\hbar} \hat H_W \right) | \xi_k \rangle \right]
\langle \bar\xi_1 |
\exp\left( -
{\e\over\hbar} \hat H_W \right) | \c \rangle
\label{disFT}
\eea
where we defined  $\bar\xi_N = \bar\h$.
Now use the midpoint rule for each matrix element
\bea
\langle \bar\xi_{k+1}| \exp\left(-{\e\over\hbar} \hat H_W\right)
|\xi_k\rangle &=& \nonumber \\ && \hskip-4cm
\int\! d\bar\j_k d\j_k\,  e^{-\bar\j_k \j_k + \bar\xi_{k+1} \j_k +
\bar\j_k \xi_k} \exp\left(-{\e\over\hbar} H(\bar\j_k,
{\j_k+\xi_k\over 2} )\right)
\eea
We can integrate out the $\bar\xi,\xi$ in (\ref{disFT})
(first the $\bar\xi$, then the $\xi$) to obtain
\bea
\langle \bar\h | \exp\left( - {\b\over\hbar} \hat H_W \right) | \c
\rangle &=& \int \prod\limits_{k=0}^{N-1}
d\bar\j_k d\j_k \exp\Biggl[ \bar\h \j_{N-1} \nonumber \\ && \hskip-2 cm
- \e
\sum\limits_{k=0}^{N-1} \left( \bar\j_k {\j_k -\j_{k-1}\over\e} +
\frac{1}{\hbar} H(\bar\j_k, {\j_k + \j_{k-1}\over 2} ) \right) \Biggr],
\eea
where we defined $\j_{-1}=\c$.
The first term in the integrand is the usual boundary term one obtains
in path integrals for coherent states (cf. \cite{textb,textb2}).
{}From this result we conclude that the action in the continuum path
integral is $\int_{-\b}^0 (\hbar\bar{\psi} \dot{\psi}+H) dt-
\hbar\bar{\psi}(0) \psi(0)$ with the boundary conditions $\bar{\psi}(0)=
\bar{\eta}$ and $\psi(-\b)=\c$. Notice that the boundary term is
essential to produce the correct equations of motion $\dot{\bar{\psi}}=
\dot{\psi}=0$. After decomposing $\psi_k$ and $\bar{\psi}_k$ into
a background piece and a quantum piece, we couple the latter
to external sources. Putting all of $H$ into $H^{{\rm int}}$, we
obtain the discretized $\bar\j \j$ propagator by inverting the kinetic
term matrix $A_{j,k}=\d_{j,k}-\d_{j,k+1}$. The result reads
\be
< \bar\j_k \j_l > = \left\{ \ba{rl} -1 & k \leq l \\ 0 & k> l \ea
\right\} = -\theta_{l,k} -\frac{1}{2} \delta_{k,l}.
\label{aux29}
\ee
If we now define $\dot\j_{k-1/2} = (\j_k -\j_{k-1})/\e$ and $\j_{k-1/2}=
(\j_k + \j_{k-1})/2$,
we obtain
\bea
< \bar\j_{k} \j_{l-1/2} > &=& -\theta_{l,k} \nonumber \\
< \bar\j_{k} \dot\j_{l-1/2} > &=& -{1\over\e} \d_{k,l}
\label{disFP}
\eea
where we recall that $\theta_{k,l}=0$ if $k<l$, $\theta_{k,l}=1/2$ if
$k=l$ and $\theta_{k,l}=1$ if $k>l$.
We can now, as in the bosonic case, write down the corresponding
continuum expressions which we will use in actual
computations. However, it is important to realize that in diagrams in
which products of $\d$ and $\theta$
functions arise, we are now able to resolve the resulting ambiguities
by returning to the discretized expressions (\ref{disFP}). In
particular, we find again the `rules' that $\theta(\s,\s) = {1\over
2}$, and $\int \! d\s d\t \, \theta(\s-\t) \theta(\t-\s) \d(\s-\t) =
{1\over 4}$. The continuum propagators read
\bea
< \bar\j^a(\s) \j^b(\t) > &=& -  \d^{ab} \theta(\t-\s) \nonumber \\
< \bar\j^a(\s) \dot\j^b(\t) > &=& -  \d^{ab} \d(\s-\t)
\label{conFP}
\eea
There is no factor of $\hbar$ because we chose to work with operators
$\hat{\psi}^a ,\hat{\psi}_b$ satisfying $\{\hat{\psi}^a,\hat{\psi}_b\}
=\delta^{a}_{b}$.
Clearly, in this derivation we could instead equally well have
started from the second line in (\ref{ferWO}) and have
introduced $\dot{\bar\j}_{k+1/2} = (\bar\j_{k+1} - \bar\j_{k})/\e$,
$\bar\j_{k+1/2} = (\bar\j_{k+1} + \bar\j_{k})/2$,
leading again to $\< \bar\j_{k+1/2} \j_l \> = -\theta_{l,k}$
for the
discretized propagators, and therefore also to identical Feynman rules
in the continuum limit.

\subsection{Weyl-ordering of $N{=}2$ and $N{=}1$ Hamiltonians}

We will now derive the Weyl-ordered Hamiltonians corresponding to the
supersymmetric $N{=}2$ and $N{=}1$ Hamiltonians.
These are the most interesting fermionic systems, and play the same
privileged role as (\ref{eqq9}) in the bosonic case.
As it turns out,
different expressions result when we take the independent fermionic
fields to have flat or curved indices, and Weyl-order with respect to
these independent fields. Let us first consider the $N{=}2$
quantum Hamiltonian \cite{alwi}
\bea
H_{N{=}2} &=& {1\over 2} g^{-1/4} \pi_i g^{1/2} g^{ij} \pi_j g^{-1/4} -
{1\over 8} \hbar^2 R_{abcd} \j^a_\a \j^b_\a \j^c_\b \j^d_\b \nonumber
\\
\pi_i &=& p_i -{i\hbar\over 2} \w_{iab} \j^a_\a \j^b_\a
\label{HN2}
\eea
where $\w_{iab}$ is the spin connection and $\a = 1,2$.
We can now define Dirac spinors in the following way
\be
\j^a = {1\over\sqrt{2}} (\j^a_1 + i \j^a_2) \quad ; \quad \bar\j^a =
{1\over\sqrt{2}} (\j^a_1 -i \j^a_2)
\label{defDS}
\ee
satisfying the anticommutation relations $\{ \bar\j^a, \j^b \} =
\d^{ab}$. In terms of those, the $N{=}2$ quantum Hamiltonian
corresponds to the following
classical action
\be
\cl=\half g_{ij}(x) \dot{x}^i \dot{x}^j +\hbar
\bar\j^a (\dot{\j}^a + \dot{x}^j \omega_{j}{}^{ab} \j_b)
-\half \hbar^2 R_{abcd}(\omega)
\bar\j^a \j^b \bar\j^c \j^d.
\ee
Its field equations read
\bea \label{ff44}
\fder{x^i}\cl & = & -g_{ij} \frac{D}{dt} \dot{x}^j
+ \hbar R(\omega)_{ijab} \dot{x}^j \bar\j^a \j^b -\half \hbar^2
D_i R(\omega)_{abcd}
\bar\j^a \j^b \bar\j^c \j^d \nonumber \\
& & - \omega_{iab} \left(
\bar \j^a \fder{\bar\j^b} \cl +
 \j^a \fder{\j^b} \cl \right), \nonumber \\
\fder{\bar\j^a} \cl & = &
 \frac{D}{dt} \j^a -\hbar R(\omega)^a{}_{bcd} \j^b \bar\j^c \j^d,
\nonumber \\
\fder{\j^a} \cl & = &
 \frac{D}{dt} \bar\j^a -\hbar R(\omega)_{dbc}{}^a  \bar\j^d \j^b \bar\j^c,
\eea
where
\be
 \frac{D}{dt} \dot{x}^j  = \ddot{x}^j+\G^j_{kl} \dot{x}^k \dot{x}^l
\ee
and
\be
 \frac{D}{dt} \j^a = \dot{\j}^a + \dot{x}^i \omega_i{}^a{}_b \j^b,
 \frac{D}{dt} \bar{\j}^a = \dot{\bar{\j}}^a + \dot{x}^i \omega_i{}^{ab}
 \bar{\j}_b.
\ee
The invariance of the action under the two rigid supersymmetries follows by
contracting the field equations with the variations
\bea \label{aaa47}
\d x^i & = & \bar\e e^i_a \j^a + \bar\j^a e^i_a \e \nonumber \\
\d \j^a & = & \dot{x}^i e^a_i \e - \d x^i \omega_i{}^{ab} \j_b
\nonumber \\
\d \bar\j^a & = &-\bar\e e^a_i \dot{x}^i  - \d x^i \omega_i{}^{ab} \bar\j_b .
\eea
All terms then cancel (using the cyclic and the Bianchi identities for
$D_i R_{abcd}$). Since this action is in Euclidean time, it is not
hermitian, nor is $(\d\j^a)^{\dagger}$ equal to $\d\bar{\j}^a$,
but it can be obtained from a hermitian action in Minkowski
space by the Wick rotation $t_M=-it_E$. The classical Noether charge
for supersymmetry reads $e_{ia}(x) \j^a_{\a} \dot{x}^i$ with
$\a=1,2$, while the quantum charge is given by $Q_{\a}=
e_{ia} (x) \j^a_{\a} g^{1/4} \pi_i g^{-1/4} =
g^{-1/4}  \pi_i g^{1/4} e_{ia} (x) \j^a_{\a} $. It is hermitian and Einstein
invariant, and $\{Q_{\a},Q_{\b} \}=2\d_{\a\b} H$. This shows that
the Hamiltonian is supersymmetric and that supersymmetry is preserved
at the quantum level. Note that the variation $\d\j^a+\d x^i
\omega_i{}^{ab} \j_b$ is covariant; the `pull-back' term
$\d x^i \omega_i{}^{ab} \j_b$ is due to the presence of
fermionic equation of motion terms in the bosonic equation of motion.

Weyl-ordering of the bosonic part of the Hamiltonian gives
rise to the usual contribution (\ref{HWeyl})
\be
{1\over 8} \hbar^2 \left( R + g^{ij} \G^k_{il} \G^l_{jk} \right)
\label{WOb}
\ee
For the Weyl-ordering with respect to the fermions, we first choose the
fermions with flat indices as our independent variables. Clearly the
terms quadratic in the fermions yield no contribution, because their
anticommutator is proportional to $\d^{ab}$, which gives zero upon
contraction with $\w_{iab}$. For the terms quartic in the fermions, we
use the identity (see e.g. \cite{AFFR})
\be
{1\over 8} \left\{ \left[ \bar\j^a, \j^b \right], \left[ \bar\j^c,
\j^d \right] \right\} = \left( \bar\j^a \j^b \bar\j^c \j^d
\right)_{W} + {1\over 4} \d^{ad} \d^{bc}
\label{WO4f}
\ee
Since the four fermion terms in the Hamiltonian have exactly the same
symmetry as the operator on the left hand side of the above identity,
we can easily deduce the fermionic contribution to the Weyl-ordered
Hamiltonian to be
\be
- {1\over 8} \hbar^2 \left( R + g^{ij} \w_{ia}{}^b \w_{jb}{}^a \right)
\label{WOf2}
\ee
Adding this term to the bosonic contribution (\ref{WOb}) we find, for
the $N{=}2$ case,
\be
{1\over 8} \hbar^2 g^{ij} \left( \G^k_{il} \G^l_{jk} - \w_{ia}{}^b
\w_{jb}{}^a \right)
\label{WO2}
\ee
Alternatively, one can take the fermions with curved indices,
namely $\j^i=\frac{1}{\sqrt{2}}(\j_1^i+i \j_2^i)$ and
$\bar{\j}_i=g_{ij} \frac{1}{\sqrt{2}} (\j^j_1-i\j^j_2)$, as
independent variables, and Weyl-order with respect to these.
One finds then that now the bosonic and fermionic contributions exactly
cancel, or, in other words, the $N{=}2$ supersymmetric Hamiltonian
expressed in these variables is already Weyl-ordered \cite{AFFR}.

The $N{=}2$ Hamiltonian cannot be interpreted as the regulator of a
corresponding quantum field theory, because each of the $2n$
$\j^a_{\a}$ ($\a=1,2$) would have to correspond to a Dirac
matrix, whereas there are only $n$ Dirac matrices in an $n$-dimensional
quantum field theory (with $x^i$ with $i=1\ldots n$). However,
it plays a role in the path integral evaluation of
the index of the $\bar{\partial}$-operator
of the Dolbeault complex \cite{alwi}.

We now consider the $N{=}1$ supersymmetric case, where only one
species of Majorana fermions is present, which makes the
generalization of the previous result non-straightforward. The
Hamiltonian in this case is equal to
\bea
H_{N{=}1} &=& {1\over 2} g^{-1/4} \pi_i g^{1/2} g^{ij} \pi_j g^{-1/4} -
{1\over 8} \hbar^2 R \nonumber \\
\pi_i &=& p_i - {i\hbar\over 2} \w_{iab} \j^a \j^b
\label{HN1}
\eea
where $a=1\ldots n$, and the fermions satisfy the usual relation $\{
\j^a, \j^b \} = \d^{ab}$.

This Hamiltonian cannot be obtained by truncation of the $N{=}2$ Hamiltonian.
For example, putting $\j_1-\j_2=0$ requires $\e_1+\e_2=0$,
see (\ref{aaa47}), but
the resulting Hamiltonian has $-\frac{1}{16} \hbar^2 R$ instead
of $-\frac{1}{8} \hbar^2 R$, and is no longer supersymmetric.
The reason is that the truncation $\j_1-\j_2=0$ is no longer consistent at the
quantum level since $\{\j_1-\j_2,\j_1-\j_2\}$ is nonzero.
The easiest way to obtain (\ref{HN1}), is to start from the $N{=}1$
action with $\cl = \half g_{ij} \dot{x}^i \dot{x}^j +\half
 \j^a \frac{D}{dt} \j^a$, to construct the Noether quantum supersymmetry
charge $Q=g^{1/4} e^i_a \j^a \pi_i g^{-1/4}$, with $\pi_i= p_i
- \frac{i\hbar}{2} \omega_{iab} \j^a \j^b$, and then to evaluate
$H=\frac{1}{2} \{Q,Q\}$. The algebra is the same as
used to evaluate
$\{\Dsl,\Dsl\}$ in (\ref{eqq101}),
and leads to (\ref{HN1}).

In order to construct coherent states, we cannot work with Dirac
brackets or Majorana spinors, but we need creation and annihilation
operators. There are two ways to achieve this: either by combining
interacting Majorana spinors or by adding free Majorana spinors.

We will first combine these Majorana fermions
into complex spinors $\c$, $\bar\c$ in the following way
\be
\c^A = {1\over\sqrt{2}} ( \j^{2A-1} + i \j^{2A}) \quad ; \quad
\bar\c^A = {1\over\sqrt{2}} ( \j^{2A-1} - i \j^{2A})
\label{defDF}
\ee
where $A=1\ldots n/2$ and $\{\c^A, \bar\c^B\}=
\delta^{AB}$. The inverse relations are given by
\bea
\j^a &=& {1\over\sqrt{2}} ( \c^{(a+1)/2} + \bar\c^{(a+1)/2}) \qquad
{\rm if} \;\; a \;\; {\rm odd} \nonumber \\
\j^a &=& - {i\over\sqrt{2}} ( \c^{a/2} - \bar\c^{a/2}) \hspace{1.5cm}
{\rm if} \;\; a \;\; {\rm even}
\label{defMF}
\eea
We substitute (\ref{defMF}) into (\ref{HN1}) and define Weyl-ordering
with respect to the
operators $\c$, $\bar\c$ in the usual way. It is easy to prove,
considering separately the cases $a,b$ odd or even, that the following
equality holds
\be
\j^a \j^b = \left( \j^a \j^b \right)_{W} + {1\over 2} \d^{ab}
\label{WOmf}
\ee
We can now derive the Weyl-ordered expression corresponding to the
$N{=}1$ Hamiltonian. The bosonic part yields the same contribution as
before, see (\ref{WOb}), and, because of the above identity, the part
quadratic in the fermions is again already Weyl-ordered. It remains to
consider the term quartic in the fermions $H^{(\rm quartic)} = -
{1\over 8} \hbar^2 g^{ij} \w_{iab} \w_{jcd} \j^a \j^b \j^c \j^d$. If
only one pair of
fermions gives rise to a non-trivial anticommutator, we obtain a
contribution proportional to $g^{ij} \w_{ia}{}^c \w_{jcb} \j^a \j^b$,
which vanishes identically for symmetry reasons.
We therefore need also the second pair of
fermions to yield a non-vanishing anticommutator in order to find a
non-zero contribution. Now there are two possibilities: the two
sets of fermions are in different sectors (have different Dirac index
$A$), in which case we only need to Weyl-order both pairs separately
using (\ref{WOmf}), leading to
\bea
g^{ij} \w_{iab} \w_{jcd} \j^a \j^b \j^c \j^d &=& \left( g^{ij}
\w_{iab} \w_{jcd}
\j^a \j^b \j^c \j^d \right)_{W} \nonumber \\ && \qquad {}+
{1\over 2} g^{ij} \w_{ia}{}^b \w_{jb}{}^a
\label{WOmf4}
\eea
where we should remember that, if we define $A=[(a+1)/2]$ etc., not
all four indices $A,B,C,D$ are identical.
The second possibility
corresponds to the case $A,B,C,D$ all identical.
In that case $a,b,c,d$ are all equal to $2k+1$ or $2k+2$, and we only need
to consider the case that two of them are equal to $2k+1$ and the other two
equal to $2k+2$. In this case the Weyl ordered expression
vanishes\footnote{To see this, take $k=0$. Any real linear combination
of $\j^1$ and $\j^2$ can be written as $\a\j+\bar{\a} \bar{\j}$
for some complex $\a$. The Weyl ordered expression of an arbitrary
combination of $\j^1$ and $\j^2$ is proportional to the sum over
all graded permutations of $\a,\b,\g,\d$ of the operator
$ (\a\j+\bar\a\bar\j) (\b\j+\bar\b\bar\j) (\g\j+\bar\g\bar\j)
(\d\j+\bar\d\bar\j)$. This equals $\a\bar\b\g\bar\d\j\bar\j+
\bar\a\b\bar\g\d\bar\j\j$, and the sum over graded permutations
of the ordinary constants $\a,\bar{\b},\g,\bar{\d}$ and
$\bar{\a},\b,\bar{\g},\d$
clearly vanishes.}.
Hence this leads to the same result, so that in fact
(\ref{WOmf4}) is valid for all $a,b,c,d$.
For instance $\j^1 \j^2 \j^2
\j^1 = {1\over 4}$ and $(\j^1 \j^2 \j^2 \j^1)_W=0$;
we end up with a coefficient ${1\over 2}$ in (\ref{WOmf4})
because we obtain the same contribution from a term $\j^1 \j^2 \j^1
\j^2 = - {1\over 4}$. Adding the bosonic and fermionic parts, we find the
total contribution from Weyl-ordering to the scalar potential
\be
{1\over 8} \hbar^2 \left( R + g^{ij} \G^k_{il} \G^l_{jk} \right)
- {1\over 16} \hbar^2 g^{ij} \w_{ia}{}^b \w_{jb}{}^a
\label{WO1}
\ee
As one might have anticipated, the Weyl-ordering of the Majorana
fermions yields half the result for Dirac fermions, see (\ref{WO2}).

The other way to construct Dirac fermions from Majorana fermions is
to add a second set of free Majorana fermions. Denoting the original
fermions $\psi^a$ by $\psi^a_1$, and the new ones by $\j^a_2$,
we again construct Dirac fermions $\c$ and $\bar\c$
(as in (\ref{defDS})), but then
we use the $N{=}2$ formulation given before. The four fermion term in the
Hamiltonian now reads
\be
-\frac{\hbar^2}{8} \j^a_1 \j^b_1 \j^c_1 \j^d_1
\omega_{iab} \omega_{jcd} g^{ij}
\ee
and we should in this case Weyl order it with respect to $\c^a$
and $\bar\c^a$. Again, (\ref{WOmf}) holds for $\j^a_1$,
 even though the definition
of $\c^a$ and $\bar\c^a$ is now different. Using the operator
identity
\bea
\j^a_1 \j^b_1 \j^c_1 \j^d_1 & = &
\frac{1}{4} (\delta^{ab} \delta^{cd} - \delta^{ac} \delta^{bd}
 + \delta^{ad} \delta^{bc} ) \nonumber \\
& & + (\frac{1}{2} \delta^{ab} (\j^c_1 \j^d_1)_W + \mbox{{\rm
five more terms}}) \nonumber \\
& & + ( \j^a_1 \j^b_1 \j^c_1 \j^d_1 )_W
\eea
we find that the two-fermion terms vanish due to anti-symmetry
while the double contractions yield the same answer as in
(\ref{WO1}).

Finally we can also Weyl-order the $N{=}1$ Hamiltonian with respect to
the fermionic variables $\c$ and $\bar\c$ in (\ref{defDF}),
but now with curved rather than flat indices. In that
case one expects to be left with a remainder
\be
{1\over 8} \hbar^2 \left( R + {1\over 2} g^{ij} \G^k_{il}
\G^l_{jk} \right).
\label{WO1c}
\ee

\subsection{Evaluation of supersymmetric transition elements}

\subsubsection{The $N{=}2$ case}

We will first compute the transition element for the $N{=}2$ case
using the path integral formulation.
We shall obtain all terms through order $\b$.
We rescale
$\t=t/\b$ to make the $\b$-dependence more explicit and
facilitate keeping track of the order in the expansion in $\b$.
First note that we can write
\be
S = S_{\rm bos} + S^{kin}_{\rm fer} + S^{int}_{\rm fer}
\label{SSplit}
\ee
Here $S_{\rm bos}$ contains all terms with only bosonic or ghost fields. Note
however that it is not identical to the action we wrote for the
bosonic path integral, as the Weyl-ordering of the fermionic terms
also gives a contribution $\sim (R + \w^2)$. The
contributions to the path integral
involving only this part of the action can trivially be found from the
bosonic case, because the latter terms can, through order $\b$, simply
be replaced by their classical expectation values. The
other two terms are given by
\be
\frac{1}{\hbar} S^{kin}_{\rm fer} = \int\limits^0_{-1} \! d\t \,
\delta_{ab} \bar\j^a \dot\j^b -
\delta_{ab} \bar\j^a(0) \j^b(0)
\label{SKinfer}
\ee
and, after integrating out the momenta $p_i$,
\be
\frac{1}{\hbar} S^{int}_{\rm fer} = \int\limits^0_{-1}
\! d\t \, \left[ \dot x^i \omega_{iab} \bar\j^a \j^b - {\hbar \over 2}
\b R_{abcd} \bar\j^a \j^b \bar\j^c \j^d \right]
\label{SIntfer2}
\ee
In the background field approach we decompose $\j(\t) = \j_{\rm bg}(\t) +
\j_{\rm qu}(\t)$, and $\bar\j(\t) = \bar\j_{\rm bg}(\t) +
\bar\j_{\rm qu}(\t)$. We
choose $\bar\j_{\rm bg}(\t)$ and $\j_{\rm bg}(\t)$ to be the solutions to
the equations of motion of the kinetic part of the action,
(\ref{SKinfer}), which satisfy the boundary conditions, i.e. we
take $\bar\j_{\rm bg}^a(\t) =\bar\eta^a$ and $\j_{\rm bg}^a(\t) =\chi^a$.
Since we
require $\bar\j^a(0)=\bar\eta^a$ and $\j^a(-1) = \chi^a$, this
implies that the quantum fields need to satisfy the boundary
conditions $\bar\j^a_{\rm qu}(0) = 0$ and $\j^a_{\rm qu}(-1) =0$.
Then (\ref{SKinfer}) simplifies to $\int_{-1}^0
\bar{\j}^a_{\rm qu} \dot\j^a_{\rm qu} d\t - \bar\h^a \c^a$, i.e., there
are no terms linear in fermionic quantum fields in (\ref{SKinfer}).
Of course, (\ref{SIntfer2}) does contain such terms.

We can now compute the transition element by
expanding $\exp \left( - {1\over\hbar}  S^{int}_{\rm fer} \right)$
and contracting the quantum fields, using
\be \label{aaa64}
\frac{1}{\hbar} S^{\rm int}_{\rm bos} =
\mbox{(\ref{path18})} + \frac{1}{8}
\hbar \b \int_{-1}^0 (-R + g^{ij} \omega_{iab} \omega_j{}^{ab})
\ee
and the propagators for the fermions given in (\ref{conFP}).
When we expand $\exp \left( -{1\over\hbar}
S^{int}_{\rm fer} \right)$ we will for the first term in this
expansion only need the contraction at equal time  $<\bar\j^a(\t)
\j^b(\t) > = - {1\over 2}\hbar\delta^{ab}$.
Hence, the first term in (\ref{SIntfer2}) will yield no
contribution, whereas the second term contributes ${1\over 2} \b \hbar
R_{ab}\bar\eta^a\chi^b-{1\over 8}\b\hbar R$.
The term $-\frac{1}{8} \b \hbar R$ cancels the $R$ term in (\ref{aaa64}).
Next consider the
term ${1\over 2} \left( {1\over\hbar} S^{int}_{\rm fer} \right)^2$,
where only terms from the square of the $\dot x \omega \bar{\j} \j$ term
contribute to this order.
When we contract one pair of fermions, we find the $\omega\omega$ term
in (\ref{SFer}).
When we contract only the four fermionic fields, or two $\dot x$
fields and two fermionic fields, one finds zero.
Finally, we can contract four fermionic and two bosonic
fields.
This yields a contribution $-{1\over 8}\b\hbar g^{ij}
\omega_{ia}{}^b \omega_{jb}{}^a$,
which cancels the $\omega\omega$ term in (\ref{aaa64}). Contractions
involving the other bosonic
fields are again of higher order and need not be considered.

Taking all contributions into account, we find for the amplitude
\bea
\SAmpl &=& (2\pi\b\hbar)^{-n/2} \exp
\left( - {1\over\hbar} S_B - {1\over\hbar}  S_F \right)
\nonumber \\ && \hskip -3.2cm
\left[ 1 - {1\over 12} \b\hbar R(z) - {1\over 12}
 R_{ij}(z) (y-z)^i (y-z)^j + {1\over 2}  \b \hbar R_{ab}(z)
\bar\eta^a \chi^b \right] \nonumber \\
\label{STraAmp}
\eea
where
\bea
S_B &=& {1\over 2\b} g_{ij}(z) (y-z)^i
(y-z)^j + {1\over 4\b} \partial_k g_{ij}(z) (y-z)^i (y-z)^j
(y-z)^k \nonumber\\ && \hskip -.8cm + {1\over 12\b} \Bigl(
\partial_k\partial_l
g_{ij}(z) - {1\over 2} g_{mn}(z) \Gamma^m_{ij}(z) \Gamma^n_{kl}(z)
\Bigr) (y-z)^i (y-z)^j (y-z)^k (y-z)^l \nonumber \\
\label{SBos}
\eea
is the expansion through order $\b$ of the length of the
geodesic joining $z$ and $y$ (cf. (\ref{trans})), and
\bea
 S_F &=& - \hbar \delta_{ab} \bar\eta^a
\chi^b - \hbar (y-z)^i \omega_{iab}(z) \bar\eta^a \chi^b
\nonumber \\ && \qquad - {1\over2} \hbar (y-z)^i (y-z)^j \Bigl( \partial_i
\omega_{jab}(z) + \omega_{ia}{}^c(z) \omega_{jcb}(z) \Bigr) \bar\eta^a
\chi^b \nonumber \\ && \qquad - {1\over 2} \b\hbar^2 R_{abcd}(z)
\bar\eta^a \chi^b \bar\eta^c \chi^d
\label{SFer}
\eea
The $\omega$ and $\partial\omega$ terms are obtained by expanding the first
term in (\ref{SIntfer2}).
In the continuum limit, $S_F$ becomes the
fermionic action, including the correct boundary term, as derived
above (\ref{aux29}),
%If we add an extra boundary term $\hbar \delta_{ab}\bar\psi^a_{\rm cl}(0)
%\psi^b_{\rm cl}(0) = \hbar \delta_{ab} \bar\eta^a \psi^b_{\rm cl}(0)$ to
%(\ref{SFer}), then the sum of the first term in (\ref{SFer})
%together with this extra term becomes the leading term in the
%fermionic action
\be
S_F = \hbar \int\limits_{-1}^0\!
d\t\,\left( \delta_{ab}
\bar\psi^a \dot\psi^b + \dot x^i \omega_{iab} \bar\psi^a \psi^b -
{1\over 2}\b \hbar R_{abcd} \bar\psi^a \psi^b \bar\psi^c \psi^d \right)
 - \hbar\delta_{ab} \bar\eta^a
\psi^b_{\rm bg}(0) .
\label{SFercon}
\ee

We can easily check that the expansion through order $\b$ of
(\ref{SFercon}) indeed equals the expression in (\ref{SFer})
when the equations of motion are imposed. The latter follow from
(\ref{ff44})
(of course we
also need the bosonic part of the action to find the full equations of
motion)
\bea
\ddot x^i + \Gamma^i{}_{jk} \dot x^j \dot x^k
- \hbar R^i{}_{jab} \dot x^j \bar\psi^a \psi^b + {1\over 2} \hbar^2 g^{ij}
\left( \partial_j R_{abcd} \right)\bar\psi^a \psi^b \bar\psi^c \psi^d =
0 \nonumber \\ \dot\psi^a + \dot x^i
\omega_i{}^a{}_b \psi^b - \hbar R^a{}_{bcd} \psi^b \bar\psi^c \psi^d = 0
\hskip 2cm \nonumber \\ \dot{\bar\psi^a} + \dot x^i
\omega_i{}^a{}_b \bar\psi^b - \hbar R^a{}_{bcd} \bar\psi^b \bar\psi^c \psi^d
= 0 \hskip 2cm
\label{eom}
\eea
where a dot denotes differentiation with respect to $t$.
We can now expand the Lagrangian in a Taylor series around its value
at $t=0$, and then do the trivial time integrations. This yields
\be
S= \b L(0) - {1\over 2} \b^2 \dot L(0) + \ldots
\label{TayL}
\ee
We thus expand all fields in the Lagrangian around their
values at $t=0$, making use of the equations of motion (\ref{eom}).
The expansions up to the order we need are given by
\bea
\dot x^i(0) &=& {(z-y)^i \over \b}
+ {\b\hbar\over 2} (z-y)^j R^i{}_{jab} \bar\eta^a
\chi^b  \nonumber \\ \bar\psi^a(0) &=& \bar\eta^a \nonumber \\
\psi^a(0) &=&
\chi^a - (z-y)^i \omega_i{}^a{}_b \chi^b \nonumber \\
\dot\psi^a(0) &=& {\psi^a(0) -\chi^a \over \b} - {1\over 2}
{(z-y)^i (z-y)^j \over \b}\left( \partial_i \omega_j{}^a{}_b -
\omega_i{}^a{}_c \omega_j{}^c{}_b \right) \chi^b \nonumber \\
\label{expter}
\eea
Inserting these expansions into $ S_F$ in (\ref{SFercon})
and (\ref{TayL})
yields the expression (\ref{SFer}).

We will now show that the final result for the transition amplitude
can again be written as the product of three factors: a term
containing only the scalar curvature which is related to the trace
anomaly, the exponent of the classical action, and the square root of,
in this case, the supersymmetric generalization of the Van Vleck
determinant. The latter we define by
\be
D_S = {\rm sdet} D_{AB} \quad ; \quad D_{AB} \equiv -
{\partial\over \partial \Phi^A}  \left( S_B +  S_F \right)
{\overleftarrow{\partial}\over \partial \Phi^B}
\label{defDS2}
\ee
where $\Phi^A = (z^i, \bar\eta^a)$ and $\Phi^B =(y^j, \chi^b)$, and
for $S_B$ and $ S_F$ we substitute the expressions
(\ref{SBos}) and (\ref{SFer}).
To evaluate $D_S$ write
\be
D_{AB} = \left( \ba{cc}  A_{ij} & B_{ib} \\
C_{aj} & D_{ab} \ea  \right)
\label{DisABCD}
\ee
We find, expanding in normal co-ordinates around $z$ to simplify the
expressions,
\bea
A_{ij} &=& {1\over\b} g_{ij}(z) +
\hbar \partial_i \omega_{jab}(z) \bar\eta^a \chi^b \nonumber \\ && \qquad
- {\hbar\over 2} \Bigl( \partial_i \omega_{jab}(z) + \partial_j
\omega_{iab}(z) + \omega_{ia}{}^c(z) \omega_{jcb}(z) +
\omega_{ja}{}^c(z) \omega_{icb}(z) \Bigr) \bar\eta^a
\chi^b \nonumber \\ B_{ib} &=& - \hbar \omega_{iab}(z) \bar\eta^a \nonumber \\
C_{aj} &=& \hbar \omega_{jab}(z) \chi^b \nonumber \\ D_{ab} &=&
\hbar \delta_{ab} +\hbar  (y-z)^i \omega_{iab}(z) + {\hbar \over
2} (y-z)^i (y-z)^j \Bigl( \partial_i \omega_{jab}(z) +
\omega_{ia}{}^c(z) \omega_{jcb}(z) \Bigr) \nonumber \\ && \qquad +\hbar  \b
\Bigl( R_{abcd}(z) - R_{adcb}(z) \Bigr) \bar\eta^c \chi^d
\label{ABCD}
\eea
We do not need terms of order $\b$ in $B$ and $C$, since $D_S =
\det A \det^{-1} \bigl( D - C A^{-1} B \bigr)$ and $A^{-1}$ is already
of order $\b$. Writing $A_{ij} = {1\over\b} g_{ik} \bigl(
\delta^k{}_j + \b\hbar a^k{}_j \bigr)$ and $D_{ab} =\hbar (\delta_{ab} +
d_{ab})$, we can write the expansion of the super Van Vleck
determinant as
\be
D_S^{1/2} = (\b\hbar)^{-n/2} g^{1/2}(z)
\left[ 1 + {1\over 2} \b\hbar\,\tr a + {1\over 2} \b\hbar\,\tr CB -
{1\over 2} \tr d + {1\over 8} ( \tr d)^2 + {1\over 4} \tr (d^2) \right]
\label{expSdet}
\ee
Multiplying by $g^{-1/4}(z) g^{-1/4}(y)$ to transform $D_S^{1/2}$ into
a bi-scalar, we obtain
\bea
\tilde D_S^{1/2} &\equiv & (\b\hbar)^{n/2}
 g^{-1/4}(z) D_S^{1/2} g^{-1/4}(y) \nonumber \\
&=&
\left[ 1 - {1\over 12} R_{ij}(z) (y-z)^i
(y-z)^j + {1\over 2} \b\hbar R_{ab} \bar\eta^a \chi^b \right]
\label{DSis}
\eea
So indeed we can write
\be
\SAmpl = (2\pi\hbar\b)^{-n/2} \tilde D_S^{1/2} \exp
\left( - {1\over \hbar} \bigl( S_B +  S_F \bigr) \right) \left[
1 - {1\over 12} \b \hbar R \right]
\label{proprew}
\ee
All terms involving the Ricci curvature in
(\ref{STraAmp}) are thus completely accounted for by the super Van
Vleck determinant, which clearly would not have been the case if we
had used the ordinary determinant. Finally we note that if we would have
rescaled all fermions by a factor of $(\b\hbar)^{-1/2}$, then all
classical terms are proportional to $1/(\b\hbar)$, while all
one-loop terms in (\ref{DSis}) are $\b\hbar$ independent and the
two-loop term in (\ref{proprew}) remains proportional to $\b\hbar$
\cite{bas}.

\subsubsection{The $N{=}1$ case}

Next we consider the $N{=}1$ case. We will first evaluate the
transition element when we double the number of Majorana fermions, and
afterwards consider the case that we combine the Majorana fermions into
half as many Dirac fermions. In the first case we add $n$ free
fermions $\j^a_2$ and combining
$1/\sqrt{2}(\j^a_1+ i\j^a_2) = \j^a$ and
$1/\sqrt{2}(\j^a_1 - i\j^a_2) = \bar\j^a$  we construct the path
integral with the corresponding coherent states. The
 kinetic part of the fermionic action is again
\be
\frac{1}{\hbar} S^{kin}_{\rm fer} = \int\limits^0_{-1} \! d\t \,
\delta_{ab} \bar\j^a \dot\j^b - \delta_{ab} \bar\j^a (0) \j^b(0)
\label{SKinFer}
\ee
and yields the free field equations $\dot\j^a=\dot{\bar\j}{}^a=0$, but
the interaction part containing the fermions is now equal to
\be
\frac{1}{\hbar} S^{int}_{\rm fer} = \int\limits^0_{-1}
\! d\t \, \left[ {1\over 2} \dot x^i \omega_{iab} \j^a_1 \j^b_1 \right]
-\bar{\eta} \c
\label{SIntfer1}
\ee
Note that the terms linear in $\psi_{\rm qu}(0)$
in (\ref{SKinFer}) cancel.

We have introduced an extra set of fermions that do not couple to any
of the other fields, this way making certain that we do not alter the
dynamics. In order to preserve local Lorentz invariance we require that
the fermions $\psi^a_2$ are inert under local Lorentz transformations.
The extra contribution to the scalar potential from Weyl-ordering the
$N{=}1$ Hamiltonian is of the form ${\hbar^2\over 8} (\G\G +R) -
{1\over 16} \w\w$
to which one should add the term $-\frac{\hbar^2}{8} R$
from (\ref{HN1}), and its integral can be replaced to order $\b$ by
$\b$ times its classical value. The purely bosonic
sector of the path integral can be evaluated exactly as before, so we
only need to consider the sector involving fermions. We use a
background field expansion as in the $N{=}2$ case, again with
constant background fields which satisfy the boundary conditions,
and substitute $\psi_1^a=\frac{1}{\sqrt{2}} (\bar{\eta}^a + \chi^a)
+ \psi^a_{1,{\rm qu}}$ into (\ref{SIntfer1}).
Since the interactions depend only on $\j_1$, it will be easier to use
the propagator for $\j_{1,{\rm qu}}$
\be
< \psi^a_{1,{\rm qu}}(\s)
\psi^b_{1,{\rm qu}}(\t) > = {1\over 2}  \delta^{ab}
\bigl( \theta(\s-\t) - \theta(\t-\s) \bigr)
\label{ferGreen}
\ee
which can trivially be found from the $\bar\j \j$ propagator in (\ref{conFP})
and the propagators $\< \j\j \> = \< \bar\j \bar\j \> = 0$.
We now evaluate $\< \exp -\frac{1}{\hbar} S^{\rm int}_{\rm fer} \>$.
We first consider terms from contractions in
$-{1\over\hbar}S^{int}_{\rm fer}$.
The equal time
contraction of $\j^a_1 \j^b_1$ in (\ref{ferGreen}) vanishes, so this
term will yield no contribution. Also the
contribution from the contraction in $\dot x^i \omega_{iab}$ in
(\ref{SIntfer1}) vanishes, as it is proportional to
$\int\limits_{-1}^0 d\t \, < \dot x^i(\t) x^j(\t) >$ which is
zero. Next we consider
the term  ${1\over 2}
\left( {1\over\hbar} S^{int}_{\rm fer} \right)^2$.
The contraction of $\dot
x^i(\s)$ with $\dot x^j(\t)$ is of order $\b$, and would
contribute, but its integral over $\s$ and $\t$ vanishes. One
$\dot x^i(\s)$ contracted with $x^j(\t)$ is also of order
$\b$, but the other $\dot{x}^j$
would leave a factor $\dot x^i_{\rm bg}$ which is of order
$\b^{1/2}$, so this term is of higher order.
The contraction of only one pair of fermions vanishes for
symmetry reasons, so there is now no $\omega\omega$ in
(\ref{Soverh}).
The
contraction of all four fermions is nonvanishing, and
produces the term ${1\over 16} (z^i-y^i) (z^j-y^j) \omega_{ia}{}^b
\omega_{jb}{}^a$ in (\ref{FerTrans}).
Finally, we can contract four fermionic fields and two bosonic fields
$\dot x^i$. This yields $-{1\over 16}\b\hbar g^{ij}
\omega_{ia}{}^b \omega_{jb}{}^a$, and this term exactly cancels a similar
noncovariant term in the scalar potential due to Weyl-ordering.
Adding all contributions we find the transition element for the
$N{=}1$ case with fermion doubling
\bea
\SAmpl &=& (2\pi\b\hbar)^{-n/2}
\exp\left( -{1\over\hbar} \bigl( S_B +  S_F \bigr) \right)
\nonumber \\ && \hskip -1.5cm \biggl[ 1 + {1\over 24}
\b\hbar R(z) \nonumber - {1\over 12} R_{ij}(z) (y-z)^i
(y-z)^j \nonumber \\ && \hskip -1.5cm \qquad + {1\over 16} (y-z)^i
(y-z)^j \omega_{ia}{}^b(z) \omega_{jb}{}^a(z) \biggr]
\label{FerTrans}
\eea
where
\bea
{1\over\hbar} S_F &=&  -
\delta_{ab} \bar\eta^a \chi^b - {1\over 4} (y-z)^i \omega_{iab}(z)
(\bar\eta^a + \chi^a) (\bar\eta^b + \chi^b)  \nonumber \\ && \qquad -
{1\over 8}
(y-z)^i (y-z)^j \partial_i \omega_{jab}(z) (\bar\eta^a + \chi^a)
(\bar\eta^b + \chi^b)
\label{Soverh}
\eea
and $S_B$ is the bosonic part of the classical action.
%One may check that this is the expansion around $z$ of
%(\ref{SKinFer}) plus (\ref{SIntfer1}).
It is the same result
as obtained directly from operator methods \cite{bas}.

The terms in $S_F$ are obtained by expanding the following
classical continuum action around $z$
\bea
\frac{1}{\hbar} S & = & -\bar{\eta}^a \j^a(0) +
 \int_{-1}^0 dt \left[ \
\frac{1}{\b\hbar} \frac{1}{2} g_{ij} \dot{x}^i \dot{x}^j +
\bar\j^a \dot\j^a  \right. \nonumber \\
& & + \left. \frac{1}{2} \dot{x}^i \omega_{iab} \j_1^a \j_1^b
%+\frac{\b\hbar}{8} g^{ij} (\Gamma_i{}^k{}_l \Gamma_j{}^l{}_k +
%\omega_{iab} \omega_j{}^{ab})
 \right].
\eea
The equations of motion read
\bea
0 & = & \ddot{x}^i + \Gamma^i_{jk} \dot{x}^j \dot{x}^k
-\b\hbar \frac{1}{4} \dot{x}^j R^i{}_{jab} (\j+\bar\j)^a
(\j+\bar\j)^b \nonumber \\
0 & = & \dot{\j}^a + \frac{1}{2} \dot{x}^i
\omega_i{}^{ab} (\j+\bar\j)^b \nonumber \\
0 & = & \dot{\bar\j}^a + \frac{1}{2} \dot{x}^i
\omega_i{}^{ab} (\j+\bar\j)^b
\eea
where $R_{ijab}=\del_i \omega_{jab} +
\omega_{iac} \omega_{jcb} - ( i \leftrightarrow j)$. From
them one derives further
\bea
\dot{x}^i(0) & = & (z-y)^i -\frac{1}{2} \Gamma^i_{jk}(z)
 (z-y)^j (z-y)^k +\frac{\b\hbar}{8} (z-y)^j R^i{}_{jab}
 (\bar\eta+\c)^a (\bar\eta+\c)^b + \ldots \nonumber \\
\psi^a(0) & = & \c^a - \frac{1}{2} (z-y)^k \omega_k{}^{ab}
 (\bar\eta+\c)^b + \frac{1}{4} (z-y)^i (z-y)^j
\del_i \omega_j{}^{ab} (\bar\eta+\c)^b+\ldots
\eea
Substituting these results into $S=-\bar\eta^a \j^a(0) +
L(0)-\frac{1}{2} \frac{d}{d\t} L(0) + \ldots$, the
contribution from
$-\frac{1}{2} \frac{d}{d\t} (\bar\j \dot\j)$ cancels the terms
in $-\bar\eta^a \j^a(0) + \bar\j^a(0) \dot{\j}^a(0) $,
and one indeed arrives at (\ref{Soverh}).

We will now show that the expression for the propagator can again be
written as the product of the super Van Vleck determinant, the
exponent of the classical action, and a term involving the scalar
curvature which, as shown in section 3.3, determines the
trace anomaly of a spin-${1\over 2}$ field. Defining the super Van
Vleck determinant as in the $N{=}2$ case, we
find from (\ref{Soverh})
\bea
A_{ij} &=& {1\over\b} g_{ij}(z) +
{\hbar\over 8} \Bigl( \partial_i \omega_{jab}(z) - \partial_j
\omega_{iab}(z) \Bigr) (\bar\eta^a + \chi^a ) (\bar\eta^b + \chi^b )
\nonumber \\ B_{ib} &=& - {\hbar\over 2} \omega_{iab}(z) ( \bar\eta^a +
\chi^a ) \nonumber \\
C_{aj} &=& {\hbar\over 2} \omega_{jab}(z) (\bar\eta^b + \chi^b ) \nonumber
\\ D_{ab}
&=& \hbar\delta_{ab} + {\hbar\over 2} (y-z)^i \omega_{iab}(z)
+ {\hbar\over 4}
(y-z)^i (y-z)^j \partial_i \omega_{jab}(z).
\label{1ABCD}
\eea
Using again (\ref{expSdet}), one finds
\be
\tilde D_S^{1/2} =  \Bigl[ 1 -
{1\over 12} R_{ij}(z) (y-z)^i (y-z)^j + {1\over 16} (y-z)^i (y-z)^j
\omega_{ia}{}^b (z) \omega_{jb}{}^a (z) \Bigr]
\label{DS1is}
\ee
where only the last term in (\ref{expSdet}) did contribute and yields
the last term in (\ref{DS1is}).
So we can indeed write
\be
\SAmpl = (2\pi\hbar\b)^{-n/2} \tilde D_S^{1/2} \exp
\left( - {1\over\hbar} \bigl( S_B +  S_F \bigr) \right) \Bigl[ 1
+ {1\over 24} \b \hbar R \Bigr]
\label{prop1re}
\ee
similarly to the bosonic and $N{=}2$ supersymmetric case.

We will now repeat the analysis for the $N{=}1$ case when we do
not introduce an extra set of fermions, but instead combine the $n$
Majorana fermions $\j^a$ into $n/2$ Dirac fermions $\Psi^A$,
$\bar\Psi^A$ as $\Psi^A = {1\over\sqrt{2}} ( \j^{2A-1} + i \j^{2A})$,
$\bar\Psi^A = {1\over\sqrt{2}} ( \j^{2A-1} - i \j^{2A})$.
The kinetic part of the fermionic action is now equal to
\be
\frac{1}{\hbar} S^{kin}_{\rm fer} = \int\limits^0_{-1} \! d\t \,
\delta_{AB} \bar\Psi^A  \dot\Psi^B -
\delta_{AB} \bar\Psi^A(0)  \Psi^B(0)
\label{SKINFer}
\ee
The interaction part containing the fermions is still equal to
\be
\frac{1}{\hbar} S^{int}_{\rm fer} = \int\limits^0_{-1}
\! d\t \, \left[ {1\over 2} \dot x^i \omega_{iab} \j^a \j^b \right]
\label{SINTfer1}
\ee
but now the $\j^a$ should be expressed in terms of $\Psi^A$ and
$\bar\Psi^A$. We again
make a background field decomposition as $\Psi^A = \chi^A +
\Psi^A_{\rm qu}$, $\bar\Psi^a = \bar\h^A + \bar\Psi^A_{\rm
qu}$. Again it will be convenient to rewrite the propagators for the
Dirac fermions in terms of the Majorana fermions. We now find
\be
< \j^a (\s) \j^b (\t) > = {1\over 2}  \d^{ab}
\bigl( \theta(\s-\t) - \theta(\t-\s) \bigr) +  K^{ab}
\label{FerGr1}
\ee
where $K^{ab} = {i\over 4} \d^{a+1,b} \Bigl( 1 - (-1)^a \Bigr) -
(a\leftrightarrow b)$ (in other words, $K$ is the matrix $-\frac{1}{2}\t_2$
in the $2\times 2$ subspaces).
Furthermore, we define the
`background' Majorana fermions
\bea
\tilde\j^a &=& {1\over \sqrt{2}} ( \chi^{(a+1)/2} +
\bar\h^{(a+1)/2} ),  \qquad a \quad {\rm odd} \nonumber \\
\tilde\j^a &=& - {i\over \sqrt{2}} ( \chi^{a/2} -
\bar\h^{a/2} ), \hspace{1.5cm} a \quad {\rm even}
\label{BCMaj}
\eea
We start again by considering contractions in
$-{1\over\hbar}S^{int}_{\rm fer}$. The equal time
contraction of $\j^a \j^b$ is now equal to $K^{ab}$. These terms
therefore contribute
\be
{1\over 2} (y-z)^i \w_{iab} K^{ab} + {1\over 4}
(y-z)^i (y-z)^j \partial_i \w_{jab} K^{ab}
\label{ctS1}
\ee
Next we consider contractions in ${1\over 2}
\left( {1\over\hbar} S^{int}_{\rm fer} \right)^2$. When we contract
one pair of fermions, we obtain the tree graph
\be
{1\over 8 } (y-z)^i (y-z)^j \w_{iab} \w_{jcd} \left(
%2 K^{ab} \tilde\j^c
%\tilde\j^d
- 4 K^{ac} \tilde\j^b \tilde\j^d \right).
\label{ctS21c}
\ee
When we contract two pairs of fermions, we find
the one-loop graph
\be
{1\over 8} (y-z)^i (y-z)^j \w_{iab} \w_{jcd} \left(
{1\over 2} \d^{bc} \d^{ad}
-2 K^{ac} K^{bd}
+ K^{ab} K^{cd}
\right)
\label{ctS22c}
\ee
The last term is a product of two one-loop graphs, but should of
course not be counted as a new two-loop graph.
When we contract the two $\dot x$ fields, we only find a
contribution if in addition we contract all fermionic fields. This
yields $-{1\over 16} \b\hbar g^{ij} \w_{ia}{}^b \w_{jb}{}^a$, which
cancels again a similar contribution from Weyl-ordering (see (\ref{WO1})).
Adding all terms,
we find for the transition element for the $N{=}1$ case without fermion
doubling
\be
\SAmpl = (2\pi\b\hbar)^{-n/2} \, E \, \exp
\left( - {1\over\hbar} \bigl( S_B +  S_F \bigr) \right) \Bigl[ 1
+ {1\over 24}  \b\hbar R \Bigr]
\label{PROP1re}
\ee
where  $ S_F$ is
equal to the background part of the fermionic action
(together with the boundary term) plus the tree graph of (\ref{ctS21c})
\bea
{1\over\hbar} S_F &=&  -
\delta_{AB} \bar\eta^A \chi^B - {1\over 2} (y-z)^i \omega_{iab}(z)
\tilde\j^a \tilde\j^b  \nonumber \\ && \qquad -
{1\over 4}
(y-z)^i (y-z)^j \partial_i \omega_{jab}(z) \tilde\j^a \tilde\j^b
\nonumber \\ && \qquad +
\frac{1}{2} (y-z)^i (y-z)^j \omega_{iab} \omega_{jcd} K^{ac}
\tilde\j^b \tilde\j^d
\label{SOverh}
\eea
The one but last term is due to expanding $\omega(z+(z-y)\t)$ around
$z$. As the notation indicates, to order $\b$ $S_F$ is equal to the
classical action (\ref{SKINFer}) and (\ref{SINTfer1}) with the
equations of motion satisfied. Their solution is, however, quite a bit more
complicated than it was in the previous cases, due to the different
boundary conditions we have to impose here.
All one-loop contributions are contained in
$E$,
\bea
E &=&  \Biggl[ 1 -
{1\over 12} R_{ij}(z) (y-z)^i (y-z)^j
\nonumber \\
&& \hskip2cm {} + \left( {1\over 2} (y-z)^i \w_{iab} + {1\over 4}
(y-z)^i (y-z)^j
\partial_i \w_{jab} \right) K^{ab}
\nonumber \\
&& \hskip2cm {} + {1\over 8} (y-z)^i (y-z)^j \w_{iab} \w_{jcd}
\biggl( {1\over 2} \d^{bc} \d^{ad} \nonumber \\ && \hskip2cm \qquad {}
 + K^{ab} K^{cd} - 2 K^{ac} K^{bd}
\biggr) \Biggr]
\label{PosD}
\eea
Comparing with (\ref{Soverh}), (\ref{DS1is}) and
(\ref{prop1re}) which yield the
transition element with fermion doubling we note three differences:
$(i)$ the background value of the $\j$ in the interactions are
defined differently, namely $\chi^a+\bar\h^a = \sqrt{2} \tilde\j_1^a$ in
(\ref{Soverh}) and $\tilde\j^a$ in (\ref{BCMaj}),  $(ii)$ the
boundary term $\bar\h \chi$ contains half as many terms in
(\ref{SOverh}) as in (\ref{Soverh}), and $(iii)$
there are extra terms in (\ref{SOverh}) and
(\ref{PosD}) proportional to $K$ and $KK$. Yet, as we shall see, these
different transition elements yield the same anomalies.

Motivated by our earlier results, we will now compare the expression $E$
to the square root of the super Van Vleck determinant.
Defining the super Van Vleck determinant as before, we find using
(\ref{SOverh})
\bea
A_{ij} &=& {1\over\b} g_{ij}(z) +
{1\over 4} \Bigl( \partial_i \omega_{jab}(z) - \partial_j
\omega_{iab}(z) \Bigr) \tilde\j^a \tilde\j^b
\nonumber \\
& & + \frac{1}{2} (\omega_{iab} \omega_{jcd} +
 \omega_{jab} \omega_{icd}) K^{ac} \tilde\j^b \tilde\j^d \nonumber \\
B_{iB} &=& - \frac{1}{2}
\omega_{iab}(z) \left( \tilde\j^a \tilde\j^b
\right) {\overleftarrow{\partial}\over \partial \chi^B}
\nonumber \\
C_{Aj} &=& \frac{1}{2}
\omega_{jab}(z) {\partial\over \partial \bar\h^A} \left(
\tilde\j^a \tilde\j^b \right)
\nonumber
\\ D_{AB}
&=& \delta_{AB} + \left[ {1\over 2} (y-z)^i \omega_{iab}(z) + {1\over 4}
(y-z)^i (y-z)^j \partial_i \omega_{jab}(z) \right. \nonumber \\ &&
\qquad \left. -\frac{1}{2} \omega_{iac} \omega_{jbd} (z-y)^i
 (z-y)^j K^{ac} \right] {\partial\over \partial \bar\h^A} \left(
\tilde\j^a \tilde\j^b \right) {\overleftarrow{\partial}\over \partial
\chi^B}
\label{1mABCD}
\eea
The expansion of the super Van Vleck determinant through order $\b$
can again be written as
\bea
\tilde D_S^{1/2} &=&  g^{1/4}(z) g^{-1/4}(y)
\biggl[ 1 + {1\over 2} \b\,\tr a + {1\over 2} \b\,\tr CB
\nonumber \\ && \hskip 2cm {} -
{1\over 2} \tr d + {1\over 8} ( \tr d)^2 + {1\over 4} \tr (d^2) \biggr]
\label{ExpSdet}
\eea
Now using identities such as
\be
\d^{AB} {\partial\over \partial \bar\h^A} \left(
\tilde\j^a \tilde\j^b \right) {\overleftarrow{\partial}\over \partial
\chi^B} = - 2 K^{ab}
\ee
and
\be
 {\partial\over \partial \bar\h^A} \left(
\tilde\j^a \tilde\j^b \right) {\overleftarrow{\partial}\over \partial
\chi^C}
 {\partial\over \partial \bar\h^C} \left(
\tilde\j^c \tilde\j^d \right) {\overleftarrow{\partial}\over \partial
\chi^A}  = 2(K^{ad}K^{bc}-K^{ac}K^{bd}) +\frac{1}{2} (\d^{ad}
\d^{bc} - \d^{ac} \d^{bd} )
\ee
we find that the super Van Vleck determinant indeed equals $E$.
Hence the same factorization which we found in the case of fermion
doubling in (\ref{prop1re}) also holds in the case of fermion
halving, even though the separate factors $\tilde{D}_S^{1/2}$ and
$S_F$ are different.

\section{Anomaly calculations}

In this section we compute the chiral anomalies due to spin-$1/2$ fermions
coupled to external gravity and external Yang-Mills fields.
Then we consider the trace anomaly. In appendix A.3 we discuss
gravitational anomalies for spin-$1/2$ and spin-$3/2$ fields.

In a quantum field theory, the anomalies can be written as the
trace over the product of a Jacobian and a regulator. Both the Jacobian
and the regulator depend on which fields one considers as the independent
fields, for example, for a scalar field $\phi$ possible choices are
$\phi$ itself and $\tilde{\phi}=\phi g^{1/4}$. Of course, the anomaly
itself should not depend on this choice of basis. It has become
common practice to take $\tilde{\phi}$ for scalars, and
$\tilde{\c}^{\a}=\c^{\a} g^{1/4}$ for spinors as basic variables,
because then (with the corresponding regulator) the absence of Einstein
anomalies becomes obvious. However, in the non-linear sigma models, one
uses another regulator. Namely, since we have taken the inner product
between $x$-eigenstates as $\<x|y\>=g(x)^{-1/2}\d(x-y)$, the momentum
operator is represented by $p_i=-\hbar i g^{-1/4} \partial_i g^{1/4}$, and this
representation is clearly obtained from $p_i=-\hbar i \partial_i$ by
a similarity transformation with $g^{-1/4}(x)$. As a result, the
regulator used in non-linear sigma models is no longer
$g^{-1/4}\partial_{\m} \sqrt{g} g^{\m\n} \partial_{\n}
g^{-1/4}$ but rather $g^{-1/2} \partial_{\m} g^{1/2} g^{\m\n} \partial_{\n}$.
In terms of momenta this reads $g^{-1/4} p_i g^{1/2} g^{ij}
p_j g^{-1/4}$. We now explain in more detail how the Einstein anomaly
vanishes if we take $\tilde{\phi}$ as an independent field.

Given a set of symmetries one wants to preserve at the quantum level in
field theories, there exists a method to construct a regulator which
yields consistent anomalies and which preserves these symmetries \cite{diaz}.
The basic idea is to construct a  mass term which is bilinear in fields
and which separately preserves the symmetries. For scalars,
an Einstein invariant mass term is clearly $M\tilde{\phi} \tilde{\phi}$
with $\tilde{\phi}=g^{1/4}\phi$. The regulator is then the kinetic
operator for these fields, i.e., $\tilde{R}=
g^{-1/4} \partial_{\m} g^{1/2} g^{\m\n} \partial_{\n} g^{-1/4}$. The
Jacobian for Einstein transformations reads
\be \label{bb103}
J(\tilde{\phi}) = 1 + \xi^{\m} \partial_{\m} +
\half(\partial_{\m} \xi^{\m}) = 1+\half(
\xi^{\m} \partial_{\m} +
\partial_{\m} \xi^{\m})
\ee
 and using an orthonormal basis $\tilde{\phi}_N(x)$ satisfying
 $\int \tilde{\phi}^{\ast}_M(x) \tilde{\phi}_N(x) dx = \d_{MN} $ the
 anomaly becomes
 \bea
 {\cal A}_n & = & \half \int \tilde{\phi}_N^{\ast}(x)(\xi^{\m}\partial_{\m}+
 \partial_{\m} \xi^{\m}) \tilde{\phi}_N(x) dx  \nonumber \\
 & = &
 \half \int \tilde{\phi}_N^{\ast}(x) \xi^{\m} \partial_{\m}
 \tilde{\phi}_N(x) dx -
 \half \int \partial_{\m} \tilde{\phi}_N^{\ast}(x) \xi^{\m}
 \tilde{\phi}_N(x) dx .
 \eea
It is obvious that this vanishes if the basis $\{\tilde{\phi}^{\ast}_N \}$
is in one-to-one correspondence with the basis $\{\tilde{\phi}_N \}$, as
is the case for e.g. plane waves. In a more invariant language, the
vanishing of the Einstein anomaly follows from the fact that the
Jacobian is real and anti-hermitian, and the regulator is real and hermitian.
The trace of the product of two such operators always vanishes, which can
be deduced from the fact that the trace of any real operator (i.e. an
operator that satisfies $A(\phi^{\ast})=A(\phi)^{\ast}$ ) is equal to the
trace of its hermitian conjugate.

%The Jacobian is a real operator, i.e. $J(\varphi^{\ast})=
%J(\varphi)^{\ast}$. Given a fixed orthonormal basis with basis vectors $|M\>$,
%real operators $A$ satisfy $\<M|A^{\dagger}|N\>=\<N|U^{\dagger}AU|M\>$, where
%the unitary matrix $U$ depends on the basis but not on $A$. In terms of
%explicit wavefunctions $\varphi_M(x)$ corresponding to $|M\>$, one has
%$\varphi_N(x)=\varphi_M^{\ast}(x) U_{MN}$. If $A$ and $B$ are real
%operators, then $\tr(AB)=\tr(U^{\dagger}ABU)=\tr(B^{\dagger}A^{\dagger})=
%\tr(A^{\dagger}
%B^{\dagger})$. In particular, if $A$ is real anti-hermitian and $B$ is
%real hermitian, then the trace satisfies $\tr(AB)=-\tr(AB)$ and hence
%$\tr(AB)=0$. Thus, since the Jacobian is real anti-hermitian, the Einstein
%anomaly vanishes at the regularized level if we regulate with a real
%hermitian operator like $\exp(-\tilde{R}/M^2)$. For this reason,
%$\tilde{\phi}$ is a preferred basis in field theory.

However, as explained above,
in order to evaluate anomalies by using equivalent non-linear
sigma models, the basis $\tilde{\phi}$ is not directly
compatible with the conventions for $\<x|y\>$ we have chosen. For chiral
anomalies or trace anomalies, it does not matter whether one uses
$\tilde{\phi}$ or $\phi$, because a similarity transformation with
$g^{1/4}$ has no effect on $\gamma_5$ or ${\bf 1}$: $g^{1/4} \gamma_5
g^{-1/4} = \gamma_5$ and $g^{1/4} {\bf 1} g^{-1/4} = {\bf 1}$.
But for gravitational anomalies, the basis $\phi$ is more convenient.
As follows from appendix A.3, one finds for Einstein anomalies
(= gravitational anomalies for $\tilde{\phi}$) in a covariant notation
\be J=\xi^{\m} D_{\m} + \half (D_{\m} \xi^{\m}) \ee
where $D_{\m}=\partial_{\m}-\half \Gamma_{\m\n}{}^\n$
and $(D_{\m} \xi^{\m})=\partial_{\m} \xi^{\mu} + \Gamma_{\m}{}^{\m}{}_{\n}
\xi^\n $. The $\Gamma_{\m\n}{}^{\n}$ terms cancel in $J$. The
regulator is
\be R=g^{-1/4} \partial_{\m} g^{1/2} g^{\m\n} \partial_{\n} g^{-1/4}=
g^{1/4} R_{cov} g^{-1/4} \ee
with $R_{cov}=g^{-1/2} \partial_{\m} g^{1/2} g^{\m\n} \partial_{\n}$
the usual covariant scalar D'alembertian. Using cyclicity of the trace,
or equivalently, making
a similarity transformation (change of basis), the Einstein
anomaly becomes
\bea {\cal A}_n({\rm Ein}) & = & {\rm Tr} \left(
J \exp(R/M^2) \right) \nonumber \\
& = &  \half {\rm Tr} \left(
g^{-1/4} (\xi^\m D_{\m} + D_{\m} \xi^{\m} ) g^{1/4} \exp(R_{cov}/M^2)
\right)
\eea
A direct evaluation of this trace using plane waves is given in
\cite{ref54}.
In the non-linear sigma model, $\frac{\hbar}{i}\partial_{\m}$ becomes
$g^{1/4} p_i g^{-1/4}$, and thus
\be
{\cal A}_n = \half {\rm Tr} (\xi^i \hat{p}_i + \hat{p}_i \xi^i)
\exp(-\b \hat{H}/\hbar) )
\ee
with $\hat{H}=g^{-1/4} p_i g^{1/2} g^{ij} p_j g^{-1/4}$. Note that $J$
comes out Weyl ordered, so that ${\cal A}_n({\rm Ein})$ can be
evaluated using the methods of the preceding sections. This explains
why $\hat{H}$ is the regulator. (Of course, the Einstein anomaly vanishes,
as we already explained, but other anomalies can be calculated with the
same regulator.)

For spin-$1/2$ fields, the choice $\tilde{\c}^{\a}=\c^{\a} g^{1/4}$
as basic field leads to the same Einstein Jacobian, the field operator
is now $g^{1/4} \Dsl g^{-1/4}$ (where $\Dsl$ is the usual covariant
Dirac operator), and, as explained in \cite{diaz}, the consistent
regulator which leads to vanishing Einstein anomalies is now the
square, namely $\tilde{R}=g^{1/4}\Dsl \Dsl g^{-1/4}$. Gravitational anomalies
are really local Lorentz anomalies, but by taking a suitable linear
combination of Einstein and local Lorentz transformations one
obtains covariantly looking expressions. For chiral spin-$1/2$ fields,
one obtains then (see appendix A.3)
\be {\cal A}_n({\rm grav},\mbox{spin-}1/2)=-\half
{\rm Tr} (\xi^\m D_{\m} + D_\m \xi^\m) \exp(\tilde{R}/M^2)
\ee
where $D_{\m}=\partial_{\m} + \frac{1}{4} \omega_{\m}{}^{ab}
\gamma_a \gamma_b  - \half \Gamma_{\m\n}{}^{\n}$, but the
$\Gamma_{\m\n}{}^{\n}$ terms in the Jacobian cancel again.
Making the similarity transformation with $g^{1/4}$ one finds
(see (\ref{eqq101}))
\be {\cal A}_n({\rm grav},\mbox{spin-}1/2)=-\half
{\rm Tr} (\xi^i \pi_i  + \pi_i  \xi^i) \exp(-\b \hat{H}/\hbar)
\ee
where $\pi_i=p_i-\half i\hbar \omega_{iab} \j^a \j^b$ and
$\hat{H}$ given in (\ref{HN1}). This explains why we took this
particular quantum Hamiltonian in the non-linear sigma models.

Finally, for spin-$3/2$ similar results hold. However, here the
Jacobian is more complicated. In \cite{alwi} an expression is given
which does not correspond to a linear combination of local symmetries
of supergravity. However, as we explain in appendix A.3, the
difference (which would be difficult to evaluate for non-linear
sigma models) vanishes if one uses as regulator the same regulator
$\Dsl \Dsl$ as for spin-$1/2$. Why this is the correct regulator is
also explained in appendix A.3.

\subsection{Chiral anomalies in gravitational couplings}

The simplest anomaly one can calculate by means of the path integral
methods we have developed, is the $\g_5$ anomaly due to a loop of a
spin-$1/2$ field coupled to external gravitational fields. As regulator
for spin-$1/2$ fermions we take
$\Dsl\Dsl$, which can be rewritten as the sum of a d'Alembertian and a
gravitational curvature term (Weitzenbock identity)
\bea \label{eqq101}
\Dsl\Dsl &=& D^i D_i + {1\over 2} \g^i \g^j \left[ D_i , D_j
\right] \nonumber \\
&=& {1\over \sqrt{g}} D_i^{(0)} \sqrt{g} g^{ij} D_j^{(0)} +
{1\over 8} \g^i \g^j R_{ij ab}(\w) \g^a \g^b \nonumber \\
&=& {1\over \sqrt{g}} D_i^{(0)} \sqrt{g} g^{ij} D_j^{(0)} +
{1\over 4} R
\eea
where $D^i D_i=g^{ij} D_i D_j = g^{ij} (D^{(0)}_i D_j -
\Gamma^k_{ij} D_k)$ and
$D_i^{(0)} = \partial_i + {1\over 4} \w_{iab} \g^a \g^b$. We
used the cyclic identity $R_{i [ j ab]} =0$ to replace $\g^j \g^a
\g^b$ by $e^{aj} \g^b - e^{bj} \g^a$, and then we used that the
Ricci tensor $R_{i a}$ is symmetric.

In the non-linear sigma model we choose the representation
\be
\partial_i \leftrightarrow {i\over \hbar} g^{1/4}  p_i g^{-1/4};
\qquad \g^i e_i^a = \g^a \leftrightarrow \sqrt{2} \j^a
\ee
Then $\left\{ \j^a , \j^b \right\} =  \d^{ab}$, and the
Hamiltonian becomes
\be \label{aaa105}
\hat H = {1\over 2} g^{-1/4} \left( p_i - {\hbar i\over 2} \w_{iab} \j^a
\j^b \right) g^{1/2} g^{ij} \left( p_j - {\hbar i\over 2} \w_{jcd} \j^c
\j^d \right) g^{-1/4}- {\hbar^2\over 8} R
\ee
The representation $ p_i = g^{-1/4} {\hbar\over i} {\partial\over
\partial x^i} g^{1/4}$ is fixed by the requirement that $ p_i$ be
hermitian and that the inner product is given by $\langle x | y \rangle =
{1\over\sqrt{g(x)}} \d(x-y)$, from which we find the completeness
relations $\int\! d^n x \, \sqrt{g(x)} | x\rangle \langle x| = \unit =
\int\! d^n p | p \rangle \langle p|$, and the inner product $\langle x
| p \rangle = (2\pi\hbar)^{-n/2} \exp \left( {i\over\hbar} p\cdot x
\right) g^{-1/4}$. In the Fujikawa approach to anomalies, one often
uses plane waves to evaluate traces in field theory. Since these
plane waves are normalized to $\int \exp(\frac{i}{\hbar} p (x-y)
)d^n p = (2\pi\hbar)^{n/2} \d(x-y)$, the regulator is in these cases
$\exp (-\tilde{\Dsl} \tilde{\Dsl} /M^2)$ with $\tilde{\Dsl}=
g^{1/4} \Dsl g^{-1/4}$. In the non-linear sigma model we have
inner products $\<x|y\>=(1/\sqrt{g(x)}) \d(x-y)$, and now the
regulator corresponds to $\exp(-\Dsl\Dsl/M^2)$.

We recall that under Einstein transformations with anti-hermitian generator
$E = {-i\over 2\hbar} \left\{ p_j , F^j(x) \right\}$
for the orbital part, $x^i \rightarrow
x^i + \left[ x^i , E\right] = x^i + F^i(x) \equiv  x'^i$, while the
momenta transform as $p_i \rightarrow p'_i = {1\over 2}
\left\{ {\partial x^j \over \partial  x'^i} , p_j \right\}$. Clearly
$\j^a$ does not transform if we take $x^i$, $p_i$ and $\j^a$ as
independent variables. It follows that also $\pi_i = p_i - {\hbar i\over 2}
\w_{iab} \j^a \j^b$ transforms as $\pi'_i = {1\over 2} \left\{
{\partial x^j \over \partial x'^i} , \pi_j \right\}$
(after adding a spin part to $E$ which completes the transformation
of $\omega_{iab}$ to that of a vector), and this
shows, in the same way as in the bosonic case, that $ H$ is
Einstein invariant. Since $ p_i$ is hermitian, $ H$ is clearly
hermitian, too.

In a similar manner we may construct the orbital part
of the Lorentz generator $L= {1\over
2\hbar} \l_{bc}( x) \j^b \j^c$ and show that $\d_L p_j =
{1\over 2i} (\partial_j \l_{bc}) \j^b\j^c$,
$\d_L \j^a = \frac{1}{\hbar} \l^a{}_c\j^c$ and so, after
adding a spin part to $L$,
$\d_L (\w_{iab} \j^a
\j^b ) = - (\partial_i \l_{ab} ) \j^a \j^b$, from which also the Lorentz
invariance of $\hat H$ follows.

%The supersymmetry of the Hamiltonian follows from writing $ H =
% - {\hbar^2\over 2} \left( \g^m e_m^\m D_\m^{(0)} \right) \left( \g^n
% e_n^\n D_\n^{(0)} \right) =
% Q  Q$, where $ Q = \g^a
%e_a^\m D_\m^{(0)}$.

In the non-linear sigma model, $ H = {1\over 2}
 Q  Q$, where $ Q = \sqrt{2} \j^a e^i_a g^{1/4}
\pi_i g^{-1/4}$
is the supersymmetry generator.
Since $[H,Q]=0$, this Hamiltonian is
supersymmetric. One can easily verify that $Q$ is Einstein
Lorentz invariant.

%and since $\left\{  Q_a,  Q_b \right\} =
%\d_{ab}  H$, ${H}$ is supersymmetric.

The chiral anomaly is given by
\be
{\cal A}_n = {\rm Tr} \g_5 e^{-\frac{\b}{\hbar} \hat{H}}.
\label{chirA}
\ee
We shall compute this expression both in the case we double the
number of Majorana fermions and in the case we combine the Majorana
fermions into half as many Dirac fermions.

We start by doubling the
number of fermions, in which case we evaluate the
trace in an artificially extended Hilbert space.
After we introduce the free fermions $\psi^a_2$ ($a=1\ldots n$), we
have (for even $n$) $2^{n/2}$ states in the $\psi_1$ sector and
$2^{n/2}$ states in the $\psi_2$ sector (combining the $n$ $\psi^a_1$
and $n$ $\j^a_2$ in $n$
pairs of creation and absorption operators).
Hence, we must divide the trace over
$\psi_1$ and $\psi_2$ by a factor $2^{n/2}$, since we really should
only take the trace in the $\psi_1$ sector.

We shall now first express $\g_5$ into $\j^a$.
We define $\g_5$ by
\be
\g_5
= (-i)^{n/2} \g_1 \g_2 \cdots \g_{n} \hspace{2cm} \hbox{$n$ even}
\label{defg5}
\ee
So, in $n=2$ one has $\g_5 =\t_3$, and in $n=4$ one has
$\g_5=-\g_1\g_2\g_3\g_4$. In general $\g_5^2=1$ and $\g_5$ is hermitian.
Identifying $\g^a = \sqrt{2}
\hat\j^a_1 = (\hat\j^a + \hat\j^{\dagger}_a)$, $a=1\ldots n$,
we can evaluate
the matrix element of $\g_5$ between fermionic coherent states
and find
\be
\langle \bar\x | \g_5 | \h\rangle = (-i)^{n/2} \langle \bar\x |
\h\rangle \prod\limits_{a=1}^n (\h^a + \bar\x^a) = (-i)^{n/2}
\prod\limits_{a=1}^n (\h^a + \bar\x^a)
\ee
The expression for the anomaly becomes (recall the factor of
$2^{n/2}$)
\bea
{\cal A}_n &=& 2^{-n/2} \int
\prod\limits_{i=1}^n dx_{0}^i \sqrt{g(x_{0})} \int
\prod\limits_{a=1}^n
d\bar\h^a d\h^a d\x^a d\bar\x^a \nonumber \\ &&
\qquad \qquad \qquad e^{\bar\x\x} \langle
\bar\x| \g_5 | \h \rangle \, e^{-\bar\h \h} \, \langle x_{0}, \bar\h |
e^{-{\b\over\hbar} \hat H} | x_{0}, \x \rangle
\eea
The last term in the above expression, the transition element,
contains a factor $e^{\bar\h \x}$, and further contributions from
loops (which depend only on the sum $(\x^a + \bar\h^a)$), but no
classical action since the trace puts the initial and final points in
$x$-space equal to each other.
Now write
\be
e^{\bar\x \x} e^{-\bar\h \h} \prod\limits_{a=1}^n (\h^a + \bar\x^a) =
e^{-{1\over 2} (\h-\bar\x) (\x-\bar\h)} \prod\limits_{a=1}^n (\h^a +
\bar\x^a)
\ee
and perform the integral over $\h$ and $\bar\x$ (rewrite the measure
$d\bar\xi^a d\h^a$
in terms of the variables $\h-\bar\x$ and $\h+\bar\x$
as $2^n d(\bar\xi^a+\h^a)d(\h^a-\bar\xi^a)$). We find
\bea
{\cal A}_n &=& {(-i)^{n/2}\over (4\pi\b\hbar)^{n/2}} \int \prod\limits_{i=1}^n
dx_{0}^i \sqrt{g(x_{0})} \int
\prod\limits_{a=1}^n d\bar\h^a d\x^a \, e^{\bar\h\x}
\prod\limits_{a=1}^n (\x^a -\bar\h^a) e^{-{1\over\hbar} S_{\rm loops}
(x_{0},\bar\h+\x)} \nonumber \\
&=& {(-i)^{n/2}\over (2\pi\b\hbar)^{n/2}}\int \prod\limits_{i=1}^n dx_{0}^i
\sqrt{g(x_{0})} \int
\prod\limits_{a=1}^n d\j^a_{\rm bg} \,
e^{-{1\over\hbar} S_{\rm loops} (x_{0},\j^a_{\rm bg})}
\label{ChirAno}
\eea
We were able to integrate out the $\j^a_2$, canceling the factor
$2^{-n/2}$, because $S_{\rm loops}$ only depends on the combination
$\j^a_{\rm bg} = {1\over\sqrt{2}} ( \x^a + \bar\h^a)$, which
is the background value of $\j^a_1$.
The factor $(2\pi \hbar \b)^{n/2}$ normalizes the leading
(classical) singularity in $\langle x|\exp (-\frac{\b}{\hbar} H )
|x\rangle$ to a
Dirac delta function. The loop contributions
$S_{\rm loops} (x_{0},\j^a_{\rm bg}) $ are defined by
\be
e^{\bar\h\xi-\frac{1}{\hbar}
S_{\rm loops} (x_{0},\j^a_{\rm bg})} =
\left\langle e^{-\frac{1}{\hbar} S_{\rm int}} \right\rangle
%\int {\cal D}x^i {\cal D}\bar\psi^a {\cal D}\psi^a {\cal D}c^i
%{\cal D}b^i {\cal D}a^i e^{ -\frac{1}{\hbar} S}
\ee
with the propagators given in (\ref{ferGreen}).
 $S_{\rm int}$ is the interaction part
of the action $S$
\bea
S &=&
-\hbar\d_{ab} \bar{\psi}^a(0) {\psi}^b(0) +
\int\limits^0_{-\b}\! dt \, \Biggl[ \frac{1}{2} g_{ij}(x)
\biggl( \dot{x}^i \dot{x}^j +\frac{1}{\b^2} b^i c^j +
\frac{1}{\b^2} a^i a^j \biggr) \nonumber \\
& & {} +  \hbar\d_{ab} \bar{\psi}^a
\dot{\psi}^b + {\hbar\over 2} \dot{x}^i \omega_{iab} \psi_1^a \psi_1^b
 + (\mbox{\rm $\omega\omega$ and
$\Gamma\Gamma$ terms})  \Biggr]
\eea
with the fields subject to the boundary conditions
$x^i(0)=x^i(-1)=x_0^i$ and $\bar{\psi}^a(0)=\bar\h^a,\j^a(-1)=\xi^a$.
We now rescale $t=\b\t$, which leads in the bosonic sector to
\be
\frac{1}{\hbar} S_{{\rm bos}}  = \frac{1}{\b\hbar}
 \int\limits^0_{-1}\! d\t \, \frac{1}{2} g_{ij}(x)
\left( \dot{x}^i \dot{x}^j + b^i c^j +
 a^i a^j \right)
\label{BosAc}
\ee
To obtain also a factor $\frac{1}{\b}$ in front of the fermionic
terms we rescale $\psi$ and $\bar{\psi}$ suitably
\be
\bar{\psi}^a =\frac{1}{\sqrt{\b}} \bar{\psi}'^a, \hspace{1cm}
\psi^a =\frac{1}{\sqrt{\b}} \psi'^a
\ee
Then
\be
\frac{1}{\hbar} S_{{\rm fer}}   =  \frac{1}{\b}
\int\limits_{-1}^0\! d\t \, \left( \d_{ab} \bar{\psi}'^a
\dot{\psi}'^b + {1\over 2} \dot{x}^i \omega_{iab} \psi'^a \psi'^b
\right) -\frac{1}{\b} \d_{ab} \bar{\psi}'^a(0) {\psi}'^b(0)
\label{FerAc}
\ee
%After this rescaling, the measure
%$ {\cal D}x^i {\cal D}\bar\psi^a {\cal D}\psi^a $ has
%become $\b$ independent, which can be seen most easily by going
%back to the discretized definition of the path integral.
The same rescaling is applied to the corresponding background values
\be
\bar{\eta}^a=\frac{1}{\sqrt{\b}} \bar{\eta}'^a, \hspace{1cm}
\chi^a=\frac{1}{\sqrt{\b}} \chi'^a
\ee
With this rescaling of the fermionic background fields, the $\b$
dependence in the measure in (\ref{ChirAno}) is also
canceled. Dropping the primes, we arrive at
\be \label{aaa118}
{\cal A}_n = {(-i)^{n/2}
\over (2\pi\hbar)^{n/2}}\int \prod\limits_{i=1}^n dx_{0}^i
\sqrt{g(x_{0})} \int
\prod\limits_{a=1}^n d\j^a_{\rm bg} \, e^{-{1\over\hbar} S_{\rm loops}
(x_{0},\j^a_{\rm bg})}
\ee
with the action the sum of (\ref{BosAc}) and (\ref{FerAc}).
The $\Gamma\Gamma$ and $\omega\omega$
term are of order $\b$ (due to the rescaling $t=\b\t$) and play
no further role in the evaluation of chiral anomalies.

Expanding
\be x^i=x_0^i+q^i, \hspace{1cm} \psi^a=\chi^a + \psi_{\rm qu}^a,
 \hspace{1cm} \bar{\psi}^a=\bar{\eta}^a + \bar{\psi}_{\rm qu}^a
\label{eq102}
\ee
we see that
\begin{itemize}
\item[(i)] all vertices are proportional to $\frac{1}{\b}$ and all
 propagators are proportional to $\b$.
\item[(ii)] hence only one-loop graphs contribute.
\item[(iii)] In a frame where $\omega_{iab}(x_0)=0$, there are no terms
linear in the quantum fields in $S$, so no tadpoles.
Therefore, only vertices with exactly
two quantum fields are relevant for the one-loop graphs in that case.
\item[(iv)] Consequently,
by expanding $g_{ij}(x)$ and $\omega_{iab}(x)$ around
$x_0$, and after substituting (\ref{eq102}) in $S$, we find
that in the frame $\omega_{iab}(x_0)=0$ the only vertices are
\be
\frac{1}{\hbar} S_{{\rm int}} =
{1\over 2\b}
(\partial_j \omega_{iab}(x_0) ) \int\limits^0_{-1}\! d\t \,
q^j \dot{q}^i \psi^a_{{\rm bg}} \psi^b_{{\rm bg}}
\ee
which can be rewritten as
\be \label{aaa121}
\frac{1}{\hbar} S_{{\rm int}} =  \frac{1}{4\b} R_{ijab} (\omega(x_0))
 \psi^a_{{\rm bg}} \psi^b_{{\rm bg}}
\int\limits_{-1}^0\! d\t \, q^i \dot{q}^j
\ee
\end{itemize}
Thus the one-loop contributions are due to a $q$ loop, with at each
vertex two $\psi^a_{{\rm bg}}$ sticking out.

The one-loop result can now easily be evaluated by noting that it is
proportional to the one-loop determinant
\be
{\cal A}_n =  \frac{(-i)^{n/2}}{(2\pi\hbar)^{n/2}}
\int\! d^nx_{0} \, \sqrt{g(x_{0})} \prod_{a=1}^n
d\psi^a_{{\rm bg}} \left[
\frac{\det \left( -\frac{d^2}{d\t^2} g_{ij}(x_0)
+ \frac{1}{2} R_{ijab}
\psi^a_{{\rm bg}} \psi^b_{{\rm bg}} \frac{d}{d\t} \right) }{
\det \left( -\frac{d^2}{d\t^2} g_{ij}(x_0) \right) } \right]^{-1/2}
\label{eq106}
\ee
The denominator in this expression is the ghost contribution to
${\cal A}_n$. The factor $i^{n/2}$ can be removed
by a rescaling of $\j^a_{\rm bg}$, in which case the only
thing that changes is that $R_{ijab}$ is replaced by $iR_{ijab}$.
With this rescaling the operator in the numerator of (\ref{eq106})
is hermitian and it becomes manifest that ${\cal A}_n$ is real.
{}From here on, one can follow Alvarez-Gaum\'e and Witten \cite{alwi}.
First one skew diagonalizes $R_{ij}$, then one computes
the eigenvalues of the operator in the determinant. The product of
the corresponding eigenvalues then immediately yields the
well-known result for the chiral anomaly as the
$\hat{A}$-genus of the manifold. An alternative derivation can be
found in appendix A.4, where the one-loop diagrams contributing to the
one-loop determinant (\ref{eq106}) are explicitly evaluated.

We will now evaluate the trace in (\ref{chirA}) in case we do not
double the number of fermions, but instead combine the Majorana
fermions into half as many Dirac fermions. In this case the expression
corresponding to $\g_5$ becomes
\be
\g_5 = (-i)^{n/2} \prod_{A=1}^{n/2} (\Psi^A +\bar{\Psi}^A) \frac{1}{i}
(\Psi^A -\bar{\Psi}^A) = \prod_{A=1}^{n/2} (1 - 2\bar{\Psi}^A\Psi^A)
\ee
where the $\Psi^A$ and $\bar\Psi^A$ are defined as in (\ref{defMF})
and we identify again $\g^a = \sqrt{2} \j^a$.
Since $P_A=\bar{\Psi}^A\Psi^A$ for fixed $A$
is a projection operator ($P_A^2 = P_A$) we
can rewrite this as
\be
\g_5 = \prod_{A=1}^{n/2} e^{-i\p\bar{\Psi}^A\Psi^A} =
\exp \biggl( -i\p \sum\limits_{A=1}^{n/2}\bar{\Psi}^A\Psi^A \biggr)
\label{eq108}
\ee
since expansion gives $\prod_A(1 + (e^{-i\p} -1)\bar{\Psi}^A\Psi^A)$.
%Note that this operator is normal-ordered as it stands.

The matrix element of $\g_5$ between coherent states simplifies even further
\bea
\<\bar\x|\g_5|\h\> &=&
e^{\bar{\x}\h} \prod_{A=1}^{n/2} (1-2\bar{\x}^{A} \h^A)
= \prod_{A=1}^{n/2} (1-2\bar{\x}^{A} \h^A)(1+\bar{\x}^{A} \h^A)
\nonumber \\ &=& \prod_{A=1}^{n/2} (1-\bar{\x}^{A} \h^A)
= \exp \biggl( - \sum_{A=1}^{n/2} \bar{\x}^{A} \h^A \biggr)
\label{eq109}
\eea

To evaluate the anomaly, we put together the expression for the Jacobian and
the transition element
\bea
{\cal A}_n &=& \int \prod_{i=1}^n dx_{0}^i \sqrt{g(x_{0})}
\int \prod_{A=1}^{n/2}
d\bar{\h}^A d\h^A d\x^A  d\bar{\x}^A
\nonumber \\
&& \qquad e^{\bar{\x} \x} \<\bar{\x}|\g_5|\h\> e^{-\bar{\h} \h}
\< x_0, \bar{\h} |e^{-\frac{\b}{\hbar} \hat{H}}|x_0,\x \>
\eea
As before, we extract a factor $\exp \bar\h \x$ from the transition
element, and write
\bea
{\cal A}_n &=& {1\over (2\pi\b\hbar)^{n/2}} \int \prod_{i=1}^n dx_{0}
\sqrt{g(x_{0})}
\int \prod_{A=1}^{n/2}
d\bar{\h}^A d\h^A d\x^A  d\bar{\x}^A
\nonumber \\
&& e^{\bar\x \x - \bar\x \h - \bar\h \h
+ \bar\h \x} \exp\biggl( -\frac{1}{\hbar}
S_{\rm loops}(x_{0},\bar{\h}^{A}, \x^A) \biggr)
\eea
We can now trivially do the integral over $\h^A$ and $\bar\x^A$, after
which the above expression equals (\ref{ChirAno}), with the
identifications
\bea
\j^a_{\rm bg} &=& {1\over\sqrt{2}} ( \x^{(a+1)/2} + \bar\h^{(a+1)/2} ) \qquad
\hbox{$a$ odd} \nonumber \\
\j^a_{\rm bg} &=& - {i\over\sqrt{2}} ( \x^{a/2} - \bar\h^{a/2} ) \qquad
\hbox{$a$ even}
\label{eq112}
\eea
Hence, also in this case, we obtain the same expression for the chiral
anomaly.

\subsection{Chiral anomalies in Yang-Mills couplings}

We consider complex spin-$1/2$ fermions coupled to external Yang-Mills
gauge fields corresponding to a group G, with the
fermions transforming under a representation $R$.
We will only consider flat space here; by combining the techniques in
this and the previous section, one can easily obtain the combined
gravitational and Yang-Mills anomalies. We leave this as an
exercise to the reader. New in this section is the treatment of
the Yang-Mills ghosts by path integrals. In \cite{alwi}, they were kept
as operators.

The Jacobian is still $\g_5$,
but the regulator is now proportional to
\be
\Dsl \Dsl = D^\m D_\m + \frac{1}{2} \g^\m \g^\n [D_\m, D_\n],
\ee
where $D_\m = \pa_\m - g A_\m^\a T_\a$,
$[T_\a, T_\b] = f_{\a \b}^{\ \ \g} T_\g$ and
$[D_\m, D_\n] = -g F_{\m \n}^\a T_\a$.
Hence, compared to the bosonic case, there is now an extra term proportional
to the Yang-Mills curvature.
We represent the matrices $(T_\a)^M_{\ N}$, $M,N=1\ldots {\rm dimR}$,
where ${\rm dimR}$ is the dimension of the representation $R$,
 by operators
\be
\hat{T}_\a =  c^*_M (T_\a)^M_{\ N} c^N,
\ee
and require the anticommuting relations
\be
\{ c^N, c^*_M \} =  \d^N_{\ M}.
\ee
Then $[\hat{T}_\a, \hat{T}_\b] = f_{\a \b}^{\ \ \g} \hat{T}_\g$.
The Dirac matrices $\g^a$ are again represented by $\sqrt{2} \j^a$
with
\be
\{ \j^a, \j^b \} =  \d^{ab} \hspace{2cm} (a,b = 1, \dots, n)
\ee
We will only represent the Majorana fermions $\j^a$
by half as many Dirac
fermions here; if one doubles the number of fermions one obtains the
same answer. This can easily be demonstrated in a similar fashion as
we did in the previous section. With half as many Dirac fermions
we had (\ref{eq108})
\be
\g_5 \rightarrow e^{-i\pi  \bar\j_A \j^A}.
\ee
The anomaly is then represented in the non-linear sigma model by
\bea
{\cal A}_n &=& {\rm Tr}' e^{- i\pi
\bar\j_A \j^A} e^{-\b H/\hbar} \label{an1} \\
\hat{H} &=&
\frac{1}{2} (p_j + \hbar i A^\a_j c^* T_\a c)(p_k +
\hbar i A^\b_k c^* T_\b c) \d^{jk}
+\frac{\hbar^2}{2} \j^a \j^b F_{ab}^\a c^* T_\a c \label{ham}
\ena
We have rescaled $A_j^\a(x)$ and $F^{\a}_{\m\n}$
such that the coupling constants have been
absorbed.

The prime on the trace, ${\rm Tr}'$, indicates that the trace is not over all
states in the Fock space (which are $|0\>, c^*_{M_1} |0\>,
c^*_{M_1} c^*_{M_2} |0\>, \dots,
c^*_{M_1} \cdots c^*_{M_{\rm dimR}} |0\>$,
but rather only over the one-particle states $c^*_M |0\>$.
Only on these states does $c^{\ast}T_{\a} c$ act like the matrix $T_{\a}$.
To still write the trace as an unconstrained trace, we introduce the
one-particle projection operator $P$.
We claim that
\be
P = :xe^{-x}:, \hspace{2cm} x = c^*_M c^M. \label{projection} \\
 \ee
Indeed, from the definition of $x$ it follows that
\be
:x^n: = (x-n+1) :x^{n-1}: = \cdots =
n!
\left( \begin{array}{c}
x \\ n
\end{array} \right). \label{normal}
\ee
Then we use a representation of the Kronecker delta $\d_{x,1}$,
\be
\d_{x,1} = \lim_{p \rightarrow 1} xp (1-p)^{x-1}
=\lim_{p \rightarrow 1} p \frac{\pa}{\pa p} [-(1-p)^x] .
\ee
Expanding $(1-p)^x$ in a power series
 and performing the limit $p \rightarrow 1$ we arrive at
\be
\sum_{n \geq 0} n (-1)^{n-1}
\left( \begin{array}{c}
x \\ n
\end{array} \right) \nonumber \\
=\sum_{n>0} \big[ n!
\left( \begin{array}{c}
x \\ n
\end{array} \right)
\big]
\frac{(-1)^{n-1}}{(n-1)!}
\ee
Using (\ref{normal}) we get the desired result, namely
\be
P=:xe^{-x}:=\d_{x,1}.
\ee

The anomaly is thus given by
\be
{\cal A}_n = {\rm Tr} \left(e^{-i\pi  \bar\j_A \j^A} :c^*_Mc^Me^{-c^*_Nc^N}:
e^{-\b H/\hbar}\right) . \label{an2}
\ee
Using complete sets of coherent states, one finds the corresponding path
integral representation
\be
{\cal A}_n = {\rm tr}_x {\rm tr}_f {\rm tr}_{gh}
\<x_0, \bar{\c}_{gh}, \bar{\c}_{f}|
e^{-i\pi \bar\j \j} P I_{gh} I_{f} e^{-\b H/\hbar}
| \c_{f}, \c_{gh}, x_0 \>, \label{an3}
\ee
where
\bea
I_{gh} &=& \int d\bar{\h}_{gh} d\h_{gh}
|\h_{gh} \> e^{-\bar{\h}_{gh} \h_{gh}} \<\bar{\h}_{gh}|, \\
I_{f} &=& \int d\bar{\h}_{f} d\h_{f}
|\h_{f} \> e^{-\bar{\h}_{f} \h_{f}} \<\bar{\h}_{f}|, \\
{\rm tr}_x & = & \int dx_0 \\
{\rm tr}_{gh} &=&
\int d\c_{gh} d\bar{\c}_{gh} e^{\bar{\c}_{gh} \c_{gh}}, \\
{\rm tr}_{f} &=&
\int d\c_{f} d\bar{\c}_{f} e^{\bar{\c}_{f} \c_{f}}.
\ena
We have used that in fermionic spaces the trace of an operator
is given by (\ref{deftrace}) and further inserted two unit
operators $I$ for which we used the decomposition in terms
of coherent states.

The trace involving $e^{- i\pi \bar{\j} \j} P$ factorizes.
The ghost dependent part of ${\cal A}_n$ reads
\be
\int d\c_{gh} d\bar{\c}_{gh} d\bar{\h}_{gh} d\h_{gh}
e^{\bar{\c}_{gh} \c_{gh}} e^{-\bar{\h}_{gh} \h_{gh}}
\<\bar{\c}_{gh}|:xe^{-x}:|\h_{gh}\>
\<\bar{\h}_{gh}|e^{-\b H}|\c_{gh}\>.
\ee
Since $P$ is a one-particle projection operator its matrix element simply
yields
\be
\<\bar{\c}_{gh}|:xe^{-x}:|\h_{gh}\>=
\bar{\c}_{gh} \h_{gh},
\ee
The integration over $\bar{\c}_{gh}$ then yields
\bea
\int d\bar{\c}_{gh}^{{\rm dimR}} \cdots d\bar{\c}_{gh}^1
\big( \sum_M \bar{\c}_{gh,M} \h_{gh}^M \big) e^{\bar{\c}_{gh} \c_{gh}}
& = & \sum_M \Big(  \Big(
\prod_{N > M} \c_{gh}^N \Big) \h_{gh}^M \Big(
\prod_{N < M} \c_{gh}^N  \Big) \Big) \nonumber \\ &
= & \sum_M \c_{gh}^{{\rm dimR}} \cdots \h_{gh}^M \cdots \c_{gh}^1,
\eea
namely, a product of all $\c^N_{gh}$ except that the M-th factor
is replaced by $\h_{gh}^M$.
The integration over $\h_{gh}$ then yields
\be
\int d\h_{gh} e^{-\bar{\h}_{gh} \h_{gh}} \sum_M
 \Big(  \Big(
\prod_{N > M} \c_{gh}^N \Big) \h_{gh}^M \Big(
\prod_{N < M} \c_{gh}^N  \Big) \Big) =
\sum_M \prod_{N \neq M} \big( \bar{\h}_{gh,N} \c_{gh}^N \big).
\ee
Clearly, this operator projects an arbitrary function of $\bar{\h}_{gh}$
and $\c_{gh}$ onto the terms with precisely one $\bar{\h}_{gh}$ and one
$\c_{gh}$.

In the fermionic sector, the matrix element of $\g_5$ is again (\ref{eq109})
\be
\< \bar{\c}_{f}|e^{-i \pi  \bar\j_A \j^A}|\h_{f}\> =
e^{-\bar{\c}_{f} \h_{f}}. \label{g5}
\ee
This leads to the integral
\be
\int d\bar{\c}_{f} d\h_f e^{\bar{\c}_{f} \c_{f}}
e^{-\bar{\h}_{f} \h_{f}} e^{-\bar{\c}_{f} \h_{f}}=
e^{-\bar{\h}_{f} \c_{f}}.
\label{eq137}
\ee

Inserting these subresults from the ghost and fermionic sectors into the
anomaly equation we arrive at
\bea
{\cal A}_n  & =  &  \int  \prod_{i, M, A}[ dx_0^i
d\c_{gh}^M d\bar{\h}_{gh,M} d\bar{\h}_{f,A} d\c_f^A ]
\nonumber \\  & & \times
\Big[
\sum_M \big( \prod_{N \neq M} \bar{\h}_{gh,N} \c_{gh}^N \big) \Big]
e^{-\bar{\h}_{f} \c_{f}}
\<x_0, \bar{\h}_{gh}, \bar{\h}_f| e^{-\b H/\hbar} |\c_f, \c_{gh}, x_0\>.
\eea
The transition element contains a factor $e^{\bar\h_f \c_f}$ and a
factor $e^{\bar\h_{gh} \c_{gh}}$, in addition to contributions from
loops. The former cancels against (\ref{eq137}), so that we obtain
\bea
{\cal A}_n &=& \int \prod_i [\frac{dx_0^i}{\sqrt{2 \p \hbar \b}}]
\int \prod_{M,A} [d\c_{gh}^M d\bar{\h}_{gh,M}
d\bar{\h}_{f,A} d\c_f^A ]
\Big[ \sum_M
\big( \prod_{N \neq M} \bar{\h}_{gh,N} \c_{gh}^N \big) \Big]
e^{\bar{\h}_{gh} \c_{gh}} \nonumber \\ & &
% &\ & \int \prod_{k=1;j,A,I}^{N-1}
% [\frac{dx_k^j}{2 \p \hbar \b} dc_k^A dc^*_{k,A} d\j_k^I d\bar{\j}_{k,I}]
\times \exp\left(-\frac{1}{\hbar} S_{\rm loops}(x_0,\bar\h_{gh},\bar\h_f,
\c_{gh},\c_f) \right),
\ena
where
\be
e^{ \bar\h_f \c_f+ \bar\h_{gh} \c_{gh}
-\frac{1}{\hbar} S_{\rm loops}(x_0,\bar\h_{gh},\bar\h_f,
\c_{gh},\c_f) } =
\left\langle e^{-\frac{1}{\hbar} S_{\rm int}} \right\rangle
%\int {\cal D}x{\cal D}\bar\j {\cal D}\j {\cal D}c^* {\cal D}c
%e^{-\frac{1}{\hbar} S}.
\label{eq199}
\ee
Here, $S_{\rm int}$ is the interaction part of the action
%The path integral on the right hand side is to be evaluated with
%the boundary conditions $x(0)=x(-\b)=x_0$, $\bar\j_f(0)=\bar\h_f$,
%$c^*(0)=\bar\h_{gh}$, $\j_f(-\b)=\c_f$, and $c(-\b)=\c_{gh}$.
%The action $S$ is given by
\bea \frac{1}{\hbar} S & = &
\frac{1}{\hbar} \int_{-\b}^0 dt \left( \frac{1}{2} \left(\frac{dx^i}{dt}
\right)^2 + \hbar \bar\j_A \frac{d\j^A}{dt}
+\hbar c^*_M \frac{d c^M}{dt} \right) - \bar\j_A(0) \j^A(0)
- c^*_M(0) c^M(0) \nonumber \\ & &
-\int_{-\b}^0 dt ( \dot{x}^j A_j^{\a} c^*_M (T_{\a})^M{}_N c^N -
 \frac{\hbar}{2} \psi^a \psi^b F^{\a}_{ab} c^*_M (T_{\a})^M{}_N c^N ),
\label{eq1444}
\eea
The couplings $\dot{x}Ac^* Tc$
result from integrating out the
momenta $p$, and
combine with the ghost kinetic term to the covariant derivative
$D_t c^M = \dot{c}^M - \dot{x}^j A^\a_j(x) T_\a{}^M{}_N
 c^N$.
The fields satisfy the boundary conditions
$x(0)=x(-\b)=x_0$, $\bar\j_f(0)=\bar\h_f$,
$c^*(0)=\bar\h_{gh}$, $\j_f(-\b)=\c_f$, and $c(-\b)=\c_{gh}$.
Decomposing all fields into background parts and quantum fluctuations as in
(\ref{eq102}), we note
that the object $S_{\rm loops}$ now contains tree graphs because
the four-fermion couplings $\j\c c^* c$
contain terms linear in quantum fields and hence
lead to tree graphs.
These tree graphs do, of course,
contribute to the classical action, as we discussed in section 2.
%\bea
%-\frac{1}{\hbar} S &=& \frac{1}{\hbar} \Big[ \sum_{k=1}^N
%\frac{-1}{2\b} (x_k - x_{k-1})^2
%-\bar{\j}_{k,I}(\j_k^I - \j_{k-1}^I)
%- c^*_{k,A}(c_k^A - c_{k-1}^A) \nonumber \\
%&+& c^*_{k,A}(T_\a)^A_{\ B}c^B_k (x_k^j - x_{k-1}^j) A_j^\a
%\big(\frac{x_k^j + x_{k-1}^j}{2} \big)\nonumber \\
%&-&\frac{1}{2} F^\a_{ab} \big(\frac{x_k^j + x_{k-1}^j}{2}\big) \j_k^a \j_k^b
%c^*_{k,A}(T_\a)^A_{\ B}c^B_k \Big].
%\ena
%We recall that $\bar{\j}_{N,I} = \bar{\h}_{f,I}$ and
%$c^*_{N,A}=\bar{\h}_{gh,A}$,
%while $\j_0^I=\c_f^I$ and $c_0^A = \c_{gh}^A$. The integrations over $N$
%factors $p_{k,j}$ gave the Feynman measure $(2 \p \hbar \b)^{-\frac{1}{2}
%%nN}$,
%and the couplings $c^*Tc\dot{x}A$ resulted from completing the $p$-squares.

In order to study the loop expansion in detail, we rescale $t=\b \t$, and
$\j \rightarrow \j' (\b\hbar)^{-1/2}$,
$\bar{\j} \rightarrow \bar{\j}' (\b\hbar)^{-1/2}$,
$\c_f \rightarrow \c_f' (\b\hbar)^{-1/2}$,
$\bar{\h}_f \rightarrow \bar{\h}_f' (\b\hbar)^{-1/2}$.
After the  rescaling of $d\c_f$
and $d\bar{\h}_f$
{\it the measure becomes completely $\b\hbar$-independent}, and
%\bea
%-\frac{1}{\hbar} S &=&
%\frac{-1}{\b \hbar} \int_{-1}^0 d\t \big(
%\frac{1}{2} (\frac{dx^i}{d\t})^2 + \bar{\j}_I \frac{d\j^I}{d\t}
%\big) \nonumber \\
%&-&\frac{1}{\hbar} \int_{-1}^0 d\t \big(
%c_A^* \dot{c}^A - \dot{x}^j A_j^\a c^*T_\a c
%+\frac{1}{2} \j^a \j^b F_{ab}^\a c^* T_\a c
%\big).
%\ena
\bea \frac{1}{\hbar} S & = &
\frac{1}{\b\hbar} \int_{-1}^0 d\t \left( \frac{1}{2} \left(\frac{dx^i}{d\t}
\right)^2 +  \bar\j_A \frac{d\j^A}{d\t}
\right) -\frac{1}{\beta\hbar} \bar\j_A(0) \j^A(0)
- c^*_M(0) c^M(0)
\nonumber \\ & &
+\int_{-1}^0 d\t (c^*_M \dot{c}^M
-\dot{x}^j A_j^{\a} c^*_M (T_{\a})^M{}_N c^N +
 \frac{1}{2} \psi^a \psi^b F^{\a}_{ab} c^*_M (T_{\a})^M{}_N c^N )
\eea

Since the $x$ and $\j$ propagators are proportional to $\b\hbar$
and all vertices and $c^*c$ propagators are $\b\hbar$-independent, we need
not consider the vertices containing the $q^i,\bar\j_{{\rm qu},A}$ or
$\j^A_{\rm qu}$ of (\ref{eq102}). Hence, we can restrict our attention
to the vertices $F\j \j c^* T c$, with the $\j$'s replaced by their
background value.  There are no
classical contributions to $S_{\rm loops}$
since we evaluate the classical action from
$y$ to $z$ with $y=z=x_0$.
So altogether we are left with
\bea
{\cal A}_n &=&  \int \prod_{i,M,A}
[\frac{dx_0^i}{\sqrt{2 \p \hbar}} d\c_{gh}^M d\bar{\h}_{gh,M}
d\bar{\h}_{f,A} d\c_f^A ]
\Big[ \sum_M
\big( \prod_{N \neq M} \bar{\h}_{gh,N} \c_{gh}^N \big) \Big]
e^{\bar{\h}_{gh} \c_{gh}} \nonumber \\
&\ & \left\<
%\int \prod_N dc^*_N dc^N
\exp \left[ -
\int_{-1}^0 d\t (c^*_M \dot{c}^M +
\frac{1}{2} \j_{\rm bg}^a \j_{\rm bg}^b F_{ab}^\a c^* T_\a c)
\right]
\right\>_{\rm loops}
 \label{eq144}
\ena
where $\j_{\rm bg}^a$ are the classical values of $\j^a$
exactly as in (\ref{eq112}).

The propagator $\<c^M(\s) c^*_N(\t) \>$ is equal to
$\d^M_N\q(\s-\t)$, hence no
closed $c$-loops can contribute (in a closed $c$-loop one always moves
somewhere backwards in time). Only terms with precisely
one $\bar{\h}_{gh}$ and one
$\c_{gh}$ can contribute due to the projection operator in the measure.
These terms are  just tree graphs, with $c^*=\bar{\h}_{gh}$ at one end and
$c=\c_{gh}$ at the other end. If we consider a
diagram with $k$ vertices, then it will contain an integration
\be
\int_{-1}^0 d\s_1 \cdots d\s_k \q(\s_1-\s_2) \cdots \q(\s_{k-1}-\s_k) =
\frac{1}{k!}.
\ee
The combinatorical factor $1/k!$ from the expansion of $e^{-S/\hbar}$
cancels because there are exactly $k!$ different ways to build a
connected tree graph from the vertices. Putting all this together
yields
\be
-\frac{1}{\hbar} S_{\rm loops} = \bar\h_{gh,M} \left(
e^{ -\frac{1}{2} F^{\a}_{ab} \j^a_{\rm bg} \j^b_{\rm bg} T_{\alpha} }
-1 \right)^M{}_N \c_{gh}^N \label{eq145a}
\ee
The background term $\exp(\bar\h_{gh,M}\c^M_{gh})$ in (\ref{eq144})
cancels the $-1$ in
(\ref{eq145a}).
The integration over $\bar\h_{gh}$ and $\c_{gh}$ in (\ref{eq144})
with $S_{\rm loops}$ replaced by (\ref{eq145a}) can now easily been done,
since it picks out the piece with only one $\bar\h_{gh}$ and $\c_{gh}$ from
$S_{\rm loops}$.
Transforming in addition variables from $\bar\h_f$
and $\c_f$ to $\j^a_{\rm bg}$ leaves us then finally with the
following result
\bea
{\cal A}_n & = & \left(\frac{-i}{2\pi}\right)^{n/2}
\int dx_0^i d\j^1_{\rm bg} \cdots d\j^n_{\rm bg}
{\rm tr} \left( e^{-\frac{1}{2}
F^{\a}_{ab} \j^a_{\rm bg} \j^b_{\rm bg} T_{\alpha} } \right)
\nonumber \\
& = & \left( \frac{i}{4\pi} \right)^{n/2} \frac{1}{(\frac{n}{2})!}
\int dx_0^i \epsilon^{a_n\cdots a_1} {\rm tr} (
F_{a_1a_2}\cdots F_{a_{n-1} a_n} )
\eea
where the trace is over the Yang-Mills indices of $T_\a$ in
$F_{ab}=F^\a_{ab}T_\a$. This is the correct anomaly.

\subsection{Trace anomalies}

We can now easily compute the trace anomalies for a spin-$0$ and
spin-$1\over 2$ field in an $n$-dimensional quantum field theory. In this
case, the Jacobian is equal to one, so we just need to evaluate the
trace of the appropriate transition element, and no additional
operators need to be inserted. The anomaly is then equal to the
$\b$-independent part of the trace. New in this section is the treatment of
fermions by path integrals; in \cite{bapvn} they were kept as operators.

For a real scalar field in $n=2$ dimensions, we can directly take the
trace in (\ref{trans}). Singling out the $\b$-independent part yields
\be
A_2^{\rm trace} = - {\hbar\over 24\pi} R
\label{trAs2}
\ee
In order to obtain the trace anomaly for a scalar field in higher
dimensions, we need to compute the terms of higher order in $\b$ in the
transition element.
Since the transition element contains a factor $(2\pi\b\hbar)^{-n/2}$,
one needs $\frac{n}{2}+1$ loops in $n$-dimensional space.

%These can most easily be found in the path
%integral formalism; for the trace anomaly in $n$ dimensions ($n$ is
%even) we have to evaluate the transition element up to $n/2+1$
%loops in order to obtain the $\b$-independent part.

We will now evaluate the trace anomaly for a spin-$1\over 2$ field in
$n$ dimensions. In this case we have to compute the trace of the ${\rm
N}{=}1$ supersymmetric transition element, and project out the
$\b$-independent part. As claimed before, both approaches to obtain
fermionic creation and annihilation operators will lead to the same
result.

Recall that when we double the number of fermions we should normalize
the trace by dividing by an extra factor $2^{n/2}$.  The trace anomaly
for a spin-${1\over 2}$ field in $n$ dimensions is therefore given by
(cf. \cite{bapvn}, equation (2.9))
\be
{\cal A}_n^{spin-{1\over 2}} = -{1\over
2^{n/2}} \lim_{\b \rightarrow 0} \int \! d\chi^a d\bar\eta^a \,
e^{\bar\eta \chi} \langle x_{0},
\bar\eta | \exp \left( - {\b\over\hbar} \hat H \right) | x_{0},
\chi \rangle \quad ; \qquad a = 1\ldots n
\label{SAnn}
\ee
with the transition element given in (\ref{FerTrans}).
For $n=2$ the trace anomaly becomes ${-\hbar\over 24\pi}R$, which is
indeed the result for a Dirac fermion; for the anomaly of a Majorana
fermion we have to divide this expression by two.

When we instead combine the Majorana fermions into half as many Dirac
fermions, we should take the trace of the transition element given in
(\ref{PROP1re}). In this case of course no extra normalization
factor is needed, and we find directly
\be
{\cal A}_n^{spin-{1\over 2}} = - \hbar \lim_{\b\rightarrow 0} \int \!
d\chi^A d\bar\eta^A \,
e^{\bar\eta \chi} \langle x_{0},
\bar\eta | \exp \left( - {\b\over\hbar} \hat H \right) | x_{0},
\chi \rangle \quad ; \qquad A = 1\ldots n/2
\label{SAnn1}
\ee
now with the transition element given in (\ref{PROP1re}). This gives
again the same result for the anomaly as above.

\section{Conclusions}

In this article we have given a complete, explicit derivation of
quantum mechanical path integrals for bosons and fermions, both for Dirac
fermions and Majorana fermions. Our main result is
that the factors $\d(\s-\t)$ in
the Feynman
rules for configuration space path integrals
should be interpreted as Kronecker delta functions,
even in the continuum case, and should not be regulated
by mode regularization.

We {\it define} our path integrals in curved space by starting from
the Hamiltonian (operator) formalism. After inserting complete sets
of states (coherent states for fermions), and Weyl-ordering the
Hamiltonian (leading to order $\hbar$ and $\hbar^2$ terms in
the path integral action), we obtained the discretized propagators and
vertices {\it in closed form}. With these ingredients one can
construct a loop expansion of the path integral in terms of
Feyman integrals. Some of the bosonic propagators contain
$\d(\s-\t)$ singularities, but adding `Lee-Yang ghosts', terms
with two or more $\d(\s-\t)$ cancel. Terms with one $\d(\s-\t)$
should then be evaluated as indicated above.

We paid particular attention to Majorana fermions. Of course, starting
with an arbitrary initial state $|A\>$ and acting on it with
products of $\j$'s, the states so obtained will span a Hilbert space
on which the $\j$'s can be represented as matrices. One can
then define matrix-valued Hamiltonians as in \cite{bapvn}.
We found it much simpler to define creation and annihilation
operators and then to use the standard formulation of coherent
states \cite{textb}. The particular way of defining creation
and annihilation operators is, of course, arbitrary, and so is
therefore the choice of vacuum, but in problems involving a
trace over the Hilbert space, this arbitrariness should cancel.
We achieved the construction of creation and annihilation operators
in two ways: either by combining the Majorana spinors pairwise
into creation and annihilation operators (`halving'), or by
adding another set of Majorana spinors (`doubling', this works
also for odd-dimensional spaces). Of course, the Hilbert spaces,
vacua etc. are different in both cases, and indeed we found
different expressions for the transition element, but the anomalies
came out the same. This confirms our claim that in traces over
the Hilbert spaces, differences created by choosing different
vacua should cancel. We verified our formalism by computing the
transition elements for a bosonic and several fermionic transition
elements through order $\b$, and comparing the results with those
from (unambiguous) operator calculations. We provided further evidence
by doing a three-loop calculation in appendix A.1.

We applied our general formalism to trace anomalies and to chiral
anomalies for spin-$1/2$ fields. The chiral anomalies were
already studied by Alvarez-Gaum\'e and Witten and the trace anomalies
in \cite{bapvn}, but we have treated the Yang-Mills ghosts and the
Majorana fermions on equal footing with the bosons and
obtained a uniform path integral treatment. Moreover,
our derivation makes a detailed and complete treatment for any anomaly
possible, including all normalizations.

On the more technical side, we have seen that the Hamiltonian $\hat{H}$
in the operator formalism and the Hamiltonian function $H$ in the
path integral are related by the formula $\hat{H}=(H)_W$, where
$(H)_W$ contains in general terms of order $\hbar$ and $\hbar^2$.
If $\hat{H}$ is Einstein invariant, the corresponding action in the
path integral is not Einstein invariant (although the transition element
is); rather there are noncovariant $\G\G$ terms. Conversely, if the
action is the naive action, $\hat{H}$ will contain $\hbar$
and $\hbar^2$ terms. In particular, the supersymmetric Hamiltonians
(whose ambiguity was fixed by requiring hermiticity and Einstein invariance)
do not lead to the usual classically supersymmetric action in the path
integral.
Rather, there are extra terms proportional to $\hbar^2$.
For chiral anomalies, though, these extra terms in the
action do not contribute, which explains why the results of
Alvarez-Gaum\'e and Witten \cite{alwi}
are correct (that they are correct can
be checked by doing loop calculations in the corresponding quantum field
theory, see \cite{alwi} and \cite{leuven}).
For trace anomalies, the extra terms do matter. Here
higher loop calculations are needed
 and we stressed that
the noncovariant vertices (of the form $\G\G$ and $\omega\omega$)
as well as our new Feynman rules {\it must} be taken into account,
even when one uses normal co-ordinates,
to obtain the correct results.

This concludes our analysis of quantum mechanical path integrals.
One might wonder whether the subtleties we have found in one dimension
have a counterpart in higher dimensions. For local field theories
this seems unlikely, because we expect that possible extra terms in
the path integral will be proportional to $\d^{n-1}(\vec{x})$, and
hence would vanish in dimensional regularization. However, in non-local
field theories, such as Yang-Mills theories in the Coulomb gauge
\cite{schw,chlee}, there might be effects. This is under study.

\noindent
{\bf ACKNOWLEDGEMENTS}

Over the years, we have discussed
several of the issues raised and solved in this
paper with many colleagues. We thank all of them, in particular R. Endo
for help in obtaining the results of appendix A.3.

\appendix

\section{Appendices}

\subsection{A 3-loop computation}

We check that there are no counterterms beyond the two-loop counterterm
in (\ref{HWeyl}) by performing a 3-loop calculation in a model which describes
a free particle, but which has been cast into the form of a non-linear
sigma model by a nontrivial co-ordinate transformation.
This extends the two-loop phase space calculation of \cite{geji}. We shall
then redo the calculation in configuration space and show that we
obtain the same result (Matthews' theorem).

We consider a free massive point particle on the interval
$-\infty <t< \infty$  with $L =
{1\over 2} \dot{q}^2 - {1\over 2} q^2$ and substitute $q = Q + {1\over
3} Q^3$.  Then the action becomes
\eq \label{y133}
L (Q, \dot{Q}) = {1\over 2} \dot{Q}^2 (1+ Q^2)^2 - {1\over 2} Q^2
(1+{1\over 3} Q^2)^2
\eqe
and the Hamiltonian is $H (Q, P) = {1\over 2} P^2 (1+Q^2)^{-2} + {1\over
2} Q^2 (1+{1\over 3} Q^2)^2$.  The first-order action in phase space
reads
 \eqa
L(Q, P) &=& (P \dot{Q} - {1\over 2} P^2 - {1\over 2} Q^2 ) + \cl_{int} (Q, P)
\nonumber\\
L_{int} (Q, P) &=& {1\over 2} P^2 (2Q^2 + Q^4) (1+ Q^2)^{-2} - {1\over
3} Q^4 - {1\over 18} Q^6 \nonumber\\
&=& P^2 Q^2 - {3\over 2}  P^2 Q^4 - {1\over 3} Q^4 - {1\over 18} Q^6 + \ldots
 \eqae
We consider the vacuum self-energy at the 3-loop level.  There are
\begin{itemize}
\item [(i)] ``clover-leaf graphs" from the $P^2 Q^4$ and $Q^6$
couplings with three loops all meeting at one point,
\item [(ii)] a string of three loops with two four-point vertices.  We call
such a
string a $PP$ string (or $PQ$ string  or $QQ$ string) if the equal-time
contractions at the ends of the string consist of two $P$ lines (or one $P$
and one $Q$ line, or two $Q$ lines.  Since an equal-time $PQ$
contraction vanishes, this exhausts all possibilities.)
\item [(iii)] watermelon graphs with 4 propagators between 2 vertices.
\item [(iv)] a one-loop contribution from the extra two-loop
interaction $\D V$; since $\D V$ is of order $\hbar^2$, this one-loop
graph contributes also at order $\hbar^3$.  To evaluate $\D V = {1\over
8} \G^i_{jk} \G^j_{i\ell} g^{kl}$ we note that $g_{ij} = (1+Q^2)^2$ and
find
 \eq \label{y135}
\D V = {\hbar^2\over 2} Q^2 (1+ Q^2)^{-3} = {\hbar^2\over 2} Q^2 + \ldots
 \eqe
There is no $R$ term since the metric is flat in this model.
For the phase-space calculation we use the propagators
 \eqa
< P (\s) P (\t) > &=& < Q (\s ) Q (\t ) > = {1\over 2} e^{-i | \s - \t |}
\nonumber\\
< P (\s ) Q (\t ) > &=& - < Q (\s ) P (\t ) > = {-i\over 2} e^{-i | \s - \t
|} \e
(\s - \t)
 \eqae
and further the equal-time contractions
 \eq
< P (\s ) P (\s ) > = < Q (\s ) Q (\s ) > = {1\over 2} ; < P(\s) Q(\s) > =
< Q (\s
) P (\s ) > = 0
 \eqe
\end{itemize}
We find the following results
\be
\begin{array}{lll}
\underline{{\rm From} \; P^2 Q^2 - {3\over 2} P^2 Q^4} &
\underline{{\rm From} \; - {1\over 3} Q^4 - {1\over 18} Q^6} &
\raisebox{-.6ex}{\underline{{\rm Cross $\;$ terms}}} \\
 {\rm clover} = - {9i \over 16} & {\rm clover} = - {5i\over 48} \\ \\
P-P \; {\rm string} = {i\over 16} & {\rm string} = {i\over 4} & QP \;
{\rm string} = {-i\over 4} \\
Q-Q \; {\rm string} = {i\over 16} & & QQ \; {\rm string} = {i\over 4} \\
P-Q \; {\rm string} = {-i\over 8} \\
{\rm watermelon}\;({\rm no}\; PQ) = {i\over 16} & {\rm watermelon}\;
= {i\over 24} &
{\rm watermelon} \; ({\rm two} \; PQ) = {i\over 4} \\  {\rm
watermelon} \; ({\rm two}\; PQ) = {i\over 4} \\
{\rm watermelon} \; ({\rm four} \; PQ) = {i\over 16} & & \underline{{\rm
{}From}\; \D V}: {-i\over 4}
\end{array} \label{3loop}
\ee
Adding all contributions, we find the correct result:  the sum of the
vacuum self-energies vanishes at the three-loop level.
This demonstrates that at the 3-loop level no further
counterterms are present in the phase space path integral.

In the configuration space path integral, there are extra vertices and
an extra term in the $\dot{Q} \dot{Q}$ propagator. According to
Matthews' theorem, the final answer should be the same as in the phase space
approach. We now check this. The $Q Q$ propagator is the same, while the
$\dot{Q} Q$ propagator in configuration space is equal to the $P Q$
propagator in phase space. The $\dot{Q} \dot{Q}$ propagator differs from
the $P P$ propagator by a Dirac delta function
\be
\< \dot{Q}(\s) \dot{Q}(\t) \> = {1 \over 2} e^{-i |\s-\t|}
+ i \d(\s-\t),
\ee
which is due to differentiation of the time-ordering $\q(\s-\t)$ and
$\q(\t-\s)$ functions. The action contains also ghosts,
\be
L(Q, \dot{Q}, b, c, a) =
{1 \over 2} (\dot{Q} \dot{Q} + b c + a a)(1+Q^2)^2 -
{1 \over 2} Q^2 (1 + {1 \over 3} Q^2)^2,
\ee
whose only role is to cancel products of Dirac delta functions.
We shall now show that the contribution from the extra vertices in the
Hamiltonian approach equals the contributions of the extra terms with one
Dirac delta function in the configuration space approach.

The extra vertices in the phase space approach are
$L_{\rm int} (P, Q) - L_{\rm int} (\dot{Q}, Q)\Big|_{\dot{Q} =P} =
-2 P^2 Q^4$. Hence we get only an extra contribution to the clover leaf graph.
Since $-3/2 P^2 Q^4$ gave $-9i/16$ according to (\ref{3loop}), we now get
$4/3$ times this contribution
\be
\hbox{contribution extra vertices in phase space} = - \frac{3i}{4}.
\label{extraph}
\ee

In the configuration space approach, all delta functions in the  clover leaf
graph cancel, since for every $\dot{Q} \dot{Q}$ contraction there is a
compensating $bc$ and $aa$ contraction. In the graph with a string of three
loops, no delta functions remain if one of the outer loops is a
$\dot{Q} \dot{Q}$ contraction.
The graph with one $\dot{Q}$ in an outer loop vanish.
Hence only the graph in which the inner loop
contains two $\dot{Q} \dot{Q}$ contractions yields an extra contribution.
This contribution comes from the graph with two $\dot{Q}^2 Q^2$ vertices.
It contains a term proportional to $\delta(\s-\t)^2$ which cancels against
the contribution from similar graphs with an internal ghost loop.
Only the terms with one delta function remain and one finds
\be
\hbox{extra contribution to string} = -{i \over 4}.
\label{extrastr}
\ee
Finally there are the watermelon graphs. We must consider graphs with two
 $\dot{Q} \dot{Q}$ propagators and graphs with one $\dot{Q} \dot{Q}$
propagator. However, the latter do not contribute because they also contain
a $\< Q(\s) \dot{Q}(\t) \>$ and a $\< \dot{Q}(\s) Q(\t)\>$ propagator, and
the integral $\int \delta(\s-\t) \epsilon(\s-\t) \epsilon(\t-\s) d\s d\t$
vanishes according to our rules.
In the graph with two $\< \dot{Q} \dot{Q} \>$ propagators we again need
the term proportional to one $\d(\s-\t)$.
\be
\hbox{extra contribution watermelon graphs} = -{i \over 2}.
\label{extrawm}
\ee
Since (\ref{extraph}) equals the sum of (\ref{extrastr}) and  (\ref{extrawm}),
Matthews' theorem is verified in this model at the three-loop level. The extra
potential in (\ref{y135}), being proportional to two Christoffel symbols,
will not contribute below the three-loop level in normal co-ordinates. However,
at the three-loop level it does contribute. Note that the usual co-ordinate
transformation from arbitrary co-ordinates to normal co-ordinates
will yield correct results in the one- and two-loop computation, but will
yield an incorrect result at the three-loop level. In our model, this is very
clear: in normal co-ordinates, the action (\ref{y133}) reverts to a free
model, but the extra potential in (\ref{y135}) would yield a spurious
contribution in the normal co-ordinates $q$.

\subsection{Higher derivative theories}

To test our rules in a higher-derivative and higher dimensional model,
we consider a massive real scalar field
$\varphi$ with rather singular interactions \cite{bern}
\begin{equation}
  {\cal L} ~=~ - {\scriptstyle \frac{1}{2}}\, \partial_{\mu}\, \varphi\,
\partial^{\mu} \, \varphi - {\scriptstyle \frac{1}{2}} m^2
\varphi^2 - {\scriptstyle \frac{1}{4}} \, \lambda
\, \left( \partial_{\mu}\, \varphi\, \partial^{\mu}\, \varphi\right)^2
\end{equation}
One novelty that appears in this example is the need to introduce `Lee-Yang'
ghosts which couple to derivatives of the scalar fields, see (\ref{y00a}).

%Because the
%scalar field
%is massive, we must also consider contractions at equal times (tadpole
%diagrams).
%We already considered tadpole diagrams in Yang-Mills theory, but if we would
%have used dimensional regularization, we could have omitted them since they
%then vanish.  In field theory massive tadpoles do not vanish in dimensional
%regularization.  In this model the extra term in the propagator contributes
%even to
%the tadpoles, a situation we did not encounter before.

The conjugate momentum is defined by
\begin{equation}
\label{2}
  \pi (x) ~=~ \frac{\delta \, {\cal L}}{\delta\, \dot{\varphi}\, (x)}~=~
\dot{\varphi} + \lambda\,\frac{\delta\,{\cal L}_{int}}{\delta\,
\dot{\varphi}}\, \left(\varphi, \vec{\bigtriangledown}\,\varphi,
\dot{\varphi}\right)~=~\dot{\varphi} +\lambda\, \dot{\varphi}\, \left[ \left(
\vec{\bigtriangledown}\,\varphi\right)^2 - \dot{\varphi}^2\right],
\end{equation}
and the Hamiltonian density is given by
\begin{eqnarray*}
  {\cal H} &=& \pi \, \dot{\varphi}  - {\cal L}\, \left(\varphi,
\vec{\bigtriangledown}\,\varphi, \dot{\varphi}\right) = \pi \, f\,
\left(\varphi, \bigtriangledown\,\varphi, \pi\, ;\, \lambda\right) - {\cal L}
\, \left(\varphi, \vec{\bigtriangledown}\,\varphi, f\right) \\
   & & \\
   &=& \sum\limits^{\infty}_{n=0} \, \lambda^n\, {\cal H}^{(n)}\,
   \left(\varphi, \bigtriangledown\,\varphi, \pi\right)
\end{eqnarray*}
It is straightforward to find the first few terms of ${\cal H}$,
\eqa
\ch^{(0)} &=&\half [\p^2 + (\vec{\nabla} \varphi)^2 ] +
\half m^2 \varphi^2 \nonumber\\
\ch^{(1)} &=& - \cl_{int} (\varphi ,\vecnab \varphi, \p ) =
{\scriptstyle{\frac{1}{4}}} [ \p^2 - (\vecnab \varphi)^2 ]^2 \nonumber\\
\ch^{(2)} &=& \half [ \frac{\d \cl_{int}}{\d \dot{\varphi}} (\varphi ,
\vecnab \varphi , \p ) ]^2  = \half \p^2 [ \p^2 - (\vecnab \varphi)^2
]^2
\eqae
The terms linear in $\l$ yield the same vertices as in the
Lagrangian approach, namely $\cl_{int} (\varphi , \vecnab \varphi
, \p)$, but with $\dot{\varphi}$ replaced by $\p$.  Extra with respect to the
Lagrangian approach are the vertices in $\ch^{(2)}$. In the interaction
picture where one uses the free field equations and in-fields
$\varphi_{in}$ and $\p_{in}$, one replaces $\p_{in}$ by
$\dot{\varphi}_{in}$.

On an infinite $t$-interval the boundary conditions on the bra and ket
vacuum are that they are the lowest energy states. For the
Feynman propagator one then obtains
 \eq < 0 | T \varphi (x) \varphi (y ) | 0 > = \int \frac{ d^4
k}{(2\p )^4} \frac{-i}{k^2 + m^2 - i\e} e^{ik(x-y)}
 \eqe
For the $\varphi \p$ propagator one finds
\eq
< 0 | T \varphi (x) \p (y) | 0 > = < 0 |T \varphi (x) \dot{\varphi} (y)| 0 > =
\frac{\del}{\del y^0} < 0 | T \varphi (x) \varphi (y) | 0>
\eqe
because $\frac{\del}{\del y^0}$ hitting $\q (x^0 - y^0)$ produces a $\d
(x^0 - y^0)$ times a commutator $[\varphi (\vec{x}, t_0), \varphi
(\vec{y}, t_0 ]$ which vanishes.  However, for the $\p \p$ propagator one
obtains an extra term
\eqa
&& < 0 | T \p (x) \p (y) | 0 > = < 0 | T \dot{\varphi} (x) \dot{\varphi}
(y) | 0 >
= \nonumber\\
&& = \frac{\del}{\del x^0} \frac{\del}{\del  y^0} < 0 |T \varphi (x) \varphi
(y)
| 0 >- \d (x^0 - y^0) [\varphi (x) , \p (y) ] \nonumber\\
&& = \frac{\del}{\del x^0 } \frac{\del}{\del y^0} < 0 | T \varphi (x)
\varphi (y)
 | 0 > - i \hbar \d^4 (x-y)
\eqae

More generally we have
\eqa
< 0 | T \del_\m  \varphi (x) \del_\n \varphi (y) | 0 > &=& \frac{\del}{\del
x^\m} \frac{\del}{\del y^\n} < 0 | T \varphi (x) \varphi (y) | 0 >
\nonumber\\
&& - i \hbar \d_\m{}^0 \d_\n{}^0 \d^4(x-y)
\eqae
In momentum space we find
\eq
< 0 | T \del_\m \varphi (x) \del_\n \varphi (y) | 0 > = \int \frac{d^4 k}{(2
\p)^4} \left( \frac{-i k_\m k_\n}{k^2 + m^2 - i \e} - i \hbar \d_\m{}
^0 \d_\n{}^0 \right) e^{ik(x-y)}.
\eqe
The same phase space propagators are obtained from the path integral
by inverting the kinetic matrix of the fields $\varphi$ and $\pi$.

We shall now consider elastic $\varphi \varphi$ scattering.  There are the
extra vertices and the extra terms in the unequal-time propagator to
use, but we shall also need equal-time contractions.  They are given by
\eq
< 0 | \del_\m \varphi (x) \del_\n \varphi (x) | 0 > = \sum_{\vec{k}}
\frac{k_\m k_\n}{2\o V} = \int \frac{d_3 \vec{k}}{(2\p)^3} \, \frac{k_\m
k_\n}{2\o}
\label{o165}
\eqe
These equal-time contractions are the same in the Lagrangian approach as
in the Hamiltonian approach.
We shall not try to regulate these divergent expressions,
but we shall only use that for $k_\m k_\n$ equal to $k_i k_j$ one may replace
$k_i
k_j$ by ${1\over 3} \d_{ij} \vec{k}^2$, while for $k_\m k_\n$ equal to $k_i
k_0$ one
obtains zero.  We shall denote the corresponding divergent integrals by
 \eqa
I(\vec{k}^2) &=& \int {d_3 \vec{k} \over (2\p)^3} \; {\vec{k}^2 \over 2\o}
\; ,\hspace{6mm}  I
(k^2_0) = \int {d_3 k \over (2 \p)^3} \; {k^2_0 \over 2\o}  \nonumber\\
I (k_i k_j) &=& {1\over 3} \d_{ij} I (\vec{k}^2) ,
\hspace{6mm} I (k_i k_0) = 0
 \eqae
(where $k_0^2 = \o^2)$.  The equal-time contractions in the Hamiltonian
approach are thus equal to those in
the Lagrangian approach.  Our aim is to show that the extra terms with $I
(\vec{k}^2)$ and $I(k_0^2)$ in the Hamiltonian approach (due to extra
vertices in
$H_{\rm int})$ cancel algebraically with similar terms in the Lagrangian
approach
(due to the term $i \d_\m^0 \d_\n^0 \d^4 (x-y)$ in the propagator).

Recalling the vertices
\eqa
\ch_{int}  &=& \l \ch^{(1)} + \l^2 \ch^{(2)} + \ldots \nonumber\\
&=& \frac{\l}{4} (\del_\m \varphi \del^\m \varphi)^2 + \half \l^2
\dot{\varphi}^2 (\del_\m \varphi \del^\m \varphi)^2 + \ldots
\eqae
there are 3 graphs to be computed, shown in figure 1.

\vspace{.20in}

\begin{figure}
\centerline{\hbox{\psfig{figure=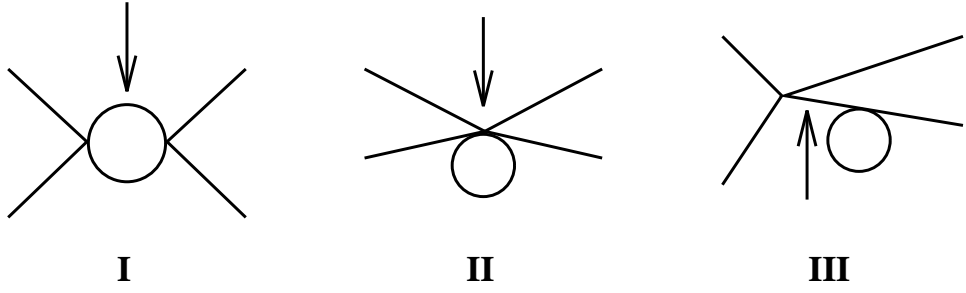,width=5in}}}
\caption{The three graphs to be computed. Arrows indicate where the
noncovariant propagator or the extra vertex is to be used.}
\end{figure}

%\eqa
%\hspace{4.0in} \begin{array}{l} {\rm The \; arrows \; indicate} \\
%{\rm where \; the \; noncovariant} \\ {\rm propagator \; or \; the \;
%extra} \\ {\rm vertex \; is \; to \; be \; used.} \end{array}
%\eqae

%(I) \hspace{1.5in} (II) \hspace{1.0in} (III)

In (I) we find the combination ($< p_\m (x)  p_\n (y)  > + i \d_\m^0
\d_\n^0 (x-y))
(<p_\r (x) p_\s (y)> + i \d_\r^0 \d_\s^0\d^4 (x-y)$) where we have written the
Feynman propagators as a sum of phase space propagators plus contact terms.
The
two cross terms yield extra terms proportional to the equal-time contractions
$<p_\m (x) p_\n (x) >$.  Since the phase space propagators are continuous
at $x=y$,
there are no subtleties in taking the equal-time limit.  In (II)  the
equal-time
contraction is the same in phase-space and in configuration space, but the
vertex
${\cal H}_{(2)}$ is extra in phase space.  Finally, in (III) the propagator
in the loop is
the same in phase space as in configuration space for the same reason, but the
propagator connecting the two vertices has a noncovariant piece.  The net
effect of
these noncovariant pieces in the propagator is to contract the propagator
to a point.
Hence, all 3 graphs become topologically of the form of figure (II), which
makes the
cancellation possible.  We now study in more detail whether the extra
contributions
from (I) and (III) in the Lagrangian approach
are equal to the extra contribution from (II) in the Hamiltonian approach.

For (I) we need the contractions in
\eqa
S_I &=& \frac{(-i)^2}{2!} T \int \l \ch^{(1)} (x) d^4 x \int \l \ch^{(1)} (y)
d^4 (y) \nonumber\\
&=& \frac{-\l^2}{32} T \int  \del_\m \varphi \del^\m \varphi
\del_\n\varphi \del^\n \varphi (x) d^4 x \int \del_\r \varphi \del^\r
\varphi \del_\s \varphi \del^\s \varphi (y)  d^4 y
\eqae
As first contraction we can take $< \del_\n \varphi (x) \del_\s  \varphi (y)
>$, it comes with a statistical factor $4\times 4=16$.  The second
contraction can then still be done in 3 different ways $(\del_\n \varphi
\del_\s \varphi , \del_\n \varphi \del_\r \varphi$ or $\del_\m \varphi
\del_\r \varphi$) ; however, to avoid double-counting, we need to multiply
with a factor 1/2.  Hence
\eqa
S_I &=& \frac{-\l^2}{32} \frac{16}{2} [ (\del_\m \varphi)^2 (x) < \del_\n
\varphi (x) \del_\s \varphi (y) > < \del^\n \varphi (x) \del^\s \varphi (y) >
(\del_\r \varphi )^2 (y) \nonumber\\
&+& 4 (\del_\m \varphi)^2 (x) < \del_\n \varphi (x) \del_\s \varphi (y) > <
\del^\n \varphi (x) \del_\r \varphi (y) > \del^\s \varphi (y) \del^\r
\varphi (y) \nonumber\\
&+& 4 \del_\m \varphi (x) \del_\n \varphi (x) < \del^\m \varphi (x)
\del_\r \varphi (y) > < \del^\n \varphi (x) \del_\s \varphi (y) > \del^\r
\varphi (y) \del^\s \varphi (y) ] \nonumber\\
&&  \hphantom{xxxxxxxxxxxxxxxxxxxxxxxxxx}
\label{o170}
\eqae

In each pair of contractions we make the substitution
 \eqa
&& < \del_\m \varphi (x) \del_\n \varphi (y) > < \del_\r \varphi (x) \del_\s
\varphi (y) >
= \nonumber\\
&&  < p_\m (x) p_\n (y) + i \d_\m^0 \d_\n^0 \d^4 (x-y) >< p_\r (x) p_\s (y)
+ i \d_\r^0 \d_\s^0 \d^4 (x-y)>
 \eqae
and retain only the two cross terms, replacing $\d^4 (x-y) p_\r (x) p_\s (y)$
by equal-time contractions for which we then use (\ref{o165}).  This yields
 \eqa
S_I &=& - {i\over 2} (\del_\m \varphi)^4 I (k_0^2 ) - 2 i (\del_\m \varphi
)^2 \del^\s
\varphi \dot{\varphi} I (k_0  k_\s ) \nonumber\\
&-& 2 i \del^\m \varphi \del^\r \varphi \dot{\varphi}^2 I (k_\m  k_\r )
 \eqae
In $S_{II}$ we find four different contractions, leading to
 \eqa
S_{II} &=& - {i\over 2} \left[    4 \dot{\varphi}^2 \del^\m \varphi \del^\n
\varphi I (k_\m k_\n) + 2 \dot{\varphi}^2 (\del \varphi)^2 I (k_\n k^\n)
\right.  \nonumber\\
 &+&  \left. 8 \dot{\varphi} \del^\m \varphi (\del \varphi)^2  I
(k_0 k_\m) + (\del \varphi)^4 I (k_0, k_0) \right]
 \eqae
Finally, in (III) there are two different contractions, yielding
 \eq
S_{III} = - i (\del \varphi)^2 \dot{\varphi}^2 I (k_\m k^\m) + 2 i (\del
\varphi)^2 \dot{\varphi}^2 I (k_0^2 )
 \eqe
We tabulate the results by decomposing each $(\del_\m \varphi )^2$ into
$(\nabla \varphi)^2 - \dot{\varphi}^2$, and for notational simplicity we write
$\vec{k}^2$ instead of $I (\vec{k}^2)$, and $k_0^2$ instead of $I (k_0^2)$.
We find then the following extra terms
  \eqa
\begin{array}{lcccccc}
& (\nabla \varphi)^4 \vec{k}^2 & (\nabla \varphi)^4 k_0^2 & (\nabla \varphi)^2
\dot{\varphi}^2 \vec{k}^2 & (\nabla \varphi)^2 \dot{\varphi}^2 k_0^2 &
\dot{\varphi}^4 \vec{k}^2 & \dot{\varphi}^4 k_0^2  \\
(I): & 0 & -1/2 & -2/3 & 1+2 & 0 & -1/2-2-2 \\(III): & 0 & 0 & -1 & 3 & 1 &
-3 \\ (II):
& 0 & -1/2 & -2/3-1 & 1+4+1 & 1 & -2-1-4-1/2
 \end{array}
  \eqae
Since the sum of (I) and (III) is equal to (II), we conclude that Matthews'
theorem is satisfied in this example.  Note that this conclusion is not based
on a particular regularization scheme;  rather the cancellation is algebraic.

Finally we must account for the terms with $[\d^4 (x-y)]^2$ in (\ref{o170}).
In the Lagrangian approach, one integrates out the momenta by a
saddle-point method.  This means that one must evaluate ${\del^2
\over \del p^2}  (p \dot{q} - H (p , q )) = - {\del^2 \over \del p^2}  H
(p, q) = - {\del^2 \over \del p^2} ( p \dot{q} (p) - L (q, \dot{q} (p))).$
Using ${\del \over \del p} L (q, \dot{q} (p)) = p {\del \dot{q}\over \del
p} $ one finds
$${\del^2 \over \del p^2} \left[ p \dot{q} - H (p, q) \right]  = - {\del
\dot{q}\over \del p} = - {1\over {\del p\over \del \dot{q}}} = - {1\over
{\del^2 L\over \del \dot{q}^2}} $$
Hence the $a,b,c$  ghosts one needs to add are  given by
$$
L ({\rm ghosts}) = b {\del^2 L \over (\del \dot{q})^2} \; c + a {\del^2 L
\over (\del \dot{q})^2} \; a $$
In our case this yields
\eq \label{y00a}
 L ({\rm ghosts}) = b \left(1+\l \left\{ (\nabla \varphi)^2 - 3
\dot{\varphi}^2 \right\} \right) c + \; {\rm  same \;  with} \; a.
\eqe
The $\d^4 (x-y)^2$ terms from (\ref{o170}) are easily seen to be proportional
to $\left[ (\nabla \varphi)^2 - 3 \dot{\varphi}^2 \right]^2$, and the
ghost loops cancel these singular terms since the $bc$ and $aa$
propagators are proportional to $\d^4 (x-y)$.

This concludes our discussion of this example and
of Matthews' theorem, which states that the
Hamiltonian and Lagrangian approaches to perturbative quantum field
theory are equivalent.  We used it to test our rules in a one-dimensional and
a four-dimensional model.  Crucial was the correct definition of equal-time
contractions.  It is often stated that equal-time contractions are ill-defined,
but if one wants to compute loop corrections in non-linear $\s$ models, one
must deal with them.  Equal-time contractions are, in fact, already needed
in a much better known area: the Ward identity for the self-energy of charged
massive scalars coupled to photons are only satisfied if one includes
equal-time contractions.  These seagull-graphs can be computed with
dimensional regularization and are nonvanishing, but since they correspond
to $< q (x) q (x) >$, they are the same in Hamiltonian and Lagrangian
formalism and are the limit of $x$ tending to $y$ of $< q (x) q (y) >$.  The
equal-time contractions we considered included $< \dot{q} (x) \dot{q}(x) >$
 and $< \dot{q} (x) q (x) >$, and these are not the same as obtained from the
limit $y$ tending to $x$.

\addtocounter{footnote}{1}
\subsection{The covariant spin-$3/2$ Jacobian${}^{15}$ }
\footnotetext{These results were obtained in collaboration with R. Endo.}

In order to obtain a covariant expression for the spin-$3/2$
transformation rule under space-time
transformations, and as a consequence a covariant expression for the
corresponding Jacobian, we must take certain linear combinations of
Einstein transformations and local Lorentz transformations.  For
spin-$1/2$ fields, this is easy and well-known, as we now show.  Afterwards
we consider the more complicated case of spin-$3/2$ fields.  To wet the
appetite of the reader for this problem, we first quote the final
results, which were obtained by Fujikawa for spin-$1/2$ \cite{Fuji},
used by Alvarez-Gaum\'e and Witten \cite{alwi} and further studied in
\cite{Fuji2,Endo}
\eqa
{\rm spin} \; 1/2: \;\; \;  \d_{AW}   \tilde{\j} &=& \c^\m D_\m \tilde{\j}
\;  + \; \half (D_\m \c^\m) \tilde{\j} \nonumber\\
 D_\m \tilde{\j}^a &=& \del_\m \tilde{\j}^a + {\scriptstyle 1\over 4}
\; \o_\m{}^{mn} (\g_m \g_n)^a{}_b \tilde{\j}^b - \half \G_{\m\n}{}^\n
\tilde{\j}^a \nonumber\\
{\rm spin} \; 3/2: \d_{AW}  \; \tilde{\j}_m &=& \c^\m D_\m \tilde{\j}_m
+ \half (D_\m \c^\m) \tilde{\j}_m + [(D_m \c^n) - (D^n \c_m) ]
\tilde{\j}_n \nonumber\\
 D_m \c^n &=&  e_m{}^\m (\del_\m \c^n  + \o_\m{}^n{}_p
\c^p ) , \c^n = e_\m{}^n \c^\m\nonumber\\
D_\m \tilde{\j}_m &=& \del_\m \tilde{\j}_m + \o_{\m m}{}^n
\tilde{\j}_n + {\scriptstyle 1\over 4} \o_\m{}^{pq} \g_p \g_q \tilde{\j}_m
-  \half \G_{\m\n}{}^\n \tilde{\j}_m
\label{five1}
\eqae
The fields $\tj$ and $\tj_m$ are world-scalar {\it densities} of weight
1/2, namely $\tj = g^{1/4} \j$ and $\tj_m = g^{1/4} \j_m$ with $\j_m
= e_m{}^\m \j_\m$.  Thus the covariant derivatives contain a term
$\G_{\m\n}{}^\n$. Further the spin-$3/2$ field $\j_m$, the so-called
gravitino field, is a Lorentz vector-spinor.
This explains the term $\o_{\m m}{}^n \tilde{\j}_n$ in its covariant
derivative.

The last two terms in the spin-$3/2$ transformation rule describe a local
Lorentz
transformation which acts on the vector indices with parameter
$D_{[\m}\c_{n]}$, whereas the first two terms contain a local Lorentz
transformation which acts on the spinor indices with parameter
$\c^\m \omega_{\mu}{}^m{}_n$. At first sight these rules do not
seem to correspond to a linear combination of the usual transformations.
We now proceed to demystify these expressions.  We begin with the
easier case of spin-$1/2$.

For spin-$1/2$, we define Einstein, local Lorentz and ``covariant"
transformations by
\eqa
\d_E (\c^\m)\tj &=& \c^\m \del_\m \tj + \half (\del_\m \c^\m) \tj
\nonumber\\
\d_{\ell L} (\l^{mn}) \tj &=& {\scriptstyle 1\over 4} \l^{mn} \g_m\g_n \tj
\nonumber\\
\d_{cov} (\c^\m) &=& \d_E (\c^{\mu}) + \d_{\ell L} ( \c^\m \o_\m{}^{mn})
\label{five1a}
\eqae
Thus for the  spinor  field $\tj$  one has
\eqa
\d_{cov} (\c^\m) \tj &=& \c^\m \del_\m \tj + \half (\del_\m \c^\m)
\tj + {\scriptstyle 1\over 4} \c^\m \o_\m{}^{mn} \g_m \g_n \tj
\nonumber\\
&=& \c^\m D_\m \tj + \half (D_\m \c^\m) \tj
\label{five2}
\eqae
In order to obtain a covariant result, we have added in the last line
two terms with Christoffel symbols whose sum cancels, as follows
from
\eqa
D_\m \c^\m &=& \del_\m \c^\m + \G_{\m\n}{}^\m \c^\n \nonumber\\
D_\m \tj &=&  \del_\m \tj + {\scriptstyle 1\over 4} \o_\m{}^{mn} \g_m
\g_n \tj - \half \G_{\m\n}{}^\n \tj
\label{five3}
\eqae

We can, of course, add further covariant terms.  A particular
combination which will play a role is
\eqa
\d_{sym} (\c^\m ) &\equiv& \d_{cov} (\c^\m) + \d_{\ell L}
(D_{[ m}\c_{n]}  ) \nonumber\\
D_{[_m} \c_{n]} &=& \half \left[ \left(D_m \c_n\right) - \left( D_n \c_m
\right) \right]
\label{five4}
\eqae
For the spinor field $\tilde{\j}$ we then obtain
\eq
\d_{sym} \left( \c^\m \right) \tilde{\j} = \c^\m D_\m \tilde{\j} + \half
(D_\m \c^\m) \tj + {1\over 4} {\left[ \left(D^m  \c^n \right)
- \left( D^n \c^m \right) \right] \over 2} \g_m \g_n \tilde{\j}
\label{five5}
\eqe

To illustrate where this particular combination of Einstein and Lorentz
transformations comes from, we evaluate $\d_{sym} $ on the vielbein.
First we compute $\d_{cov}$ on the vielbein
\eqa
\d_{cov} (\c^\n) e_m{}^\m &=& \c^\n \del_\n e_m{}^\m - (\del_\n
\c^\m ) e_m{}^\n + \c^\n \o_{\n m}{}^n e_n{}^\m \nonumber\\
= - \del_m \c^\m &+& \c^\n (\del_\n e_m{}^\m + \o_{\n m}{}^n
e_n{}^\m )  \nonumber\\
= - \del_m \c^\m &+& \c^\n (-\G_{\n\r}{}^\m  e_m{}^\r) \; \equiv -
D_m \c^\m
\label{five6}
\eqae
where we defined $\del_m = e_m{}^\n \del_\n$ and used the vielbein
postulate.  The transformation law defined in (\ref{five4}) then yields
 \eqa
\d_{sym} (\c^\n) e_m{}^\m &=& - D_m \c^\m + \half (D_m \c^n - D^n
\c_m) e_n{}^\m \nonumber\\
&=& - \half (D_m \c^\m + D^\m \c_m )
\label{five7}
\eqae
Hence, if the vielbein would have been symmetric in its two indices $m$
and $\m$ to begin with, then a transformation with $\d_{sym}$ keeps
that symmetry.  For this reason one might call the transformation
$\d_{sym}$ a ``symmetric Einstein transformation", namely an Einstein
transformation with parameter $\c^\m$ which, as far as it acts on the
vielbein, has been made symmetric by suitable local Lorentz transformations
whose parameter also depends on $\c^\m$.  For spin-$1/2$ fields, both
$\d_{cov}$ in
(\ref{five2}) and $\d_{sym}$ in (\ref{five5}) are manifestly covariant,
and at this
point it is not clear which one to choose
for the computation of gravitational anomalies.
As we shall see, $\d_{AW}$ is
actually
equivalent to both.  For spin-$3/2$, only one of them is covariant, as we
now discuss.

To understand the spin-$3/2$ transformation rule in (\ref{five1}),  we first
compute $\d_{cov}$ on a spin-$3/2$ field $\tilde{\j}_m$.  We find
\eqa
\d_{cov} \tj_m &=& \c^\m \del_\m \tj_m + \half (\del_\m \c^\m)
\tj_m + \c^\m \o_{\m m}{}^n \tj_n +{\scriptstyle 1\over 4} \c^\m
\o_\m{}^{pq} \g_p \g_q \tj_m\nonumber\\
&=& \c^\m D_\m \tj_m + \half (D_\m \c^\m)\tj_m
\label{five8}
\eqae
where again in the last line the two Christoffel symbols cancel.  Hence,
the proposed spin-$3/2$ law can be rewritten as
\eq
\d_{AW} \tj_m = \left[ \d_{cov} (\c^\m) + 2 \d_{\ell L}^{(1)} \left(D_{[m}
\c_{n]} \right) \right] \tj_m
\label{five9}
\eqe
where the superscript (1) indicates that this Lorentz transformation
only acts on the spin-$1$ (vector) index of $\tj_m$, but not on its spinor
index.  Compare this now with the symmetric spin-$1/2$ rule
\eq
\d_{sym} \tj = \left[ \d_{cov} (\c^\m) + \d_{\ell L}{}^{(1/2)} (D_{[m}
\c_{n]})\right] \tj
\label{five10}
\eqe
where the superscript $({1\over 2}) $ indicates that this Lorentz
transformation acts on the
spinor indices.
These rules for the spin-$1/2$ and spin-$3/2$ fields do not seem
to agree at all.  However, they are nevertheless equivalent, due to a
surprising identity found by Fujikawa, Toniya, Yasuda and Endo
\cite{Fuji,Endo}.  Namely, {\it
if one uses} $\rld \rld$  {\it as regulator} both for the spin-$1/2$ case and
for the spin-$3/2$ case, the anomalies due to $\half \d_{cov} (\c^\m)$ are
equal to minus those coming from $\d_{\ell L}{}^{(1/2)} (D_{[m}
\c_{n]}):$
\eq
\half \d_{cov} (\c^\m) \sim - \d_{\ell L}{}^{(1/2)} (D_{[m} \c_{n]} )
\label{five11}
\eqe
Using this equivalence, the proposed transformation rules for the
spin-$1/2$ and the spin-$3/2$ field in (\ref{five1}) turn out to be the same
combination of
Einstein and local Lorentz transformations, after all
\eqa
{\rm spin} \; 1/2: \d_{AW} &=& \d_{cov} = 2\d_{cov} + 2 \d_{\ell
L}{}^{1/2} (D \c ) = 2 \d_{sym} \nonumber\\
{\rm spin} \; 3/2: \d_{AW} &=& \d_{cov} + 2 \d_{\ell L}{}^{(1)} (D\c)
\nonumber\\
&=& 2\d_{cov} + 2 \d_{\ell L}{}^{1/2} (D\c) + 2 \d_{\ell L}{}^{(1)} (D\c)
\nonumber\\
&=& 2 \d_{sym}
\label{five12}
\eqae
Note the relative factor 2 between $\d_{cov}$ and $\d_{sym}$ \cite{Endo2}.

It remains to prove that the anomalies coming from $\half \d_{cov}
(\c)$ are the same as those from $-\d_{\ell L}{}^{(1)} (D\c)$.  We begin
with the spin-$1/2$
fields.  We take $\tj$ and $\tj_m$ to be left-handed, whereas the fields
$\tilde{\bar{\j}}$ and $\tilde{\bar{\j}}_m$ are independent
right-handed fields in Euclidean space.  (So $\bar{\j}$ is not
$\j^\dagger$ in Euclidean space.   In Minkowski space $\bar{\j}=
i\j^\dagger \g^0$).  We expand the spin-$1/2$ fields as follows:
\eq
\tj = \sum_{n } a_n \; \tj_{n, L} + \sum_\a c_\a
\tilde{\chi}_{\a, L} \quad ; \quad \tilde{\bar{\j}} = \sum_{n}  b_n \;
\tj_{n, R}{}^\dagger  + \sum_{\b} d_\b \tilde{\z}_{\b, R}{}^\dagger
\label{five13}
\eqe
where $\tilde{\j}_n = g^{1/4} \j_n$ and $\j_n$ are an orthonormal set of
solutions of the Dirac equation with non-vanishing eigenvalue
\eqa
&& i \rlap{\,/}D \j_n = \l_n \j_n \quad ; \quad \j_{n, L} = {1+\g_5 \over
\sqrt{2}} \j_n \quad ; \quad \j_{n, R} = {1-\g_5 \over \sqrt{2}} \j_n
\nonumber\\
&& \int \tj_m{}^\dagger (x) \tj_n (x) dx = \d_{mn} \; , \; i \tilde{\rld}
\tj_n = \l_n \tj_n \; , \; \tilde{\rld} = g^{1/4} \rld g^{-1/4}
\label{five14}
\eqae
Of course, the chiral anomaly is equal to the number of $\c$'s minus
the number of $\z$'s, but we will rewrite this expression in a way where
the difference between $\j$, $\c$ and $\z$ disappears.

For every eigenfunction $\j_n$ of the Dirac equation with
nonvanishing eigenvalue $\l_n$, there is another one, $\g_5
\j_n$, with eigenvalue $-\l_n$.  Hence, the ``massive modes"
come in pairs, which can be combined on a chiral basis into
$\j_{n, L}$  and $\j_{n, R}$, and
\eq
\int \tilde{\j}_{mL}{}^\dagger \tilde{\j}_{nR}  dx = 0 \; , \; \int
\tilde{\j}_{mL}{}^\dagger  \tilde{\j}_{nL} dx  = \int
\tilde{\j}_{mR}{}^\dagger
\tilde{\j}_{nR}  d x = \d_{mn}
\label{five15}
\eqe
where we used the orthogonality of $\j_m$ and $\g_5 \j_n$.
(They are orthogonal because they are eigenfunction with different
eigenvalues).   In addition there are some zero modes, solutions of
$\rlap{\,/}D \j = 0$.  These we can always take to be left- and/or
right-handed.  They also appear in the expansion of $\tj$ and
$\tilde{\bar{\j}}$, respectively, but they need not come in pairs.  We
have denoted them by $\chi_{\a, L}$ and $\z_{\b, R}$ in (\ref{five13}).

The Jacobian $J$ for an infinitesimal transformation $\d \tj$ and $\d
\tilde{\bar{\j}}$ follows from
\eq
\d a_n  = \int dx \tj_{n, L}{}^\dagger \d \tj \; ,\; \d
b_n = \int dx \d \tilde{\bar{\j}} \tj_{n, R}
\label{five16}
\eqe
and similarly for $\d c_\a$ and $\d d_\b$.
For $\d_{AW}$ we obtain then the following Jacobian
\eqa
J_{AW} - 1 = &-& \int dx \left\{ \sum_n \tj_{nL}{}^\dagger \bigg[
\c^\m D_\m + \half (D_\m \c^\m ) \bigg] \tj_{n,L}
\right. \nonumber \\
&&+ \sum_\a \tilde{\chi}_{\a L}{}^\dagger \bigg[ \c^\m D_\m +
\half \bigg( D_\m \c^\m \bigg) \bigg] \tilde{\chi}_{\a L}
\nonumber\\
&&+ \sum_n \tilde{\j}_{n, R}{}^\dagger \left[
\stackrel{\leftarrow}{D}_\m \c^\m + \half (D_\m \c^\m)
\right] \tilde{\j}_{n, R}
\nonumber\\
&& + \sum_\b \tilde{\z}_{\b, R}{}^\dagger
\bigg[ {\stackrel{\leftarrow}{D}_\m} \c^\m + \half (D_\m
\c^\m ) \bigg] \tilde{\z}_{\b, R}  \bigg\}
\label{five17}
 \eqae
where $\j \stackrel{\leftarrow}{D}_\m = \del_\m \j^\dagger - {1\over 4}
\j^\dagger \o_\m{}^{mn} \g_m \g_n$.  The minus sign always
appears in Jacobians for fermions.  We now rewrite the massive modes
in terms of $\tilde{\j}_n$ and $\g_5 \tilde{\j}_n$, but the
zero modes we keep as they appear, except that we add a factor
$\g_5$ or $-\g_5$ to the left-handed or right-handed zero modes,
respectively.  (These factors $\g_5$ or $-\g_5$ equal unity, of
course).  This yields
\eqa
J_{AW} -1 = &-& \int dx \bigg\{ \sum_n \tilde{\j}_n{}^\dagger
\left[ \c^\m D_\m + \stackrel{\leftarrow}{D}_\m \c^\m +
(D_\m \c^\m ) \right] \tilde{\j}_n \nonumber\\
&& + \sum_n \tilde{\j}_n{}^\dagger \g_5 \left[ \c^\m D_\m
- \stackrel{\leftarrow}{D}_\m \c^\m \right] \tilde{\j}_n
\bigg\}\nonumber\\
&-& \int dx \left\{ \sum_\a \tilde{\chi}_{\a L}{}^\dagger
\left[ \c^\m D_\m + \half (D_\m \c^\m) \right] \g_5
\tilde{\chi}_{\a L} \right. \nonumber\\
&& +\sum_{\b} \tilde{\z}_{\b R}{}^\dagger \left[
\stackrel{\leftarrow}{D}_\m \c^\m + \half (D_\m \c^\m ) \right]
\g_5 \tilde{\z}_{\b R} \bigg\}
\label{five17-2}
\eqae
Partially integrating, all massive modes without factor $\g_5$ cancel,
while the rest yields
\be
\begin{array}{l}
J_{AW} - 1 = - \int dx \bigg[ \sum_n \tilde{\j}_n{}^\dagger
\g_5 \left[ 2 \c^\m D_\m + (D_\m \c^\m ) \right] \tj_n
\\
+ \sum_{\a} \tilde{\chi}_{\a L}{}^\dagger  \g_5 \left[ \c^\m D_\m +
\half \left(D_\m \c^\m \right)\right] \chi_{\a L} + \sum_\b \tilde{\z}_{\b
R}{}^\dagger \g_5  \left[ \c^\m D_\m + \half (D_\m
\c^\m ) \right] \tilde{\z}_{\b R} \bigg]
\end{array}
\label{five18}
\ee
We would like to recognize this as a sum over a complete set of
states.  At first sight there is a mismatch in the first term which
seems a factor 2 too large
with respect to the other terms.  However, the
complete set of massive solutions contains not only $\tj_n$ but also
$\g_5 \tj_n$, and rewriting $2 \tj_n \g_5 \tj_n$ for the massive
modes as $\tj_n \g_5 \tj_n + \tj_{-n} \g_5 \tj_{-n}$ with $\tj_{-n}
\equiv \g_5 \tj_n$ by definition, we arrive at
\eq
J_{AW} - 1 = - \int dx \sum_{N} \tilde{\varphi}_N{}^\dagger \g_5 \bigg[ \c^\m
D_\m + \half (D_\m \c^\m ) \bigg]  \tilde{\varphi}_N
\label{five19}
\eqe
where $\tilde{\varphi}_ N$ is the complete set of eigenfunctions, so that
$N$ runs over
all eigenfunctions:  $n>0$, $n<0$, and all zero modes.

To regulate the infinite sum over $N$, we add to each term a factor
$\exp (- \l^2_N/M^2)$ and will let $M$ tend to infinity at the end.  Using
\eq
e^{-\l^2_N/M^2} \varphi_N = e^{\rlap{\,/}D \rlap{\,/}D/M^2}
\varphi_N
\label{five20}
\eqe
we obtain
\eq
J_{AW} - 1 = - \int dx \sum_N \tv_N{}^\dagger \g_5 \bigg[ \c^\m
D_\m + \half (D_\m \c^\m ) \bigg] e^{\tilde{\rld} \tilde{\rld}/M^2} \tv_N
\label{five21}
\eqe
where we recall the definition $\tilde{\rld} = g^{1/4} \rld g^{-1/4}$,
with $\Dsl$ without $\Gamma_{\m\n}{}^{\n}$ term. All terms with
$\Gamma_{\m\n}{}^{\n}$ cancel in (\ref{five21}), so from now on no
explicit $\Gamma_{\m\n}{}^{\n}$ are present.

Since the set $\varphi_N$ is a complete set
$\sum_N \tv_N (x) \tv_N{}^\dagger (y) = \d (x-y)$, we have
%Using that $Tr J = \sum_N < N| J|N> = \sum_{N,x,y} <N|x><x|J|y> <y|N>$ where
%$<y|N> = \varphi_N (y), < N|x> = \varphi_N^\dagger (x)$ while $<x |J|y> =
%j(x) \d
%(x-y)$, %In general, the trace of a matrix $M(x,y)$ is defined by
%\eq
%Tr M (x,y) = \int M (x,y) \d (x-y) dxdy
%\label{five23}
%\eqe
%In our case, $M (x,y)$ equals $\del \d \tj (x) / \del \tj (y)
% = - \left[ \c^\m \td_\m + \half (\td_\m \c^\m ) \right] \d (x-y) ,$
%whose trace we already equated to $-\int dx \sum_N \tv_N{}^\dagger
%\g_5 \left[ \c^\m \td_\m + \half (\td_\m \c^\m) \right] \tv_N. $ We
%can thus identify the Jacobian matrix as
%\eq
%M (x,y) \; = \; - \sum_N \tv_N{}^\dagger (y) \g_5 \left[ \c^\m \td_\m +
%\half (\td_\m \c^\m ) \right] \tv_N (x) \; ,
%\label{five24}
%\qe
%and to take a regulated trace we multiply this by
%\eq
%\d_{\rm reg} (x-y) = \sum_N e^{-\l_N{}^2/M^2} \tv_N (x)
%\tv_N{}^\dagger (y)
%\label{five25}
%\eqe
%we can then rewrite the Jacobian as a trace
\eqa
J_{AW} -1 &=& -{\rm Tr}\left( \g_5  {1\over 2}  ( \c^\m D_\m + D_\m \c^\m )
e^{\tilde{\rld}  \tilde{\rld} / M^2} \right)
\nonumber\\
D_\m &=& \del_\m + {1\over4} \o_\m{}^{mn} \g_m \g_n
\label{five26}
\eqae
where the trace $\Tr$ indicates integrating over space and summing
over spinor indices.

This trace has been evaluated using plane waves. Due to the $g^{\pm 1/4}$
in $\tilde{\Dsl}$, the Einstein anomaly indeed did cancel \cite{ref54},
see below (\ref{bb103}).
However, by using the cyclicity of the trace we can rewrite this as
\be J_{AW}-1=-{\rm Tr}(\half\gamma_5(
\chi^{\m} g^{-1/4} D_{\m} g^{1/4} + g^{-1/4} D_{\m} g^{1/4} \chi^{\m})
e^{{\rld}  {\rld} / M^2} )
\ee
In the corresponding non-linear sigma model this becomes
\be J_{AW}-1=\frac{1}{2\pi \hbar} {\rm Tr}(\gamma_5(
\chi^i \pi_i + \pi_i \chi^i) \exp(-\b\hat{H}/\hbar) )
\ee
with $\hat{H}$ given in (\ref{aaa105}) and $\pi_i=p_i-\frac{\hbar i}{2}
\omega_{iab}
\j^a \j^b$. Note that the Jacobian is Weyl-ordered. One can from here
compute the gravitational anomalies.

For local Lorentz transformations,
\eq
\d \tj = {\scriptstyle 1\over 4} \l^{mn} \g_m \g_n \tj \;  , \;
\d \tilde{\bar{\j}} = - { 1\over 4} \l^{mn} \tilde{\bar{\j}}  \g_m \g_n
\label{five28}
\eqe
one may proceed in a similar manner.  One finds then for chiral spinors
$\tj$ the following Lorentz anomaly
 \eq
J_{\ell L} -1 = - \Tr \left(\g_5
 \left[ {\scriptstyle 1\over 4} \l^{mn} \g_m \g_n
\right]
e^{\tilde{\rld} \tilde{\rld} /M^2} \right)
\label{five29}
\eqe
Using cyclicity of the trace, all factors of $g^{1/4}$ and $g^{-1/4}$
now cancel, and one obtains
\be
J_{\ell L}-1= - {\rm Tr} (\frac{1}{4} \g_5 \lambda^{mn}
\g_m \g_n  \exp(-\b \hat{H}/\hbar) )
\ee

Now that we have found expressions for the Einstein and Lorentz
anomalies, we can prove the relation $\d_{cov} \sim - \half \d_{\ell
L}$ for the spin-$1/2$ fields.  We use the identity
\eq
\int \Tr \left( \g_5 \bigg(\rlap{\,/}\c \rld + \rld \rlap{\,/}\c
\bigg)
e^{\tilde{\rld} \tilde{\rld} /M^2} \right)
= 0
\label{five30}
\eqe
which  follows from cyclicity of the trace  and $\rld \g_5 = - \g_5 \rld$.
Since the
difference between $\rld \rld$ as regulator and $\trld \trld$ involves
terms which
are proportional to $M^{-2}$ in the path integral, and only the $M$-independent
terms yield  the anomaly while there are no singular terms in $M$ for chiral
anomalies, we may replace $\trld \trld$ by $\rld \rld$ in (\ref{five30}).
Then we
can indeed use the cyclicity of the trace.  (Chiral anomalies are rather
insensitive
to the details of the regulator.  For trace anomalies, such details do
matter, but for
trace anomalies fortunately we do not need (\ref{five30})). Hence
 \eqa 0 &=& \int \Tr \left( \g_5 \bigg( \c^\m D_\m + D_\m \c^\m +
\g^{[\m} \g^{\n]} \left( \c_\m D_\n + D_\m \c_\n \right) \bigg)
e^{{\rld} {\rld} /M^2}\right) dx \nonumber\\
&& = \int \Tr \left( \bigg[
\g_5 \left( \c^\m D_\m + D_\m \c^\m \right) + \g_5
\g^{[\m} \g^{\n]}   \left( \c_\m D_\n - D_\n \c_\m\right) \bigg]
e^{{\rld} {\rld} /M^2}\right)  dx \nonumber\\
&& = \int \Tr \left( \bigg[
\g_5 \left( \c^\m D_\m + D_\m \c^\m \right) + \g_5
\g^{[\m} \g^{\n ]} (D_\m \c_\n) \bigg]
e^{{\rld} {\rld} /M^2}\right)  dx \nonumber\\
&& =  - 2 {\cal A}_{n,cov} (\c^{\m} ) - 4
{\cal A}_{n,\ell L} (D_{[m} \c_{n]} )
\label{five31}
 \eqae
This concludes the proof of the identity in (\ref{five11}) for spin-$1/2$.

For spin-$3/2$ we use the same identity in (\ref{five30}), and all steps in
(\ref{five31}) are
the same.  Note that the local Lorentz transformations with $D_{[m}
\c_{n]}$ only act on the spin-$1/2$ indices of the spin-$3/2$ field.  Thus
(\ref{five11}) is also proven for spin-$3/2$.   The regulator $\tilde{\rld}
\tilde{\rld}$ now acts in a
combined spinor-vector space, but it is diagonal in the vector indices,
i.e., it acts on $\j_m$ the same for all $m$.

To see whether this regulator is obtained from the gauge-fixed
spin-$3/2$ action, we note that the spin-$3/2$ action in $d$ dimension reads
\eq
\cl_0 = - {1\over 2} \bar{\j}_\m \g^{[\m} \g^\n \g^{\r ]} \; D_\n \j_\r
\label{five31a}
\eqe
After adding a gauge fixing term \cite{Nieu}
\eq
\cl_{\rm fix} = {d-2\over 8} \bar{\j}_\m \g^\m \g^\n \g^\r D_\n \j_\r
\label{five32}
\eqe
and choosing a new basis for the spin-$3/2$ fields
\eq
\chi_\m = \j_\m - {1\over 2} \g_\m \g \cdot \j
\label{five33}
\eqe
the action becomes a sum of Dirac actions \cite{Endo2}
\eq
\cl_0 + \cl_{\rm fix} = \bar{\chi}_\m \rld \chi^\m = \bar{\chi}_m
\rld \chi^m
\label{five34}
\eqe
The field $\chi_m$ transforms of course in the same way as $\j_m$
under space-time transformations and the regulator for the spin-$3/2$ field
$\chi_m$ is thus
${\rld} {\rld}$, for the same reasons as for the spin-$1/2$ field.

\subsection{Chiral anomalies in gravitational couplings from Feynman diagrams
in quantum mechanics}

In this appendix we will derive the chiral anomaly in
gravitational couplings using Feynman diagrams and the Feynman rules
developed in the text. Starting point
will be equations (\ref{aaa118}) and (\ref{aaa121}) derived
in section 3.1. As explained there, we only have to consider
one-loop diagrams. An arbitrary connected one-loop
diagram consists of $k$ vertices from (\ref{aaa121}) and
$k$ propagators. There are in principal many different ways to
contract the vertices with each other, but using partial
integration (in this case one can verify it is allowed)
and the anti-symmetry of $R_{ijab}$ in the
indices $i,j$, one can always move the time derivative in
(\ref{aaa121}) from $q^j$ to $q^i$. This implies that all
different contractions of the vertices yield the same
contribution. The total number of contractions is
$2^{k-1} (k-1)!$, which combines with the factor $1/k!$
from expanding (\ref{aaa121}) in the total symmetry factor
$2^{k-1}/k$ for each diagram. If we always contract a $\dot{q}$
from one vertex with a $q$ from another vertex, we find
the following expression for $S_{\rm loops}$,
the set of connected one-loop diagrams
\be
-\frac{1}{\hbar} S_{\rm loops} =
\sum_{k=1}^{\infty} \frac{2^{k-1}}{k} \left(
\frac{-R}{4\b} \right)^k (-\b\hbar)^k a_k
\ee
where $R^k$ is $R_{i_1}^{i_2} R_{i_2}^{i_3} \ldots
R_{i_k}^{i_1}$ with $R^i_j=g^{ik}(x_0) R_{kjab}(\omega(x_0))
\j_{\rm bg}^a \j_{\rm bg}^b $, and $a_k$ denotes the
integral
\be \label{aaa230}
a_k = \int_{-1}^0 d\t_1 \ldots \int_{-1}^0 d\t_k
(\t_2+\theta(\t_1-\t_2))\cdots
(\t_k+\theta(\t_{k-1}-\t_k))
(\t_1+\theta(\t_k-\t_1))
\ee
This expression is cyclically symmetric.
To proceed, we compute $\sum_{k=1}^{\infty} \frac{y^k}{k} a_k$. This can
be rewritten as
\bea \label{a4e1}
&& \frac{y}{2}+\int_{0}^1 \frac{d\a}{\a} \left( \a \frac{d}{d\a} \right)
\left( \sum_{k=1}^{\infty} \frac{y^k}{k}
 \int_{-1}^0 d\t_1 \ldots \int_{-1}^0 d\t_k \right. \nonumber \\ & &
\left. \qquad \times
(\a\t_2+\theta(\t_1-\t_2))\cdots
(\a\t_k+\theta(\t_{k-1}-\t_k))
(\a\t_1+\theta(\t_k-\t_1)) \right) .
\eea
Here, we introduced a parameter $\a$ in the expression for
$a_k$, which will turn out to make things much simpler.
Notice that if we put this parameter equal to zero,
we are left with an integral over a product of theta functions,
which vanishes, except when $k=1$, and then it equals $1/2$ in
view of $\theta(0)=1/2$, and this explains the first
term $y/2$ in (\ref{a4e1}). The derivative $d/d\a$ in
(\ref{a4e1}) can act on any of the $k$ factors in the integral,
but by cyclic symmetry these all give the same contribution
and we choose it to hit only the last factor and put an extra
factor of $k$ in front. This yields
\be \label{a4e2}
\frac{y}{2}+\int_{0}^1 \frac{d\a}{\a}
\left( \sum_{k=1}^{\infty} y^k
 \int_{-1}^0 d\t_1 \ldots \int_{-1}^0 d\t_k
(\a\t_1) (\a\t_2+\theta(\t_1-\t_2))\cdots
(\a\t_k+\theta(\t_{k-1}-\t_k))
 \right).
\ee
Note that we now have broken the cyclic symmetry of (\ref{aaa230}).
If we expand the brackets inside the integral, we get a large sum
of terms, each of which factors into a product of integrals of the
form
\be \label{bb002}
 \int_{-1}^0 d\t_m \ldots \int_{-1}^0 d\t_{m+l}
(\a\t_m)
\theta(\t_{m}-\t_{m+1})
\cdots
\theta(\t_{m+l-1}-\t_{m+l}).
\ee
For $l=0$, the integrand is just $\alpha \t_m$. This integral depends only
on $l$, enabling us to write (\ref{a4e2}) as
\be \label{a4e3}
\frac{y}{2}+\int_{0}^1 \frac{d\a}{\a}
\left( \sum_{p=1}^{\infty}\left[ \sum_{k=1}^{\infty} y^k
 \int_{-1}^0 d\t_1 \ldots \int_{-1}^0 d\t_k
(\a\t_1) \theta(\t_1-\t_2)\cdots
\theta(\t_{k-1}-\t_k) \right]^p
 \right).
\ee
The integral in (\ref{bb002}) equals
$-\a/(l+2)!$, and we find
\bea
\sum_{k=1}^{\infty} \frac{y^k}{k} a_k & = &
\frac{y}{2}+\int_{0}^1 \frac{d\a}{\a}
\left( \sum_{p=1}^{\infty}\left[ \sum_{k=1}^{\infty} y^k
\frac{-\a}{(k+1)!}
\right]^p
 \right) . \nonumber \\
 & = &
\frac{y}{2}+\int_{0}^1 \frac{d\a}{\a}
\left( \sum_{p=1}^{\infty}\left[
\frac{-\a}{y}(e^y-1-y)
\right]^p
 \right). \nonumber \\
 & = &
\frac{y}{2}+\int_{0}^1 \frac{d\a}{\a}
\left(
\frac{1}{1+
\frac{\a}{y}(e^y-1-y)}-1
 \right). \nonumber \\
& = & \log \left( \frac{y/2}{\sinh(y/2)} \right)
\eea
This shows that
\be
\frac{-1}{\hbar} S_{\rm loops} =
\frac{1}{2} \log \left(
 \frac{\hbar R/4}{\sinh(\hbar R/4)} \right)
\ee
Thus, after a rescaling of the fermions, we find that the
anomaly is given by
\be \label{ahat}
{\cal A}_n = \int \prod_{i=1}^n dx_0^i
\sqrt{g(x_0)} \int \prod_{a=1}^n d\j^a_{\rm bg}
\exp \left[ \frac{1}{2} \log \left(
 \frac{-iR/8\pi}{\sinh(-iR/8\pi)} \right)
 \right]
\ee
Clearly, there is only a gravitational $\g_5$ anomaly in $n=4k$
dimensions.
In principle, this expression is ambiguous, since it is
not clear whether $R^4$ means $\tr(R^2)\tr(R^2)$ or
$\tr(R^4)$. However, the precise meaning follows from the
derivation above. First one has to write down a Taylor
series for the logarithm, then one has to replace $R^m$
by $\tr(R^m)$ everywhere, and only then should one take
the exponential. The advantage of writing the anomaly
in the form (\ref{ahat}) is that it is given directly
in terms of $R$, rather than in terms of its eigenvalues \cite{alwi}.


\begin{thebibliography}{99}

\bibitem{paper1} J. de Boer, B. Peeters, K. Skenderis and P. van Nieuwenhuizen,
hep-th/9504097, {\it Nucl. Phys.} {\bf B446} (1995) 211.

\bibitem{alwi} L. Alvarez-Gaum\'e and E. Witten, {\it Nucl. Phys.} {\bf B234}
 (1989) 269; L. Alvarez-Gaum\'e, in `Supersymmetry', eds. K. Dietz
 et al., Plenum Press, 1984.

\bibitem{witt}  B. DeWitt, `Supermanifolds', 2nd edition, Cambridge
 University Press, 1992. Sections (6.5), (6.6) give a non-lattice
derivation. To deal with difficulties due to products of momentum
operators at different times, DeWitt proposes to use quasi-Cartesian
frames (his eq. (6.5.15)).

\bibitem{sch} L. Schulman, `Techniques and applications of path
 integration', John Wiley and Sons, New York, 1981. Chapter
 24 gives a review.

\bibitem{Dirac} P.M. Dirac, {\it Phys. Z. Sovjetunion} {\bf 3} (1933) 64.

\bibitem{Feyn} R. Feynman, \RMP{20}{1948}{367}; {\it Phys. Rev.}
 {\bf 80} (1950) 440.

\bibitem{witt2} B. DeWitt,
 {\it Rev. Mod. Phys.} {\bf 29} (1957) 377.

\bibitem{DeWia} B. DeWitt, {\it in} ``Relativity, Groups and Topology'',
eds. C. DeWitt and B. DeWitt, Les Houches 1963; {\it Phys. Rep.} {\bf 19}
(1975) 295.

\bibitem{bapvn}
F. Bastianelli, {\it Nucl. Phys.} {\bf B376} (1992) 113;
F. Bastianelli and P. van Nieuwenhuizen, {\it Nucl. Phys.}
 {\bf B389} (1993) 53.

\bibitem{LeeYang} T.D. Lee and C.N. Yang, {\it Phys. Rev. }
{\bf 128} (1962) 885. See also E.S. Abers and B.W. Lee,
{\it Phys. Rep.} {\bf 9C} (1973), page 67, eq. (11.23).

\bibitem{Trot} H. Trotter, \PAMS{10}{1959}{545}.

\bibitem{Gava} G. Gavazzi, {\it J. Math. Phys.} {\bf 30}, (1989) 2904.

\bibitem{schw} J. Schwinger, {\it Phys. Rev.} {\bf 127} (1962) 324;
 Phys. Rev. {\bf 130} (1963) 406.

\bibitem{geji} J.-L. Gervais and A. Jevicki, {\it Nucl. Phys.} {\bf 110}
 (1976) 93. See also S.F. Edwards and Y.V. Guluaev, {\it Proc. Roy. Soc.
London} {\bf A279} (1964) 269, and D.W. Laughlin and L.S. Schulman,
{\it J. Math. Phys.} {\bf 12} (1971) 2520.

\bibitem{chlee} N.H. Christ and T.D. Lee, {\it Phys. Rev.} {\bf D22}
(1980) 939.

\bibitem{Matt} P. T. Matthews, {\it Phys. Rev.} {\bf 76} (1949) 686;
Y. Nambu, {\it Progr. Theor. Phys.} {\bf 7}, 131 (1952).

\bibitem{Fuji} Fujikawa, {\it Phys. Rev. Lett.} {\bf 42} (1979) 1195;
{\it Phys. Rev.} {\bf D21} (1980) 2848, {\bf D22} (1980) 1499;
{\it Phys. Rev. Lett.} {\bf 44} (1989) 1733.

\bibitem{diaz} A. Diaz, W. Troost, P van Nieuwenhuizen and A. van Proeyen,
 {\it Int. J. Mod. Phys.} {\bf A4} (1989) 3959;
M. Hatsuda, P. van Nieuwenhuizen, W. Troost and
A. van Proeyen, {\it Nucl. Phys.
} {\bf B335} (1990) 166.

\bibitem{colo} J.F. Colombeau, {\it Bull. A.M.S.} {\bf 23} (1990) 251, and
J. F. Colombeau, `Multiplication of distributions',
Lecture Notes in Mathematics
1532, Springer-Verlag.

\bibitem{klau} J. Klauder, {\it Ann. Phys.} {\bf 11} (1960) 123.

\bibitem{textb} L. Faddeev and A. Slavnov, ``Gauge Fields: an
Introduction to Quantum Theory'', 2nd ed., Addison-Wesley, Redwood
City, 1991.

\bibitem{soho} P. Solomonson and J. van Holten, \NPB{196}{1982}{509}.

\bibitem{gra} R. Graham, {\it Z. Phys.} {\bf B26} (1977) 281.

\bibitem{bas} B. Peeters and P. van Nieuwenhuizen, `The Hamiltonian
 approach and phase space path integration for non-linear sigma models
 with and without fermions', ITP-SB-93-51, hep-th/9312147.

\bibitem{Vlec} J. Van Vleck, {\it Proc. Natl. Acad. Sci.} {\bf 24}
(1928) 178; C. Morette, {\it Phys. Rev.} {\bf 81} (1951) 848.

\bibitem{wey1} H. Weyl, `Theory of groups and quantum mechanics',
 Dover, New York, 1950.

\bibitem{miz} M. Mizrahi, {\it J. Math. Phys.} {\bf 16} (1975) 2201.

\bibitem{sato} M. Sato, {\it Progr. Theor. Phys.} {\bf 58} (1977) 1262.

\bibitem{wey2} For an elementary discussion of Weyl ordering, see e.g.
T.D. Lee, `Particle physics and introduction to field
 theory', Harwood Academic Publishers,
 New York, 1981 (Contemporary concepts in physics, 1).

\bibitem{omo2} M. Omote, {\it Nucl. Phys.} {\bf B120} (1977) 325.

\bibitem{ber} F.A. Berezin, {\it Theor. Math. Phys.} {\bf 6} (1971) 194.

\bibitem{lasch} D. McLaughlin and L.S. Schulman, {\it J. Math. Phys.}
 {\bf 12} (1971) 2520.

\bibitem{Deck} H. Decker, \Phy{\bf 103A} {1980}{586}.

\bibitem{ref4} see e.g. C. Itzykson and J.-B. Zuber, `Quantum field
 theory', McGraw-Hill, New York, 1980, page 136.

\bibitem{textb2} M. Henneaux and C. Teitelboim, `Quantization of
 gauge systems', Princeton University Press, Princeton, 1992.

\bibitem{AFFR} V. de Alfaro, S. Fubini, G. Furlan and M. Roncadelli,
\NPB{296}{1988}{402},
\PLB{200}{1988}{323};
 V. de Alfaro and M. Gavazzi, \NPB{335}{1990}{655}.

\bibitem{ref54} A. Ceresole, P. Pizzochero and P. van Nieuwenhuizen,
{\it Phys. Rev.} {\bf D39} (1989) 1567.

\bibitem{leuven} P. van Nieuwenhuizen, `Anomalies in Quantum Field Theory',
Leuven Notes in Mathematical and Theoretical Physics, series B, volume 3,
1989.

\bibitem{bern} C. Bernard and A. Duncan, {\it Phys. Rev. D} {\bf 11},
(1975) 848.

\bibitem{Fuji2}  K. Fujikawa, M. Toniya and O. Yasuda, {\it Z. Phys. C} {\bf
289} (1985) (this paper essentially contains the identity
$\d_{cov} \sim - {1\over2} \d_{\ell L}$.
It focuses mostly on the symmetric stress tensor).

\bibitem{Endo}  R. Endo, {\it Prog. Theor. Phys.} {\bf 78} (1987) 440, in
particular the footnote on page 448 (this paper discusses the
identity $\d_{cov} - \half \d_{\ell L}$ in detail).

\bibitem{Endo2} R. Endo and M. Takao, {\it Phys. Lett.} {\bf 161B} (1985),
155 (this paper obtains the Dirac action for spin-$3/2$ fields).

\bibitem{Nieu} P. van Nieuwenhuizen, {\it Phys. Rep.} {\bf 68}
(1981) 189.

\end{thebibliography}
\end{document}